%% file: main_arXiv.tex
\documentclass[10pt,journal,compsoc]{IEEEtran}

\ifCLASSOPTIONcompsoc
  \usepackage[nocompress]{cite}
\else
  \usepackage{cite}
\fi

\ifCLASSINFOpdf
\else
\fi

\usepackage[utf8]{inputenc} 
\usepackage[T1]{fontenc}    
\usepackage{url}            
\usepackage{booktabs}       
\usepackage{makecell}
\usepackage{multirow}
\usepackage{multicol}
\usepackage{amsfonts}       
\usepackage{amsmath}
\usepackage{amssymb}
\usepackage{pifont}
\usepackage{mathtools}
\usepackage{tabularx}
\usepackage{algorithm}
\usepackage{algorithmic}

\usepackage{xcolor}
\usepackage{graphicx}
\usepackage{graphbox}
\usepackage{subfigure}

\usepackage{nicefrac}       
\usepackage{paralist}
\usepackage[breaklinks=true,colorlinks,bookmarks=false,citecolor=black,linkcolor=black]{hyperref}

\include{macro}

\renewcommand{\textcolor}[2]{#2}
\newcommand{\revised}[1]{#1}

\begin{document}
%
\title{Toward Real-World Super-Resolution\\ via Adaptive Downsampling Models}
%
%
%
%


\input{sections/0_authors}


%
%

\markboth{IEEE Transactions on Pattern Analysis and Machine Intelligence}%
{Son \MakeLowercase{\textit{et al.}}: Toward Real-World Super-Resolution: Adaptive Data Loss for Downsampling Model}
%



\input{sections/0_abstract}

\maketitle
 
\IEEEdisplaynontitleabstractindextext

%
\IEEEpeerreviewmaketitle

\input{sections/1_intro_arxiv}
\input{sections/2_related}
\input{sections/3_method}
\input{sections/4_experiment}
\input{sections/5_conclusion}

\section*{Acknowledgement}
This work was partly supported by IITP grant funded by the Korea government [No. 2021-0-01343, Artificial Intelligence Graduate School Program (Seoul National University)].

\ifCLASSOPTIONcaptionsoff
  \newpage
\fi



%


{
    \small
    \bibliographystyle{ieeetr}
    \bibliography{egbib}
}

\input{bio/bio}

\input{sections/s_appendix}

\end{document}

%% file: macro.tex
\long\def\ignorethis#1{}

\newcommand{\etal}{\textit{et al.}}

\newcommand{\img}[1]{\mathbf{I}_{\text{#1}}}
\newcommand{\paren}[1]{\left( #1 \right)}

\newcommand{\normtwo}[1]{\left\lVert #1 \right\rVert_2^2}
\newcommand{\normone}[1]{\left\lVert #1 \right\rVert_1}

\newcommand{\Paragraph}[1]{\vspace{1mm}\noindent\textbf{#1}}

\newcommand{\fakeref}[1]{\textcolor{blue}{#1}}
\newcommand{\fakeeqref}[1]{\textcolor{blue}{(#1)}}

\newcommand{\ignore}[1]{}   

\newcommand{\tabspace}{\vspace{-3mm}}
\newcommand{\tabxspace}{\vspace{-5mm}}
\newcommand{\figspace}{\vspace{-3mm}}
\newcommand{\figxspace}{\vspace{-5mm}}
\newcommand{\best}[1]{\textbf{#1}}
\newcommand{\secondbest}[1]{\underline{#1}}

\DeclareMathOperator*{\argmin}{argmin}

%% file: sections/0_authors.tex
\author{Sanghyun Son*,~\IEEEmembership{}
Jaeha Kim*,~\IEEEmembership{}
Wei-Sheng Lai,~\IEEEmembership{}
Ming-Hsuan Yang,~\IEEEmembership{}
and Kyoung Mu Lee~\IEEEmembership{Fellow,~IEEE,}
\IEEEcompsocitemizethanks{
\IEEEcompsocthanksitem$^\ast$Authors contributed equally.
\IEEEcompsocthanksitem S. Son, J. Kim, and K. M. Lee are with the Department
of ECE \& ASRI, Seoul National University, Seoul,
Korea, 08826.\protect\\
E-mail: \{thstkdgus35, hyjkim2, kyoungmu\}@snu.ac.kr 
\IEEEcompsocthanksitem W.-S. Lai is with  Google. E-mail: wslai@google.com
\IEEEcompsocthanksitem M.-H. Yang is with Google, UC Merced, and Yonsei University.\protect\\
E-mail: minghsuan@google.com
}
}

%% file: sections/0_abstract.tex
\IEEEtitleabstractindextext{%
\begin{abstract}
Most image super-resolution (SR) methods are developed on synthetic low-resolution (LR) and high-resolution (HR) image pairs that are constructed by a predetermined operation, e.g., bicubic downsampling.
As existing methods typically learn an inverse mapping of the specific function, they produce blurry results when applied to real-world images whose exact formulation is different and unknown.
Therefore, several methods attempt to synthesize much more diverse LR samples or learn a realistic downsampling model.
However, due to restrictive assumptions on the downsampling process, they are still biased and less generalizable.
This study proposes a novel method to simulate an unknown downsampling process without imposing restrictive prior knowledge.
We propose a generalizable low-frequency loss (LFL) in the adversarial training framework to imitate the distribution of target LR images without using any paired examples.
Furthermore, we design an adaptive data loss (ADL) for the downsampler, which can be adaptively learned and updated from the data during the training loops.
Extensive experiments validate that our downsampling model can facilitate existing SR methods to perform more accurate reconstructions on various synthetic and real-world examples than the conventional approaches.
\end{abstract}

\begin{IEEEkeywords}
Image super-resolution, image downsampling, unsupervised learning
\end{IEEEkeywords}}

%% file: sections/1_intro_arxiv.tex
\ifCLASSOPTIONcompsoc
\IEEEraisesectionheading{\section{Introduction}\label{sec:introduction}}
\else
\section{Introduction}
\label{sec:introduction}
\fi

\IEEEPARstart{I}{mage} super-resolution (SR), which aims to reconstruct a high-resolution~(HR) image from a low-resolution~(LR) input, plays an essential role in computer vision and digital photography.
There exist numerous applications, including enhancing the details and photorealism of an image~\cite{sr_srgan}, high-quality editing~\cite{sr_explorable}, and breaking the sensor limitation of mobile cameras~\cite{wronski2019handheld}.
Recently, a plethora of SR methods have been developed on the basis of deep CNNs~\cite{sr_srcnn, sr_lapsrn, sr_srgan, sr_edsr} and large-scale datasets~\cite{data_div2k, sr_edsr}.
However, state-of-the-art methods~\cite{sr_esrgan, sr_rcan, sr_dbpn, sr_rdn} do not generalize well to the real-world inputs even they perform relatively well on synthesized, e.g., bicubic-downsampled, LR images.
In overcoming this issue, few recent approaches~\cite{sr_zllz, sr_camera, sr_realworld, sr_cdc} have collected high-quality pairs of real-world LR and HR examples to learn their SR models.
Nevertheless, such an acquisition process remains to be challenging due to outdoor scene dynamics and spatial misalignments~\cite{sr_zllz}.

Conventional SR methods synthesize various LR samples $\img{LR}$ from ground-truth HR images $\img{HR}$ by the following:
\begin{equation}
    \img{LR} = \paren{ \img{HR} \ast k }_{\downarrow s} + n,
    \label{eq:linear}
\end{equation}
where $k \in \mathbb{R}^{2}$ is a 2D degradation kernel, $\ast$ is a spatial convolution, $_{\downarrow s}$ is a decimation with a stride $s$, and $n$ is a noise term.
The decimation operator corresponds to direct downsampling mentioned in the super-resolution literature~\cite{sr_srmd}.
With a specific assumption of blur kernels, e.g., variants of Gaussian~\cite{sr_ikc, sr_blindsr, sr_srmd}, LR and HR pairs can be synthesized to train the following SR models.
However, such prior typically limit the kernel space, and the synthesized LR images may not reflect the distribution of real-world inputs~\cite{sr_camera}.
Therefore, the learned SR models become less generalizable toward arbitrary real-world input images.

\input{sections/figure/teaser_arxiv}

On the other hand, recent unsupervised methods simulate real-world LR samples that contain unknown noise~\cite{sr_deg, sr_unpaired_pesudo} and artifacts~\cite{sr_unsupervised, sr_gandeg}.
Without using a paired dataset, they first learn a downsampling model under adversarial training frameworks~\cite{gans} to imitate the distribution of real-world images.
The following SR models can then be trained in a supervised manner on the simulated dataset to deliver accurate reconstruction results on the real-world inputs.
One of the challenges in such methods arises from preserving image content across different scales, i.e., HR and LR, while learning the downsampling model.
Existing approaches deal with this problem using a predetermined downsampling operator, e.g., bicubic downsampling, in their objective functions and constrain the simulated LR images not to deviate much from the known formulations.
However, the manual selection of the operator can introduce a bias in the unsupervised learning framework, which can also act as a restrictive prior if the ground-truth downsampling model is much different from the used one.
%
%
While KernelGAN~\cite{sr_kernelgan} alleviates the issue by estimating a low-dimensional downsampling kernel $k$ in \eqref{eq:linear} from an LR image $\img{LR}$, various regularization terms need to be applied to restrict the diversity of the possible kernel space.

Therefore, we propose an effective way of imitating the real-world LR samples of an unknown distribution to address the aforementioned issues.
Similar to the previous unsupervised methods~\cite{sr_unsupervised, sr_kernelgan}, we also train a downsampling CNN to simulate the LR images in our target distribution.
However, rather than formulating the objective function with a handcrafted downsampling operator, we propose a novel and generalizable low-frequency loss (LFL) that does not pose substantial bias.
Our LFL facilitates the downsampling model to learn much more diverse and precise functions without being constrained to a specific prior assumption.
Furthermore, we develop an adaptive data loss (ADL) that iteratively adjusts the training objective for the given dataset and stabilizes the learning process.
As shown in Fig.~\ref{fig:teaser}, our unsupervised learning framework is straightforward, effective, and generalizable to arbitrary downsampling models.
Extensive experiments validate that the SR models learned on our downsampled images perform favorably on synthetic and real-world LR images.
The contributions of this study can be organized threefold:
\begin{compactitem}[$\bullet$]
\item We present a novel unsupervised learning framework to learn an unknown downsampling process without using any HR and LR image pairs.
\item We propose LFL and ADL to simulate accurate and realistic LR samples from HR images without relying on any predetermined downsampling operators.
\item We demonstrate that the proposed method can be easily integrated with existing SR frameworks and achieve much better results on synthetic and real-world images.
\end{compactitem}

%% file: sections/figure/teaser_arxiv.tex
\begin{figure}
    \vspace{-10mm}
    \centering
    \subfigure[Bicubic]{\includegraphics[width=0.19\linewidth]{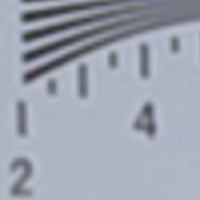}}
    \subfigure[\cite{sr_esrgan}]{\includegraphics[width=0.19\linewidth]{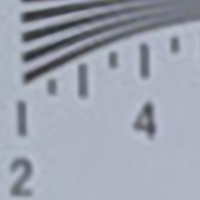}}
    \subfigure[\cite{sr_kernelgan} + \cite{sr_zssr}]{\includegraphics[width=0.19\linewidth]{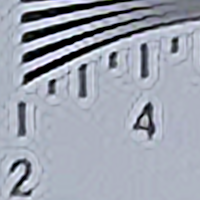}}
    \subfigure[\textbf{Ours}]{\includegraphics[width=0.19\linewidth]{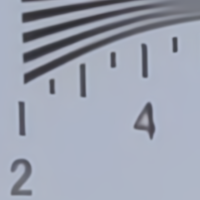}}
    \subfigure[GT]{\includegraphics[width=0.19\linewidth]{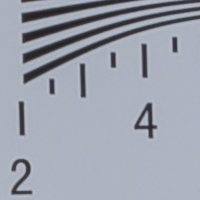}}
    \figspace
    \caption{
        \textbf{$\times 4$ SR results on a real-world LR image.}
        (a) LR image magnified with bicubic interpolation.
        (b) Result of RRDB~\cite{sr_esrgan}.
        (c) Result of KernelGAN~\cite{sr_kernelgan} + ZSSR~\cite{sr_zssr}.
        (d) Our unsupervised approach (ADL + RRDB) reconstructs a sharp and visually pleasing output without artifacts and aliasing compared with the existing methods.
        (e) Ground-truth patch from RealSR-V3~\cite{sr_realworld}.
        Images are cropped from `\emph{Canon/045.png}.'
    }
    \label{fig:teaser}
    \figxspace
\end{figure}

%% file: sections/2_related.tex
\section{Related Work}
\label{sec_background}
\subsection{SR on bicubic downsampled images}
With the success of SRCNN~\cite{sr_srcnn}, several CNN-based methods have been developed for image SR.
As one of the most influential studies, VDSR~\cite{sr_vdsr} has proposed a novel residual learning strategy to train a deep network and inspired lots of following methods~\cite{sr_srgan, sr_edsr, sr_carn, sr_msrn}.
Earlier works primarily focus on improving the network designs, such as pixel shuffling~\cite{sr_espcn}, 
progressive upsampling~\cite{sr_lapsrn, lai2018fast, wang2018fully}, 
dense connections~\cite{sr_dense, sr_rdn, ir_rdn}, 
recursive structures~\cite{sr_drcn, sr_ebrn, sr_feedback},
and back-projection~\cite{sr_dbpn}.
Recent approaches utilize the attention~\cite{sr_rcan, sr_san, sr_han}, while designing architectures for efficient inference~\cite{sr_idn, sr_pi, sr_lattice} is considered essential as well.
From the perspective of image realism, several methods introduce perceptual loss~\cite{others_perceptual, sr_enhancenet, per_contextual, feat_deep} to synthesize photorealistic textures~\cite{sr_srgan, sr_esrgan, sr_rank, sr_srobb}.
However, the existing methods are typically trained on synthesized image pairs in which LR inputs are generated using conventional bicubic interpolation from HR targets.
While state-of-the-art algorithms perform impressively well when training and test distributions are matched, i.e., test inputs are also downsampled with the same operator, they cannot be fully generalized to arbitrary in-the-wild LR images~\cite{sr_zssr, sr_correction}.
%

\subsection{Synthesizing diverse LR images for SR}
For practical SR application, it is essential to determine how to generate LR images~\cite{sr_bsrgan} so that a supervised SR model can be trained without real-world LR and HR image pairs.
Several approaches have synthesized diverse LR images with multiple degradations to train their SR algorithms, assuming that the generalization on such examples can improve SR performance on arbitrary inputs.
SRMD~\cite{sr_srmd} considers the formation of LR images under various downsampling kernels $k$ in \eqref{eq:linear}.
It can reconstruct HR images from diverse types of LR inputs, using off-the-shelf methods~\cite{tsr_blindsr, rs_deblurring_l0, sr_kernelgan} to predict a candidate kernel $k$ from a given LR input.
USRNet~\cite{rs_usrnet} further allows diversity to the downsampling kernel $k$ and can deliver clean SR results even when LR inputs are corrupted with motion blur and noise.

Furthermore, recent methods~\cite{sr_ikc, sr_blindsr, sr_udvd} present unified frameworks to jointly estimate the kernel $k$ and reconstruct visually pleasing results from an arbitrary LR image.
However, since considering all possible forms of downsampling operation is not practical, the candidate kernels in such methods are often simplified to variants of 2D Gaussian.
Recent studies have demonstrated that such approximations may not hold for actual LR images~\cite{sr_camera, sr_kmsr} in the wild, thus making the abovementioned SR algorithms less generalizable.
In this study, we demonstrate that existing approaches~\cite{sr_ikc, sr_blindsr} do not perform well on inputs from unknown downsampling kernels or real-world images, and our method provides better generalization.

\subsection{Learning to simulate real-world LR images}
Instead of synthesizing LR images from some handcrafted formulations, numerous approaches~\cite{sr_gandeg, sr_deg, sr_unsupervised, sr_unpaired_pesudo} have adopted adversarial training~\cite{gans} to simulate the unknown distributions of real-world LR images using downsampler CNNs.
These methods have shown impressive performance when dealing with unknown noise~\cite{sr_deg, sr_unpaired_pesudo} and artifacts~\cite{sr_gandeg, sr_unsupervised} in real-world LR images.
Considering the definition of downsampling, one of the required characteristics of such methods is to preserve the contents of HR input and generate a feasible LR image.
Therefore, predetermined downsampling operators~\cite{sr_gandeg, sr_unsupervised, sr_unpaired_pesudo} and cyclic architecture~\cite{others_cycle} are used to guide the generated LR images not deviating much from the desired outputs.
However, a significant limitation of this formulation is that the necessity of estimating an accurate \emph{downsampling} process is often considered less important.
In particular, the handcrafted operators may significantly differ from the unknown downsampling function and bias the following downsampler, making the model less effective in estimating the actual operators rather than noise and artifacts.

On the other hand, KernelGAN~\cite{sr_kernelgan} is designed to directly predict the degradation kernel $k$, which is used to generate the given LR image $\img{LR}$.
The estimated kernel is then used to synthesize LR and HR pairs for the following SR model~\cite{sr_zssr}.
In addressing the ill-posed problem of finding the kernel $k$ in \eqref{eq:linear}, several optimization constraints are assumed, such as patch recurrences~\cite{tsr_blindsr} in a single image, deep linear generator, and various prior knowledge on physically meaningful kernels.
However, this approach may not handle practical cases in which such strong assumptions do not hold.
While Ji~\etal~\cite{sr_rwsr} have extended the approach to a set of LR images, several prior terms for the appropriate degradation kernel still act as a bottleneck for generalization.
On the contrary, our LFL and ADL are designed to reduce inherent bias from adopting a specific downsampling operator or strong kernel priors.

\subsection{Paired datasets for real-world SR}
Limitations of existing SR methods arise from difficulties in constructing the real-world dataset.
Few approaches capture the paired dataset~\cite{sr_camera, sr_realworld, sr_cdc} by precisely manipulating camera parameters in which images captured from long and short focal lengths are labeled as HR and LR samples, respectively.
Zhang~\etal~\cite{sr_zllz} introduces SR-RAW based on raw images and contextual bilateral loss to handle misalignments in the real-world pairs.
Xu~\etal~\cite{sr_real_raw} utilizes raw and color images jointly in their model for effective real-world SR.
Those image pairs can be used to learn the real-world SR models to some extent.
Nevertheless, they still suffer from a lack of scene diversity, misalignments, dynamic motions, and scalability issues.
To overcome the several challenges in acquiring realistic SR datasets, we synthesize accurate HR and LR pairs from unpaired examples.
\revised{While we assume that a set of LR images undergo the same or similar formation process, the data collection is much easier since careful alignment and delicate post-processing are not required.}

%% file: sections/3_method.tex
\section{Learning to Downsample}
\label{sec_l2d}
In conventional frameworks, mismatches between the handcrafted kernel space and real-world downsampling model~\cite{sr_kmsr, sr_camera} makes the following SR networks less generalizable.
Thus, we develop an unsupervised learning framework to accurately simulate the LR samples $\img{LR} \in \mathcal{I}_\text{LR}$ from the \emph{unpaired} HR images $\img{HR} \in \mathcal{I}_\text{HR}$.
The following SR model can then be trained to reconstruct the HR results from the given LR dataset $\mathcal{I}_\text{LR}$.
For simplicity, we assume that the LR and HR images have spatial resolutions of $H \times W$ and $sH \times sW$, respectively, for a downsampling factor $s$.

\subsection{Learning an unknown downsampling process}
\label{ssec_l2d}
We synthesize LR images under the generalized formulation as $\img{LR} = \mathcal{D}^{\ast} \paren{ \img{HR}^{\ast} }$, where $\img{HR}^{\ast} \in \mathcal{I}_\text{HR}^{\ast}$ is the latent HR samples from the distribution $\mathcal{I}^{\ast}_\text{HR}$ and $\mathcal{D}^{\ast}$ is an unknown downsampling operator.
The goal is to learn an SR model $\mathcal{S}$, which can reconstruct a high-quality HR image from the given LR image $\img{LR} \in \mathcal{I}_\text{LR}$.
However, it is not straightforward to learn the upsampling function directly as the corresponding ground-truth HR images $\mathcal{I}_\text{HR}^\ast$ are unavailable.
Thus, we first learn a downsampling model $\mathcal{D}$ in an unsupervised manner, so that the distribution of the synthesized images $\img{Down} = \mathcal{D} \paren{ \img{HR} }$ is close to the distribution of the target LR samples $\mathcal{I}_\text{LR}$.
By using the generated pairs of $\paren{ \img{Down}, \img{HR} }$, our SR model can be trained to reconstruct HR images from the given LR distribution $\mathcal{I}_\text{LR}$ in a fully supervised manner.

To learn the downsampling function $\mathcal{D}$, we adopt adversarial training framework~\cite{gans} to jointly optimize the downsampler CNN $\mathcal{D} \left( \cdot; \theta_\mathcal{D} \right)$ and the discriminator CNN $\mathcal{F} \left( \cdot; \theta_\mathcal{F} \right)$.
Then, we formulate the downsampling and discriminator objectives, $\mathcal{L}_\mathcal{D}$ and $\mathcal{L}_\mathcal{F}$, as follows:
\begin{equation}
    \begin{split}
        \mathcal{L}_\mathcal{D} &= \alpha \mathcal{L}_\text{data} + \mathcal{L}_\text{adv} = \alpha \mathcal{L}_\text{data} - \mathbb{E} \left[ \log{ \mathcal{F} \paren{ \img{Down} } } \right], \\
        \mathcal{L}_\mathcal{F} &=
        - \mathbb{E} \left[ \log{ \mathcal{F} \paren{ \img{LR} } } \right]
        + \mathbb{E} \left[ \log{ \left( 1 - \mathcal{F} \paren{ \img{Down} } \right) } \right], \\
    \end{split}
    \label{eq:ds_obj}
\end{equation}
where $\mathcal{L}_\text{data}$ is the data loss, $\mathcal{L}_\text{adv}$ is the adversarial loss~\cite{gans}, and $\alpha$ is a hyperparameter.
If the learned downsampling model can accurately synthesize LR images from $\mathcal{I}_\text{HR}$, i.e., the distribution of $\img{Down}$ and $\mathcal{I}_\text{LR}$ are approximately the same, the following SR model can be generalized on $\mathcal{I}_\text{LR}$ by learning from a set of training pairs $\paren{\img{Down}, \img{HR}}$.
Fig.~\ref{fig:abstaction} shows the overall pipeline of our method, which learns the downsampling and super-resolution models consecutively.

For simplicity, we assume that the target LR images $\img{LR}$ are not corrupted with noise, where the term $n$ in \eqref{eq:linear} is ignored.
The primary reason is that the noise can be a discriminative feature between the real LR and downsampled images in adversarial training.
Since we do not include randomness in our downsampler architecture, such behavior also prevents the proposed method from learning an accurate downsampling function.
In Section~\ref{ssec:real}, we discuss the effect of real-world noise in the proposed framework.

\input{sections/figure/two_stage}

\subsection{Data constraint in the downsampling model}
\label{ssec:data_term}
In practice, the actual formulation of the given LR images, i.e., the ground-truth downsampling model, is unknown.
Thus, we introduce the adversarial loss $\mathcal{L}_\text{adv}$ to enforce the downsampled images $\img{Down}$ to follow a target distribution $\mathcal{I}_\text{LR}$ without using ground-truth LR images.
However, unlike the other image generation tasks~\cite{gans_dc}, appropriate constraints are also required to generate faithful LR samples to the given HR counterparts and preserve input contents.
In particular, low-level information of a given image, e.g., pixel colors and edge structures, should not be changed during the downsampling, as shown in Fig.~\ref{fig:data_loss}(a) and (b).
Thus, the appropriate formulation of the data term $\mathcal{L}_\text{data}$ in \eqref{eq:ds_obj} plays a critical role in preserving the image content across different scales.
A widely-used approach is to define the data loss $\mathcal{L}_\text{data}$ with a known operator $\mathcal{R}_\text{HR}$, such as bicubic downsampling or $s \times s$ average pooling~\cite{sr_gandeg}, as follows:
\begin{equation}
    \begin{split}
        \mathcal{L}_\text{data}
        &= \normone{ \mathcal{R}_\text{HR} \paren{ \img{HR} } - \mathcal{D} \paren{ \img{HR} } }, \\
        &= \normone{ \mathcal{R}_\text{HR} \paren{ \img{HR} } - \img{Down} }.
    \end{split}
    \label{eq:data}
\end{equation}
%
%
%
That is, a reference example $\mathcal{R}_\text{HR} \paren{ \img{HR} }$ constrains the generated LR sample $\mathcal{D} \paren{\img{HR}}$ to be a feasible downsampled image.
A recent method from Maeda~\cite{sr_unpaired_pesudo} has also combined the bicubic downsampling operator $\mathcal{B}$ and image-to-image translation CNN $\mathcal{G}$ in their downsampling model so that $\mathcal{D} = \mathcal{G} \circ \mathcal{B}$.
Under the such configuration, the translator network $\mathcal{G}$ is trained to maintain the consistency between its input and output which corresponds to $\mathcal{R}_\text{HR} = \mathcal{B}$ in \eqref{eq:data}.

\input{sections/figure/why_data_loss_matters}

%
In \eqref{eq:data}, the data term $\mathcal{L}_\text{data}$ enforces the downsampled images $\img{Down}$ to be close to references $\mathcal{R}_\text{HR} \paren{ \img{HR} }$.
Such a formulation contributes to preserve the image content and facilitate the training process for generating LR images.
Nevertheless, optimizing the data term $\mathcal{L}_\text{data}$ in \eqref{eq:ds_obj} may bias the learned model toward the used operator $\mathcal{R}_\text{HR}$.
The bias may conflict with the adversarial training objective $\mathcal{L}_\text{adv}$ if the distribution of the downsampled images $\mathcal{R}_\text{HR} \paren{ \img{HR} }$ deviate significantly from the target distribution $\mathcal{I}_\text{LR}$.

Fig.~\ref{fig:data_loss} illustrates an example to demonstrate the negative effect of using a predetermined downsampling operator, e.g., bicubic kernel $\mathcal{B}$, in the data term $\mathcal{L}_\text{data}$.
The target LR images $\img{LR}$ are generated using a different kernel $k$, where $\mathcal{B} \paren{ \img{HR} } \neq \paren{ \img{HR} \ast k }_{\downarrow s}$ for an arbitrary HR image $\img{HR}$ shown in Fig.~\ref{fig:data_loss}(a)--(c).
Then, we jointly minimize the data and adversarial loss terms in \eqref{eq:ds_obj} under different $\alpha$ values so that the downsampling model can be close to the target distribution $\mathcal{I}_\text{LR}$.
Fig.~\ref{fig:data_loss}(e)--(h) illustrate differences between the actual LR and downsampled image $\lvert \img{LR} - \img{Down} \rvert$ with a varying $\alpha$.
If the data term $\mathcal{L}_\text{data}$ is not used, i.e., $\alpha = 0$, the adversarial loss $\mathcal{L}_\text{adv}$ is solely optimized in the training so that $\mathcal{D} \paren{ \img{HR} } \in \mathcal{I}_\text{LR}$.
As shown in Fig.~\ref{fig:data_loss}(e), the downsampled image does not preserve the original colors and becomes inconsistent with the input $\img{HR}$ in such case.

On the other hand, if we increase weight $\alpha$ to retain the input content, the resulting downsampled images will more likely resemble $\mathcal{B} \paren{ \img{HR} }$ rather than the desired output $\paren{ \img{HR} \ast k}_{\downarrow 2}$, as shown around edge and corner regions of Fig.~\ref{fig:data_loss}(f)--(h).
The tradeoff between preserving image contents and synthesizing an accurate distribution of the LR images occurs due to the inherent conflict between the predetermined downsampling operator $\mathcal{R}_\text{HR}$ and the adversarial loss $\mathcal{L}_\text{adv}$.
While the data term $\mathcal{L}_\text{data}$ is necessary to learn an appropriate downsampling function, it also operates as a restrictive prior and prevents an accurate simulation of the target LR images.
Therefore, an SR method developed with the biased downsampler may not perform well on the target distribution $\mathcal{I}_\text{LR}$, as conventional bicubic SR algorithms cannot be generalized on real-world images.

\input{sections/figure/data_loss_comparison}

\subsection{Data loss over low-frequency components}
\label{ssec_ap}
We propose an effective and generalizable formulation of the data term $\mathcal{L}_\text{data}$ to address the limitations of the existing approaches.
Similar to the previous methods, our downsampler also takes an input HR image $\img{HR}$ and generates a downsampled image $\img{Down}$.
However, to preserve image contents and low-level structures in the downsampling process, we first define the operator $\text{LPF}_{m}: \mathbb{R}^{H \times W} \rightarrow \mathbb{R}^{\nicefrac{H}{m} \times \nicefrac{W}{m}}$ as a combination of low-pass filtering and subsampling by $m$, which reduces the resolution of a given image by a factor of $m > 1$.
Then, we rewrite the data loss in \eqref{eq:ds_obj} with a low-frequency loss (LFL) $\mathcal{L}_\text{data} = \mathcal{L}_\text{LFL}$ as follows:
\begin{equation}
    \mathcal{L}_\text{LFL} = \normone{ \text{LPF}_{ms} \left( \img{HR} \right) - \text{LPF}_{m} \left( \img{Down} \right) },
    \label{eq:la}
\end{equation}
where $s$ is a scaling factor.
Since the HR image $\img{HR}$ is $s$ times larger than the downsampled one $\img{Down}$, sizes of $\text{LPF}_{ms} \left( \img{HR} \right)$ and $\text{LPF}_{m} \left( \img{Down} \right)$ are the same.
We adopt two different low-pass filters: the box and Gaussian, to formulate the loss term.
As the HR and downsampled images have different resolutions, we adjust the filter weights proportionally so that the same context can be covered from the images with different scales.
By default, we use $32 \times 32$ and $16 \times 16$ box filters for $\text{LPF}_{ms}$ and $\text{LPF}_{m}$, respectively, with the scaling factor $s = 2$.
We provide more details and ablations regarding the low-pass filters in Appendix~\fakeref{A}.

Fig.~\ref{fig:data} illustrates the differences between the existing formulation and the proposed loss term.
As shown in Fig.~\ref{fig:data}(a), the handcrafted operator constrains each pixel of the downsampled image $\img{Down}$ to be a predetermined function of the input HR image $\img{HR}$.
The primary limitation of such an approach is that the HR image $\img{HR}$ and operator $\mathcal{R}_\text{HR}$ are both kept unchanged throughout the entire learning process.
Therefore, the reference image $\mathcal{R}_\text{HR} \paren{ \img{HR} }$ is also fixed for each HR image, which can bias the learning process.
Thus, even with the adversarial training objective, the learned downsampler $\mathcal{D}$ can be biased toward the predetermined operator $\mathcal{R}_\text{HR}$ rather than the desired downsampling model, especially when the weight $\alpha$ in \eqref{eq:ds_obj} is large.

Our motivation is that we only need to preserve the low-frequency components of the image contents and structures.
Fig.~\ref{fig:data}(b) demonstrate that the downsampled image $\img{Down}$ is no longer constrained to be a specific function of its HR counterpart with our LFL.
Instead, we adopt a relaxed objective designed to match low-frequency structures between input and output of the downsampler.
By doing so, the adversarial loss can play a significant role in rendering the unknown types of LR images.
Our LFL is not a restrictive constraint for a general downsampling model and can be generalized well on various synthetic and real-world images.
In other words, we can minimize the new data term $\mathcal{L}_\text{LFL}$ without causing notable conflict with the adversarial loss for LR images $\mathcal{I}_\text{LR}$ from an arbitrary downsampling model.
More details are described in Section~\ref{ssec:ablation}.

\Paragraph{Scale transfer learning.}
The proposed LFL does not include any scale-specific formulation and can be generalized to larger scales, e.g., $\times 4$.
However, directly optimizing a high-scale downsampler may cause less stable behaviors due to the significant differences in the HR and downsampled images.
To ensure stability, we learn a $\times 2$ model $\mathcal{D}^{\times 2}$ on the desired distribution $\mathcal{I}_\text{LR}$ and repeat it $n > 1$ times~\cite{sr_kernelgan} to obtain the $\times 2^{n}$ downsampling models $\mathcal{D}^{\times 2^n}$ by following:
\begin{equation}
    \mathcal{D}^{\times 2^n} = \mathcal{D}^{\times 2^{n - 1}} \circ \mathcal{D}^{\times 2} \quad \paren{n > 1}.
\end{equation}

\subsection{Adaptive data loss}
\label{ssec:adl}
Our LFL is designed to reduce the bias from selecting a predetermined downsampling operator for the data loss $\mathcal{L}_\text{data}$.
While this formulation enables LFL to be generalized well across various unknown degradations, several limitations exist.
For example, an inherent ambiguity in LFL makes it challenging to solve the optimization problem because the LR images from the different downsampling processes may share similar low-frequency components, i.e., $\text{LPF}_m \paren{ \img{a} } = \text{LPF}_m \paren{ \img{b} }$ for $\img{a} \neq \img{b}$.
Considering that our goal is to simulate the unknown downsampling model $\mathcal{D}^{\ast}$ with a CNN-based downsampler $\mathcal{D}$, the ideal data loss $\mathcal{L}_\text{data}$ should be zero only if the condition $\mathcal{D} \paren{ \img{HR} } \equiv \mathcal{D}^{\ast} \paren{ \img{HR} }$ satisfies.
The primary limitation of LFL is that it is designed to maintain consistency between HR and LR images, not to simulate structures of LR images in the target distribution.
Since LFL only considers low-frequency components in the image, optimizing the term is an ill-posed problem where numerous possible $\mathcal{D}$ exist.
In particular, minimizing LFL allows the downsampler to generate valid LR images, while it is not guaranteed that the learned downsampler achieves our desired behavior.
Therefore, the definition of LFL is generalizable but cannot be an optimal one for any arbitrary downsampling model due to the ambiguity.

Moreover, LFL can be problematic when the downsampled image $\img{Down}$ is corrupted with high-frequency noise, which is suppresssed after low-pass filtering.
Since the proposed LFL cannot reject noisy estimations, it is challenging to generate clean and accurate LR samples of the desired distribution.
Consequently, the downsampler heavily relies on adversarial loss to simulate an accurate distribution of $\mathcal{I}_\text{LR}$, which may not be very stable in practice~\cite{gans_dc}.

Therefore, we propose an adaptive data loss (ADL) to complement the limitations of LFL.
The primary motivation is that the LFL-based downsampler $\bar{\mathcal{D}}$ can serve as a dataset-specific objective if $\bar{\mathcal{D}}$ and the ground-truth downsampling model $\mathcal{D}^\ast$ are similar to some extent.
To formulate the ADL, we first reduce the noise in the pre-trained model $\bar{ \mathcal{D} }$.
Rather than introducing a new objective term in \eqref{eq:ds_obj} for regularization, we retrieve a low-rank approximation of the learned network with a simple function.
From the observation that a proper downsampling function consists of low-pass filtering and decimation~\cite{tsr_blindsr, sr_ikc, sr_blindsr, sr_kernelgan}, we linearize the learned downsampling model $\bar{ \mathcal{D} }$ to a corresponding 2D kernel $\bar{k}$:
\begin{equation}
        \bar{k} = \argmin_{k} \sum_{i = 1}^{N} {\normtwo{ \left( \img{HR}^i \ast k \right)_{\downarrow s} - \bar{\mathcal{D}} \paren{ \img{HR}^{i} }} }, \\
    \label{eq:kernel}
\end{equation}
where $\img{HR}^{i}$ denotes an $i$-th example to estimate the kernel and $N$ is the total number of samples that have been used, respectively.
We note that there exists a closed-form solution for the least-squares in \eqref{eq:kernel}.
Since \eqref{eq:kernel} can be interpreted as an average of the possibly noisy downsampling network over $N$ inputs, the kernel $\bar{k}$ is a regularized representation of the pre-trained network $\bar{ \mathcal{D} }$.
With the estimated kernel $\bar{k}$, a novel ADL for data term $\mathcal{L}_\text{data} = \mathcal{L}_\text{ADL}$ is defined as follows:
\begin{equation}
    \mathcal{L}_\text{ADL} = \normone{ \paren{ \img{HR} \ast \bar{k} }_{\downarrow s} - \img{Down} }.
    \label{eq:adl}
\end{equation}

\input{sections/figure/example_kernels}

While \eqref{eq:adl} looks identical to the data terms with handcrafted downsampling in \eqref{eq:data}, we can deduce several merits from the ADL formulation. 
Unlike the predetermined operators $\mathcal{R}_\text{HR}$ or $\mathcal{B}$, the kernel $\bar{k}$ is adaptively learned from the training data and shows less conflict to the adversarial loss $\mathcal{L}_\text{adv}$.
In other words, the linear downsampling process in \eqref{eq:adl} is less likely to deviate considerably from our desired downsampling model $\mathcal{D}^{\ast}$.
Compared with the LFL formulation, the ADL term can provide a stable training objective and prevent the downsampler from learning false-negative cases.
Moreover, the learned downsampling model $\mathcal{D}$ is not constrained to be a deep linear network~\cite{sr_kernelgan}, as we jointly optimize the adversarial loss $\mathcal{L}_\text{adv}$ with the ADL.

Also, we introduce two modifications to utilize our ADL effectively in practice.
First, the downsampler $\mathcal{D}$ has been observed to simulate a target downsampling model $\mathcal{D}^{\ast}$ to some extent under the LFL, even with few training iterations.
Rather than using a fully pre-trained model $\bar{ \mathcal{D} }$ for the kernel estimation, we start from scratch and train the downsampler for $t_\text{warm-up}$ iterations with the LFL.
We then replace our data term with the ADL, in which the kernel $\bar{k}$ is calculated from the downsampler $\mathcal{D}$ after $t_\text{warm-up}$ updates.
Second, we periodically adjust the kernel $\bar{k}$ to prevent our downsampling model from being biased toward a fixed operator.
Similar to \eqref{eq:data}, our ADL may also bias the training process unless $\mathcal{D}^{\ast} \paren{ \img{HR} } \equiv \paren{ \img{HR} \ast \bar{k} }_{\downarrow s}$ holds.
%
Thus, we periodically update the kernel $\bar{k}$ by retrieving it from the currently learned downsampler.
Even if the initial estimation $\bar{k}$ is less accurate, such periodic updates allow the kernel to be adaptively adjusted during learning loops.
The training pipeline of our downsampler $\mathcal{D}$ with the modified ADL formulation is summarized in Algorithm~\ref{alg:adl}.

\input{sections/algorithm/adl}

\subsection{Image super-resolution}
\label{ssec:sr}
To learn the SR model $\mathcal{S} \paren{\cdot; \phi}$, we first generate the LR images $\img{Down}$ from HR images with the learned downsampler $\mathcal{D}$ to construct a training set of $\left( \img{Down}, \img{HR} \right)$ pairs.
A downsampling-specific SR model can be trained in a supervised manner by optimizing the $L_1$ loss~\cite{sr_edsr, sr_lapsrn}:
\begin{equation}
    \begin{split}
        \mathcal{L}_{\mathcal{S}}
        &= \normone{ \img{HR} - \img{SR} } \\
        &= \normone{ \img{HR} - \mathcal{S} \paren{ \img{Down} }}
        = \normone{ \img{HR} - \mathcal{S} \paren{ \mathcal{D} \paren{ \img{HR} } } },
    \end{split}
    \label{eq:sr}
\end{equation}
where $\img{SR} = \mathcal{S} \paren{ \img{Down} }$ refers to a super-resolved image.
As shown in \eqref{eq:sr}, our approach does not require any paired examples, i.e., LR image $\img{LR}$, to learn the SR model for $\mathcal{I}_\text{LR}$.

One of our contributions is that the downsampling and SR models can be learned independently.
For instance, it is straightforward to introduce perceptual objective~\cite{others_perceptual, sr_srgan, sr_esrgan, feat_deep} for the SR network, which can be used to reconstruct photo-realistic results.
To reconstruct more realistic textures from the real-world LR images, we jointly optimize $\mathcal{L}_\mathcal{P}$ and $\mathcal{L}_\mathcal{G}$ to learn the perceptual SR model $\mathcal{P} \paren{\cdot; \theta_\mathcal{P}}$ and the discriminator network $\mathcal{G} \paren{\cdot; \theta_\mathcal{G}}$ respectively:
\begin{equation}
    \newcommand{\vgg}[1]{\mathcal{V}_\text{54} \paren{#1}}
    \begin{split}
        \mathcal{L}_{\mathcal{P}}
        &= \normone{ \vgg{ \img{HR} } - \vgg{ \img{SR-p} } }
        + \beta \mathcal{L}_{\text{adv-}\mathcal{P}}, \\
        \mathcal{L}_\mathcal{G}
        &= -\mathbb{E}\left[ \log{ \mathcal{G} \paren{ \img{HR} } } \right]
        - \mathbb{E}\left[ \log{ \paren{ 1 - \mathcal{G} \paren{ \img{SR-p} } } } \right],
    \end{split}
    \label{eq:sr_percep}
\end{equation}
where $\mathcal{V}_\text{54}$ is features of the pre-trained VGG-19~\cite{net_vgg, sr_srgan, sr_esrgan} network after the {\fontfamily{pcr}\selectfont conv5\_4} layer, $\img{SR-p} = \mathcal{P} \paren{ \img{LR} }$ is a super-resolved image, $\mathcal{L}_{\text{adv-}\mathcal{P}} = -\mathbb{E} \left[ \log {\mathcal{G} \paren{ \img{SR}} } \right]$ is adversarial loss, and $\beta = 0.02$ is a hyperparameter, respectively.

%% file: sections/figure/two_stage.tex
\begin{figure}
    \vspace{-3mm}
    \centering
    \subfigure[Downsampling]{\includegraphics[width=0.47\linewidth]{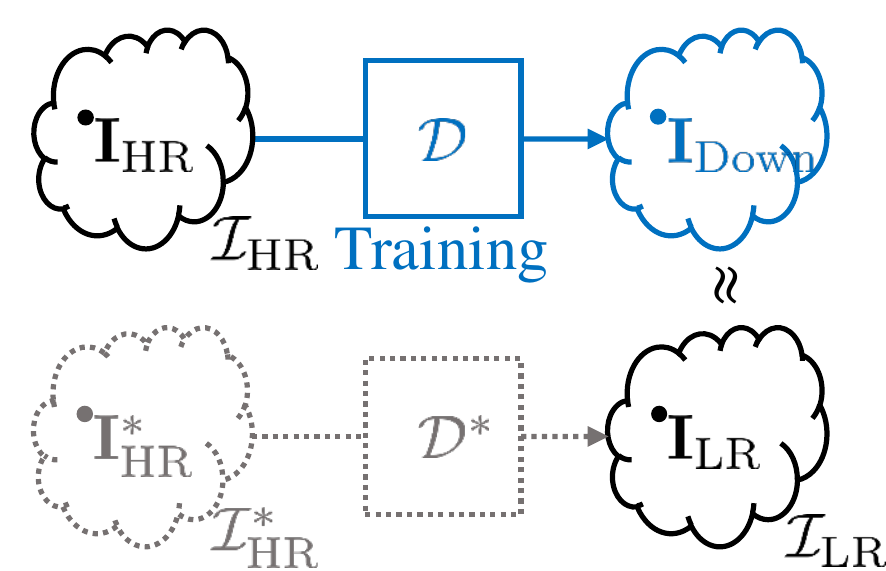}}
    \subfigure[Super-resolution]{\includegraphics[width=0.47\linewidth]{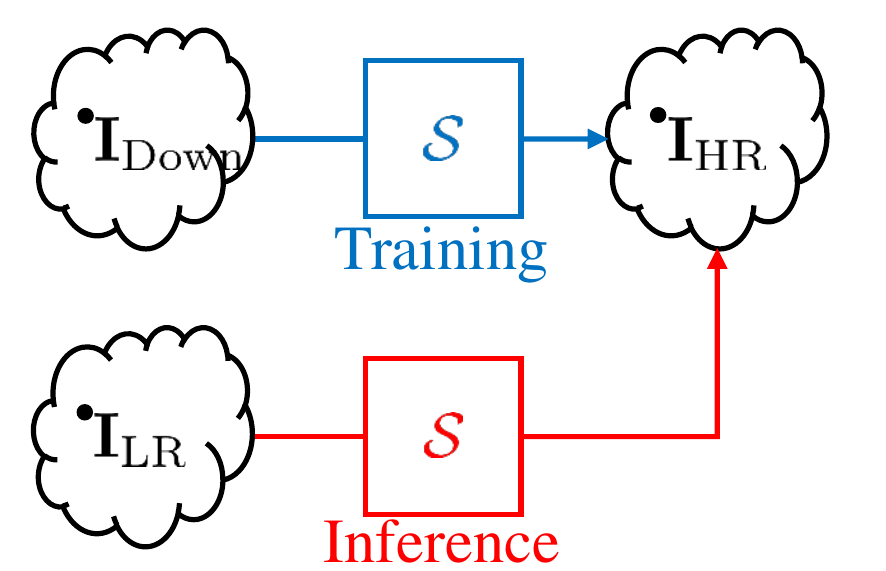}}
    \\
    \figspace
    \caption{
        \textbf{Our two-stage approach for unpaired SR.}
        (a) We first optimize a downsampling model $\mathcal{D}$ to synthesize $\img{LR}$ from $\img{HR}$.
        %
        %
        The primary goal is to learn the distribution of downsampled images rather than a proper downsampling function.
        (b) Using generated pairs, we train the SR model $\mathcal{S}$, which can also be generalized to the target LR images $\img{LR}$.
        Dotted lines represent latent components that are not available in the entire learning process.
        \textcolor{blue}{Blue} items show learned elements in each stage, and \textcolor{red}{red} elements denote the actual goal we want to achieve.
    }
    \label{fig:abstaction}
    \figxspace
\end{figure}

%% file: sections/figure/why_data_loss_matters.tex
\begin{figure}
    \centering
    \renewcommand{\wp}{0.22}
    \subfigure[$\img{HR}$]{\includegraphics[width=\wp\linewidth]{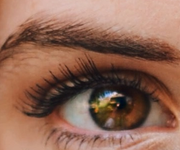}}
    \subfigure[$\img{LR}$]{\includegraphics[width=\wp\linewidth]{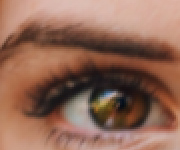}}
    \subfigure[$\mathcal{B} \paren{ \img{HR} }$]{\includegraphics[width=\wp\linewidth]{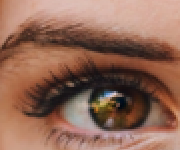}}
    \subfigure[\textbf{Ours}]{\includegraphics[width=\wp\linewidth]{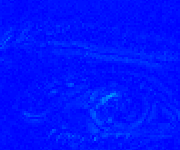}}
    \subfigure{\includegraphics[width=0.044\linewidth]{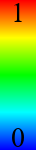}}
    \\
    \vspace{-2mm}
    \addtocounter{subfigure}{-1}
    \subfigure[$\alpha = 0$]{\includegraphics[width=\wp\linewidth]{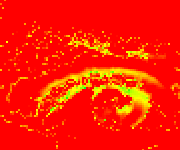}}
    \subfigure[$\alpha = 1$]{\includegraphics[width=\wp\linewidth]{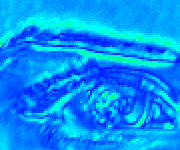}}
    \subfigure[$\alpha = 10$]{\includegraphics[width=\wp\linewidth]{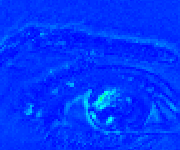}}
    \subfigure[$\alpha = 100$]{\includegraphics[width=\wp\linewidth]{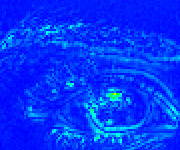}}
    \subfigure{\includegraphics[width=0.045\linewidth]{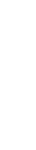}}
    \\
    \figspace
    \caption{
        \textbf{Differences between the ground-truth and learned LR images under various configurations.}
        (a) A reference HR.
        (b) A ground-truth LR patch $\img{LR} = \paren{ \img{HR} \ast k}_{\downarrow 2}$ we want to synthesize, where the kernel $k$ is unknown.
        (c) The corresponding bicubic-downsampled LR which is \emph{different} from $\img{LR}$.
        (d) The absolute difference between $\img{LR}$ and generated LR from our downsampling model is visualized with color-coding, where red pixels indicate large differences.
        (e)-(h) Difference between $\img{LR}$ and outputs from the learned downsampler under \eqref{eq:ds_obj}, where the bicubic downsampling operator is used for $\mathcal{L}_\text{data}$.
        Difference maps are normalized for better visualization.
        See more details in Section~\ref{ssec:eval_lr}.
    }
    \label{fig:data_loss}
    \figxspace
\end{figure}

%% file: sections/figure/data_loss_comparison.tex
\begin{figure*}[t]
    \centering
    \subfigure[Data loss from a predetermined kernel]{\includegraphics[width=0.28\linewidth]{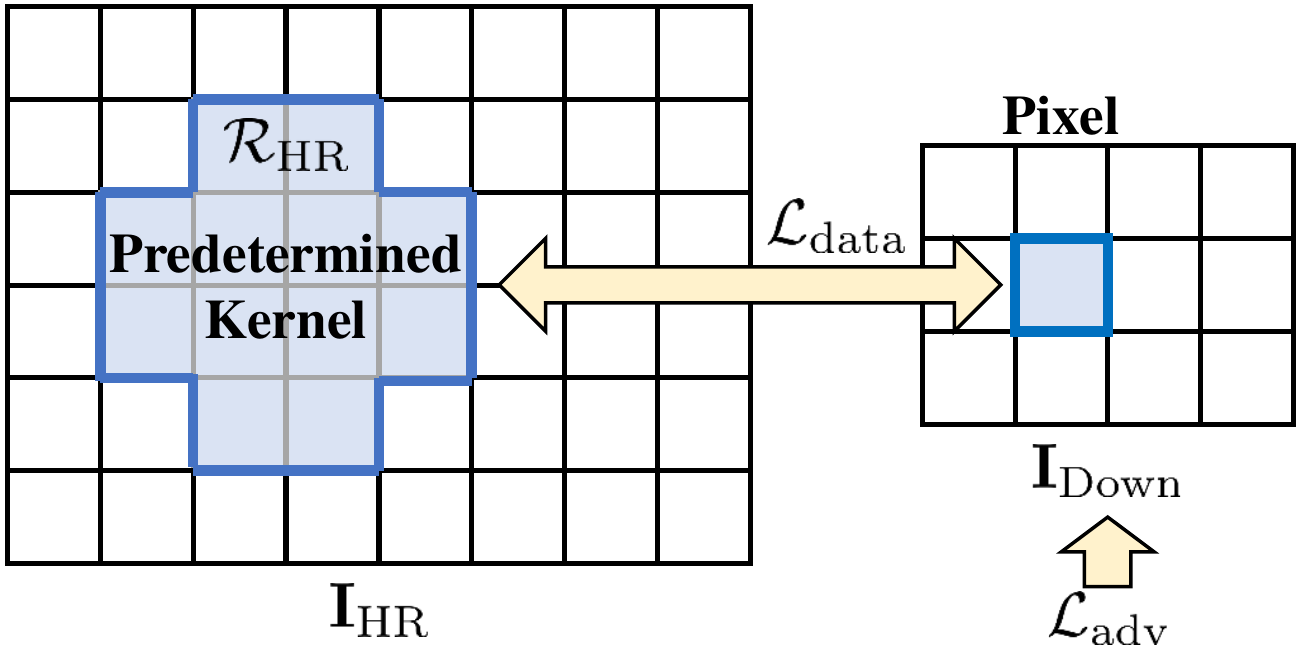}}
    \subfigure[\textbf{LFL~(Proposed)}]{\includegraphics[width=0.28\linewidth]{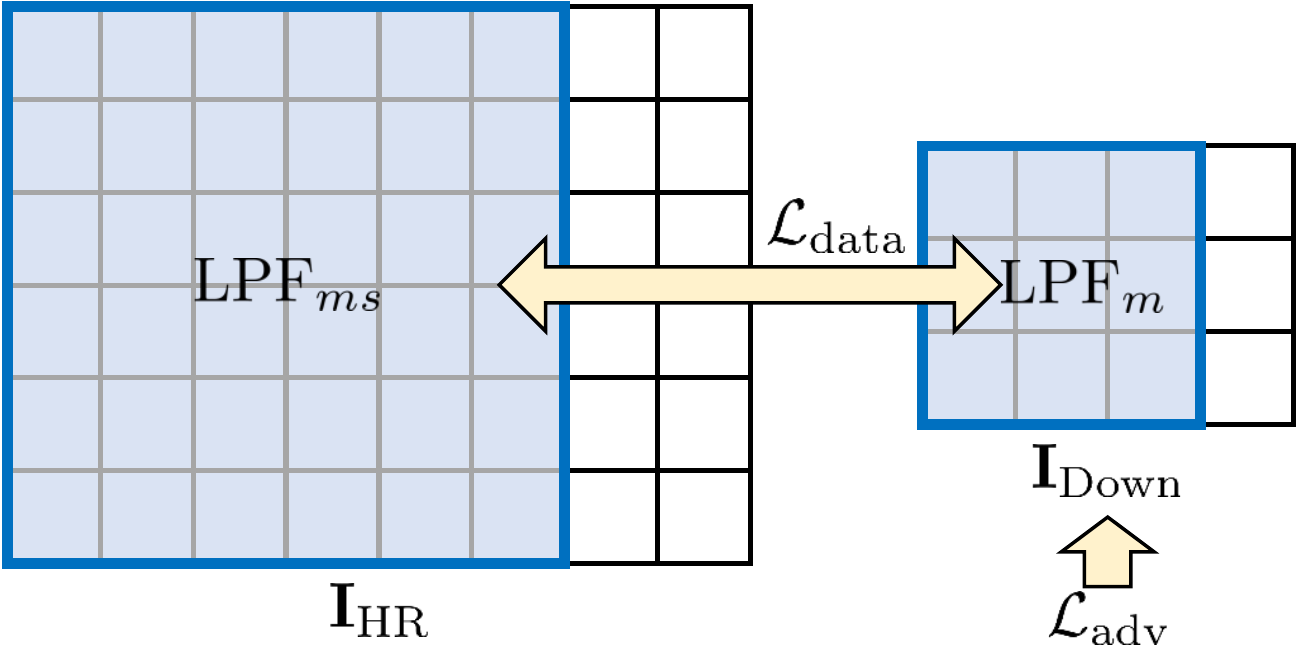}}
    \subfigure[\textbf{ADL~(Proposed)}]{\includegraphics[width=0.28\linewidth]{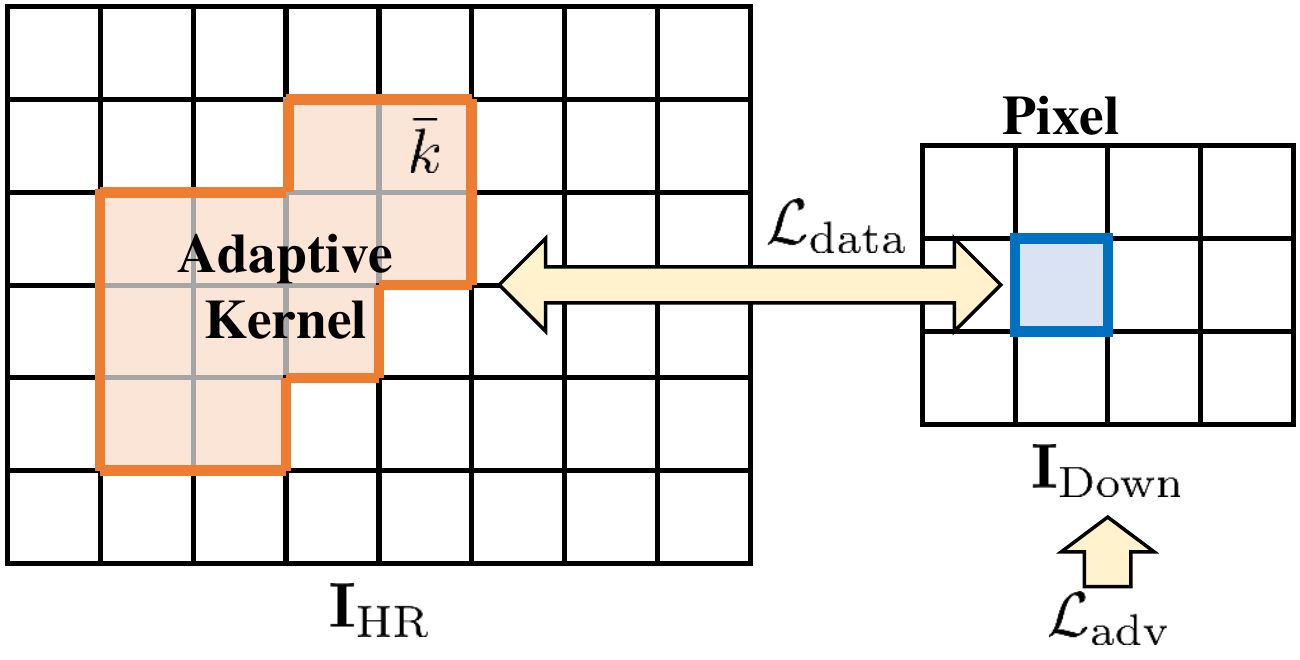}}
    \figspace
    \caption{
        \textbf{Different formulation for the data term}.
        We visualize how pixels in $\img{Down}$ is constrained to $\img{HR}$ depending on the data term $\mathcal{L}_\text{data}$.
        (a) Data loss from a predetermined kernel.
        (b) In the proposed LFL, we apply low-pass filters to HR and downsampled images so that image contents can be preserved across different scales regardless of the downsampling model.
        (c) In our adaptive data term, the orange kernel is learned from training samples and iteratively adjusted inside the training loops rather than handcrafted.
    }
    \label{fig:data}
    \figxspace
\end{figure*}

%% file: sections/figure/example_kernels.tex
\begin{figure*}[t!]
    \centering
    \renewcommand{\wp}{0.080}
    \subfigure[$\img{HR}$]{
        \includegraphics[width=\wp\linewidth]{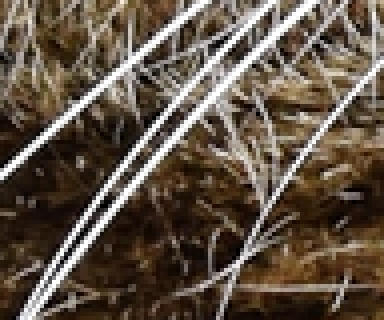}
    }
    \hfill
    \subfigure[$k_0$ and $\left( \img{HR} \ast k_0 \right)_{\downarrow 2}$]{
        \includegraphics[width=\wp\linewidth]{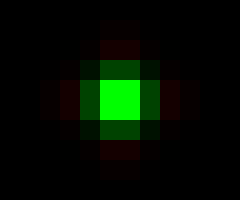}
        \hfill
        \includegraphics[width=\wp\linewidth]{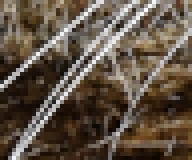}
    }
    \hfill
    \subfigure[$k_1$ and $\left( \img{HR} \ast k_1 \right)_{\downarrow 2}$]{
        \includegraphics[width=\wp\linewidth]{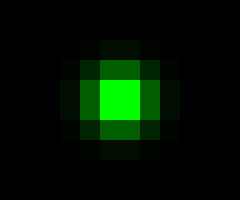}
        \hfill
        \includegraphics[width=\wp\linewidth]{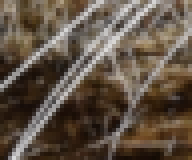}
    }
    \hfill
    \subfigure[$k_2$ and $\left( \img{HR} \ast k_2 \right)_{\downarrow 2}$]{
        \includegraphics[width=\wp\linewidth]{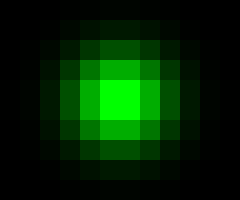}
        \hfill
        \includegraphics[width=\wp\linewidth]{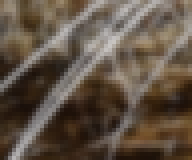}
    }
    \hfill
    \subfigure[$k_3$ and $\left( \img{HR} \ast k_3 \right)_{\downarrow 2}$]{
        \includegraphics[width=\wp\linewidth]{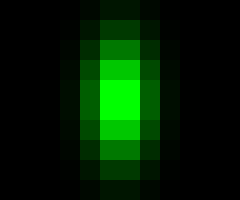}
        \hfill
        \includegraphics[width=\wp\linewidth]{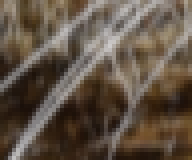}
    }
    \hfill
    \subfigure[$k_4$ and $\left( \img{HR} \ast k_4 \right)_{\downarrow 2}$]{
        \includegraphics[width=\wp\linewidth]{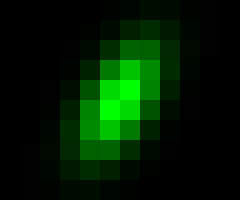}
        \hfill
        \includegraphics[width=\wp\linewidth]{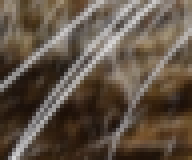}
    }
    \\
    \vspace{-2mm}
    \addtocounter{subfigure}{-1}
    \subfigure{
        \includegraphics[width=\wp\linewidth]{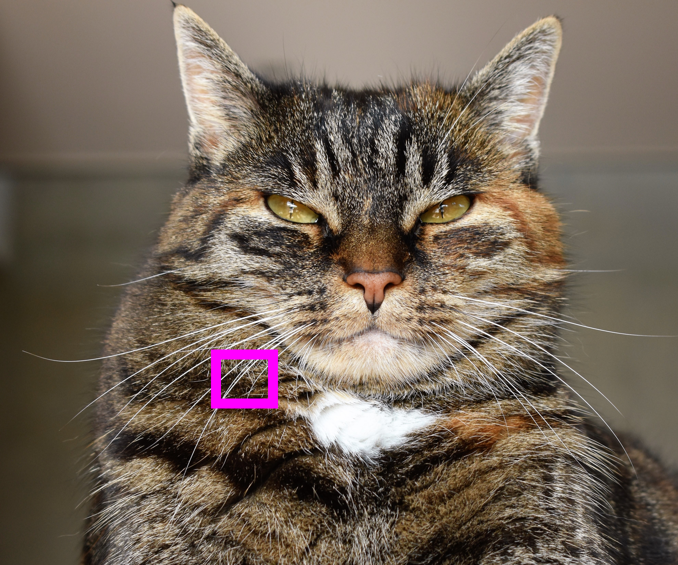}
    }
    \hfill
    \subfigure[$\bar{k}_0$ and $\mathcal{D}_0 \paren{ \img{HR} }$]{
        \includegraphics[width=\wp\linewidth]{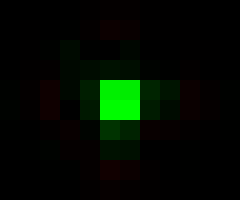}
        \hfill
        \includegraphics[width=\wp\linewidth]{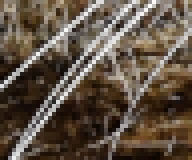}
    }
    \hfill
    \subfigure[$\bar{k}_1$ and $\mathcal{D}_1 \paren{ \img{HR} }$]{
        \includegraphics[width=\wp\linewidth]{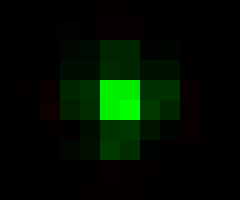}
        \hfill
        \includegraphics[width=\wp\linewidth]{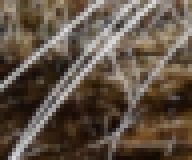}
    }
    \hfill
    \subfigure[$\bar{k}_2$ and $\mathcal{D}_2 \paren{ \img{HR} }$]{
        \includegraphics[width=\wp\linewidth]{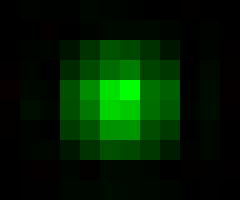}
        \hfill
        \includegraphics[width=\wp\linewidth]{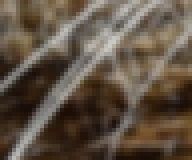}
    }
    \hfill
    \subfigure[$\bar{k}_3$ and $\mathcal{D}_3 \paren{ \img{HR} }$]{
        \includegraphics[width=\wp\linewidth]{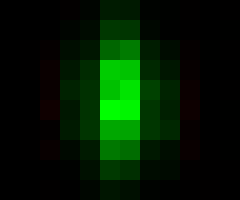}
        \hfill
        \includegraphics[width=\wp\linewidth]{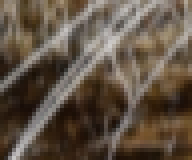}
    }
    \hfill
    \subfigure[$\bar{k}_4$ and $\mathcal{D}_4 \paren{ \img{HR} }$]{
        \includegraphics[width=\wp\linewidth]{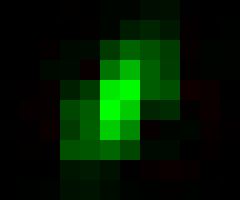}
        \hfill
        \includegraphics[width=\wp\linewidth]{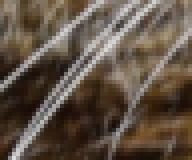}
    }
    \\
    \figspace
    \caption{
        \textbf{Examples of bicubic and randomly selected Gaussian kernels with corresponding $\times 2$ LR images.}
        (a) We use DIV2K~\cite{data_div2k} `\emph{0869.png}' for the HR image $\img{HR}$.
        (b)-(f) We note that the bicubic kernel $k_0$ contains positive (\textcolor{green}{\textbf{green}}) and small negative (\textcolor{red}{\textbf{red}}) values together. The former two Gaussian kernels $k_1$ and $k_2$ are isotropic, while later kernels $k_3$ and $k_4$ are anisotropic.
        We note that there exist subtle differences between images from different downsampling kernels.
        (g)-(k) We also visualize downsampled images $\img{Down}$ from the proposed ADL formulation.
        Here, $\mathcal{D}_i$ and $\bar{k}_i$ refer to the downsampling CNN $\mathcal{D}$ and approximated kernel $\bar{k}$ in Algorithm~\ref{alg:adl} that are learned on the synthetic DIV2K dataset from $k_i$.
        Kernel boundaries are cropped for better visualization.
        Best viewed with digital zoom.
    }
    \label{fig:kernels}
    \figxspace
\end{figure*}

%% file: sections/algorithm/adl.tex
\begin{algorithm}[t!]
    \caption{ADL for learning our downsampler $\mathcal{D}$}
    \renewcommand{\algorithmiccomment}[1]{\bgroup\hfill//~#1\egroup}
    \begin{algorithmic}[1]
    \REQUIRE
        Set of HR patches $\mathcal{I}_\text{HR}$,
        set of unpaired LR patches $\mathcal{I}_\text{LR}$,
        warm-up interval $t_\text{warm-up}$,
        update interval $t_\text{update}$,
        total training iterations $T$,
        and learning rate $\eta$.
    \ENSURE
        Downsampler parameters $\theta_\mathcal{D}$ and discriminator parameters $\theta_\mathcal{F}$. \\
    \STATE $\theta_\mathcal{D}, \theta_\mathcal{F} \leftarrow \mathcal{N} \left( 0, 0.02^2 \right)$.
    \COMMENT{Parameter initialization~\cite{gans_dc}.} \\
    \STATE $\bar{ k }$ = None. \\
    \FOR {$i = 1:T$}
        \STATE $\img{LR} \sim \mathcal{I}_\text{LR}$, $\img{HR} \sim \mathcal{I}_\text{HR}$.
        \COMMENT{Sample training batches.} \\
        \STATE $\img{Down} = \mathcal{D} \left( \img{HR}; \theta_\mathcal{D} \right)$. \\
        \STATE $\theta_\mathcal{F} \leftarrow \theta_\mathcal{F} - \eta \nabla_{\theta_\mathcal{F}} \mathcal{L}_\mathcal{F}$.
        \COMMENT{Update $\theta_\mathcal{F}$ by \eqref{eq:ds_obj}.} \\
        \IF {$i < t_\text{warm-up}$}
            \STATE Calculate $\mathcal{L}_\text{data}$ with \eqref{eq:la}. \\
        \ELSE
            \IF {$\bmod \paren{i, t_\text{update}} == 0$ or $\bar{ k }$ is None}
                \STATE Calculate $\bar{ k }$ from $\mathcal{D}$.
                \COMMENT{Retrieve the kernel.} \\
            \ENDIF
            \STATE Calculate $\mathcal{L}_\text{data} = \mathcal{L}_\text{ADL}$ with \eqref{eq:adl}. \\
        \ENDIF
        \STATE $\theta_\mathcal{D} \leftarrow \theta_\mathcal{D} - \eta \nabla_{\theta_\mathcal{D}} \mathcal{L}_\mathcal{D}$.
        \COMMENT{Update $\theta_\mathcal{D}$ by \eqref{eq:ds_obj}.} \\
    \ENDFOR
    \end{algorithmic}
    \label{alg:adl}
\end{algorithm}

%% file: sections/4_experiment.tex
\section{Experiments}
\label{sec_experiment}
%
%
We implement our method based on the PyTorch framework.
More results can be further provided in our Appendix and project site: https://cv.snu.ac.kr/research/ADL.
We will also release the source code and pre-trained models.

\subsection{Experimental setups}
\label{ssec:configs}
\Paragraph{Dataset.}
To validate whether our method can simulate an unknown distribution of LR images accurately, we construct a synthetic dataset by using a bicubic kernel ($k_0$) and the Gaussian kernels ($k_1$--$k_4$) with random shapes~\cite{sr_ikc, sr_blindsr}.
Then, we obtain LR inputs for the test from HR images by following \eqref{eq:linear}.
We visualize the different $\times 2$ degradation kernels $k_i$ used in our experiments and the corresponding LR images in Fig.~\ref{fig:kernels}.
For the $\times 4$ configurations, we use two times the enlarged versions of the $\times 2$ kernels.
More details about the selected kernels are described in Appendix~\fakeref{B}.

We construct unpaired training data by dividing 800 HR images from the DIV2K~\cite{data_div2k} training that is split by half.
For each degradation kernel, we assign 400 HR samples (`\emph{0001.png}'--`\emph{0400.png}') to $\mathcal{I}_\text{HR}$.
%
The remaining 400 images (`\emph{0401.png}'--`\emph{0800.png}') are used to synthesize LR samples and allocated to $\mathcal{I}_\text{LR}$.
The images do not overlap between $\mathcal{I}_\text{HR}$ and $\mathcal{I}_\text{LR}$.
With the proposed LFL and ADL formulation, the downsampler $\mathcal{D}$ can learn to simulate the distribution of LR samples $\mathcal{I}_\text{LR}$ by using the given HR images $\mathcal{I}_\text{HR}$.
For fair evaluations, we use another 100 images from the DIV2K~\cite{data_div2k} validation set to generate test inputs for different kernel $k_i$.

\Paragraph{Evaluation metrics.}
We evaluate our downsampling methods in two aspects.
As the primary goal of our methods is generating training examples to learn SR models on an unknown distribution of LR images, we generate pairs of $\left( \img{Down}, \img{HR} \right)$ using 400 HR images in $\mathcal{I}_\text{HR}$ with the learned downsampler $\mathcal{D}$ for each test degradation $k_i$.
Then, we train the SR model as described in Section~\ref{ssec:sr}, and report the PSNR values between the reconstructed images $\img{SR}$ and the reference HR images $\img{HR}$.
We note that generating more accurate LR images allows the following SR model to improve generalization on the inputs from unknown degradation.
In addition to the SR task, we also measure the PSNR values between the downsampled images $\img{Down}$ and ground-truth LR images to quantitatively evaluate the performance of the learned downsampling models $\mathcal{D}$.
All PSNR values are calculated using RGB channels rather than luminance.

\Paragraph{Model architecture.}
We use the patch-based discriminator~\cite{others_pix2pix} with the instance normalization~\cite{others_in} for training. 
For the SR task, we use a small EDSR~\cite{sr_edsr} model as the baseline with 1.5M parameters.
To demonstrate that our method is orthogonal to the selection of the SR backbone, we also introduce a larger RRDB~\cite{sr_esrgan} architecture with 16.7M parameters.
The details regarding our downsampling and discriminator CNNs are described in Appendix~\fakeref{E}.

\Paragraph{Hyperparameters.}
In all experiments, we use a $32 \times 32$ box filter for $\text{LPF}_{ms}$ and the one with $16 \times 16$ spatial size for $\text{LPF}_{m}$ with a scale factor of 2.
The ablation studies about the filter selection and relevant hyperparameters are described in Section~\ref{ssec:ablation}.
In training, one epoch consists of 1,741 iterations, which is proportional to the number of total training patches.
More details are provided in Appendix~\fakeref{F}.

\begin{table}[t!]
    \centering
    \renewcommand{\arraystretch}{1.05}
    \caption{
        \textbf{Evaluation of LR images from our unsupervised downsampler.} \protect\\
        We evaluate PSNR~(dB) between downsampled and ground-truth LR images on the synthetic DIV2K dataset for each kernel $k_i$.
        The best and second-best methods are \textbf{bolded} and \underline{underlined}, respectively.
    }
    \label{tab:eval_lr}
    \tabspace
    \begin{tabularx}{\linewidth}{p{2.8cm} >{\centering\arraybackslash}X >{\centering\arraybackslash}X >{\centering\arraybackslash}X >{\centering\arraybackslash}X >{\centering\arraybackslash}X >{\centering\arraybackslash}X}
        \toprule
        \multirow{2}{*}{$\mathcal{L}_\text{data}$} & \multirow{2}{*}{$s$} & \multicolumn{5}{c}{PSNR$^{\uparrow}$ between $\img{LR}$ and $\img{Down}$} \\
        & & $k_0$ & $k_1$ & $k_2$ & $k_3$ & $k_4$ \\
        \midrule
        $\normone{ \mathcal{B} \paren{ \img{HR} } - \img{Down} }$ & \multirow{4}{*}{$\times 2$} & 40.70 & 39.64 & 35.79 & 37.88 & 37.22 \\
        $\normone{ \text{AP}^{s} \paren{ \img{HR} } - \img{Down} }$ & & 40.33 & 38.54 & 36.03 & 37.35 & 35.13 \\
        $\mathcal{L}_\text{LFL}$ in \eqref{eq:la} (\textbf{Proposed}) & & \secondbest{43.16} & \secondbest{42.03} & \secondbest{43.30} & \secondbest{43.69} & \secondbest{43.75} \\
        $\mathcal{L}_\text{ADL}$ in \eqref{eq:adl} (\textbf{Proposed}) & & \best{45.83} & \best{45.61} & \best{46.34} & \best{44.86} & \best{46.41} \\
        \midrule
        $\normone{ \mathcal{B} \paren{ \img{HR} } - \img{Down} }$ & \multirow{4}{*}{$\times 4$} & 26.36 & 26.91 & 26.85 & 25.54 & 25.67 \\
        $\normone{ \text{AP}^{s} \paren{ \img{HR} } - \img{Down} }$ & & 24.18 & 25.63 & 25.40 & 26.64 & 26.70 \\
        $\mathcal{L}_\text{LFL}$ in \eqref{eq:la} (\textbf{Proposed}) & & \secondbest{31.13} & \secondbest{34.28} & \secondbest{39.66} & \secondbest{38.31} & \secondbest{37.17} \\
        $\mathcal{L}_\text{ADL}$ in \eqref{eq:adl} (\textbf{Proposed}) & & \best{38.24} & \best{38.12} & \best{43.54} & \best{39.08} & \best{41.10} \\
        \bottomrule \\
    \end{tabularx}
    \tabxspace
\end{table}

\begin{table}[t!]
    \centering
    \renewcommand{\arraystretch}{1.05}
    \caption{
        \textbf{Training configurations of different SR methods.} \protect \\
        All the other hyperparameters are kept fixed to train those SR models.
        We note that the downsampler $\mathcal{D}$ is learned for each specific degradation in an unsupervised manner.
    }
    \label{tab:data_synth}
    \tabspace
    \begin{tabularx}{\linewidth}{p{1.5cm} >{\centering\arraybackslash}X >{\centering\arraybackslash}X}
        \toprule
        Method & Training input & Training target \\
        \midrule
        Bicubic & $\mathcal{B} \paren{ \img{HR} } = \paren{ \img{HR} \ast k_0 }_{\downarrow s}$ & \multirow{3}{*}{$\img{HR}$} \\
        Oracle & $\paren{ \img{HR} \ast k_i }_{\downarrow s}$ & \\
        Proposed & $\mathcal{D} \paren{ \img{HR} }$ & \\
        \bottomrule
    \end{tabularx}
    \tabxspace
\end{table}

\subsection{Evaluating simulated LR images}
\label{ssec:eval_lr}
The primary contribution of our LFL and ADL is that they do not make a conflict with the adversarial loss, which guides downsampled images to resemble LR samples from an unknown distribution.
To demonstrate the advantages of the proposed framework when simulating an arbitrary downsampling process, we compare our LFL and ADL to data terms using predetermined operators.
Maeda~\cite{sr_unpaired_pesudo} proposes to utilize bicubic downsampled images in an unsupervised downsampling model, especially for cycle consistency and identity loss terms.
While the unsupervised learning approach from Maeda~\cite{sr_unpaired_pesudo} is not the same as our formulation, the objective between the generated LR and bicubic downsampled images can be interpreted as $\mathcal{R}_\text{HR} = \mathcal{B}$ in \eqref{eq:data}.
Similarly, Bulat~\etal~\cite{sr_gandeg} used an $s \times s$ average pooling ($\text{AP}^{s}$) for the resizing operator $\mathcal{R}_\text{HR}$ in the data term $\mathcal{L}_\text{data}$, where $s$ corresponds to a scaling factor.

\begin{table*}[t]
    \centering
    \renewcommand{\arraystretch}{1.05}
    \caption{
        \textbf{Blind super-resolution results on synthetic LR images.} \protect\\
        We show PSNR~(dB) between ground-truth HR and SR images from various methods on the synthetic DIV2K test dataset.
        Performance is not reported~($-$) if the pre-trained model is available only for a specific scale or cannot generate output images.
        $k_0$ refers to the bicubic kernel.
    }
    \label{tab:ablation_synthetic}
    \tabspace
    \begin{tabularx}{\linewidth}{p{3.6cm} >{\centering\arraybackslash}X >{\centering\arraybackslash}X >{\centering\arraybackslash}X >{\centering\arraybackslash}X >{\centering\arraybackslash}X >{\centering\arraybackslash}X >{\centering\arraybackslash}X >{\centering\arraybackslash}X >{\centering\arraybackslash}X  >{\centering\arraybackslash}X}
        \toprule
        \multirow{2}{*}{Method} & \multicolumn{5}{c}{PSNR$^{\uparrow}$ for $\times 2$ SR} & \multicolumn{5}{c}{PSNR$^{\uparrow}$ for $\times 4$ SR} \\
        & $k_0$ & $k_1$ & $k_2$ & $k_3$ & $k_4$ & $k_0$ & $k_1$ & $k_2$ & $k_3$ & $k_4$ \\
        \midrule
        EDSR~\cite{sr_edsr} (Bicubic) & 34.61 & 31.51 & 27.76 & 27.91 & 27.95 & 28.92 & 26.35 & 24.06 & 24.21 & 24.21 \\
        RRDB~\cite{sr_esrgan} (Bicubic) & $-$ & $-$ & $-$ & $-$ & $-$ & 29.45 & 26.44 & 24.07 & 24.22 & 24.22 \\
        \midrule
        EDSR (Oracle) & 34.61 & 34.44 & 33.64 & 33.23 & 33.27 & 28.92 & 28.73 & 28.02 & 27.79 & 27.84 \\
        RRDB (Oracle) & $-$ & $-$ & $-$ & $-$ & $-$ & 29.45 & 29.28 & 28.39 & 28.08 & 28.62 \\
        \midrule
        KernelGAN~\cite{sr_kernelgan} + ZSSR~\cite{sr_zssr} & 22.32 & 26.42 & 30.44 & 29.10 & 29.12 & 20.11 & 24.67 & 25.85 & 25.21 & 25.36 \\
        IKC~\cite{sr_ikc} & $-$ & $-$ & $-$ & $-$ & $-$ & \best{28.59} & \secondbest{28.07} & \best{27.65} & 24.15 & 25.12 \\
        BlindSR~\cite{sr_blindsr} & 26.56 & $-$ & 26.62 & 26.49 & $-$ & $-$ & $-$ & $-$ & $-$ & $-$ \\
        \midrule
        LFL + EDSR (\textbf{Proposed}) & \secondbest{33.91} & \secondbest{33.26} & \secondbest{31.38} & \secondbest{31.48} & \secondbest{31.57} & 27.45 & 27.31 & 26.69 & 26.65 & 26.33 \\
        ADL + EDSR (\textbf{Proposed}) & \best{34.07} & \best{33.68} & \best{32.51} & \best{32.08} & \best{32.05} & 28.16 & 28.04 & 27.08 & \secondbest{26.82} & \secondbest{26.97} \\
        ADL + RRDB (\textbf{Proposed}) & $-$ & $-$ & $-$ & $-$ & $-$ & \secondbest{28.55} & \best{28.49} & \secondbest{27.51} & \best{27.00} & \best{27.19} \\
        \bottomrule \\
    \end{tabularx}
    \tabxspace
\end{table*}

We note that direct comparisons between ours and the existing generation-based methods~\cite{sr_unpaired_pesudo, sr_gandeg}, including Lugmayr~\etal~\cite{sr_unsupervised}
are not conducted due to several reasons.
First, we explicitly find the unknown downsampling operator, while previous approaches focus on modeling noise and artifacts in real-world LR images.
In addition, as those methods do not provide source code, evaluation on diverse synthetic kernels cannot be carried out for fair comparisons. 
Therefore, we train multiple downsampling networks under different data terms $\mathcal{L}_\text{data}$ on different synthetic kernels ($k_0$--$k_4$) and scales ($\times 2$ and $\times 4$).
Then, we compare how the proposed data term outperforms the previous formulations in terms of the feasibility of the synthesized samples.

Table~\ref{tab:eval_lr} illustrates the average PSNR between the generated LR images from each downsampler and ground-truth.
%
%
When the predetermined operator is well-matched with a ground-truth downsampling function, e.g., using $\mathcal{B}$ to estimate $k_0$, the unsupervised models effectively simulate target LR images.
However, if the predetermined functions (bicubic and average pooling) are not overlapped with the unknown degradation kernel ($k_1$--$k_4$), the data term $\mathcal{L}_\text{data}$ biases the training objective and conflicts with the adversarial loss.
Table~\ref{tab:eval_lr} demonstrates that the conflict affects the learned downsampling model in a negative way and makes the SR model less generalizable, even with synthetic kernels.
On the other hand, the proposed LFL and ADL terms can be generalized better and facilitate the downsampler $\mathcal{D}$ to generate accurate LR images for various configurations.

\subsection{SR on the synthetic examples}
\label{ssec:synth}
%
%
Using generated LR images from our downsampler, we train baseline EDSR~\cite{sr_edsr} and RRDB~\cite{sr_esrgan} and evaluate them on each degradation kernel $k_i$ individually on three different configurations described in Table~\ref{tab:data_synth}.
In the bicubic configuration, bicubic-downsampled images are used to train the SR model as those in existing approaches~\cite{sr_srcnn, sr_vdsr, sr_srgan}.
We note that the bicubic models are shared across different setups.
In contrast, our method first learns a degradation-specific downsampling model $\mathcal{D}$ from unpaired LR and HR images and leverages it to generate training samples for the SR model.
We also introduce an oracle for each degradation $k_i$, where the SR model can fully utilize the ground-truth kernel to synthesize training images.
As the distributions of the training and test images are matched, the oracle serves as an upper bound for a specific degradation kernel $k_i$.

Table~\ref{tab:ablation_synthetic} compares various SR methods on the synthetic DIV2K dataset.
EDSR and RRDB trained on bicubic LR images perform well when the inputs are also formed by the bicubic kernel ($k_0$).
However, they do not generalize well when the inputs are downsampled by different kernels ($k_1$--$k_4$), as the distribution of the test images deviates significantly from that of the training samples.
Also, the larger RRDB network does not bring any advantages over the smaller EDSR model, showing that the bicubic SR models cannot be generalized to unknown types of LR images.

\input{sections/table/realsr}

On the other hand, EDSR and RRDB achieve significant performance gains over the other approaches when the training LR images are generated from the proposed LFL and ADL.
As discussed previously in Section~\ref{ssec:eval_lr}, our approach can generate a set of \emph{faithful} LR and HR training pairs so that the following SR models can achieve much better performance on LR images from some unknown downsampling process.
Since LFL does not bias the learned downsampler $\mathcal{D}$ to a specific downsampling operator, e.g., bicubic or average pooling, the respective EDSR and RRDB generalize well across various kernels ($k_0$--$k_4$) and scales ($\times 2$ and $\times 4$).
Furthermore, our downsampling model with ADL generates more accurate training LR images for the SR models and brings significant improvements to EDSR and RRDB across all kernel configurations consistently.

Interestingly, a larger RRDB model with ADL achieves better performance and even comparable to the oracle EDSR, especially on the $\times 4$ SR task with $k_0$ and $k_1$.
If the distribution of the downsampled images, i.e., $\mathcal{D} \paren{ \img{HR} }$, deviates much from that of the target LR images, then a better fitting to the training data may worsen the performance on the test images. 
Therefore, the performance gain of the RRDB model demonstrates that our unsupervised downsampling framework can faithfully simulate the distribution of the target LR images to a certain extent.
We also note that our data generation process is orthogonal to the architecture of the SR models.
Therefore, integrating the proposed downsampling models with state-of-the-art SR architectures~\cite{sr_dbpn, sr_rcan} can directly improve performance.

We also apply the existing approaches to various synthetic degradation kernels.
The combination of KernelGAN~\cite{sr_kernelgan} and ZSSR~\cite{sr_zssr} first estimates an input-specific degradation kernel from a single image~\cite{sr_kernelgan} and applies the zero-shot SR model~\cite{sr_zssr} to deal with the arbitrary LR images.
Compared with bicubic EDSR and RRDB, the single image approach achieves better performance on $k_2$--$k_4$, demonstrating the importance of estimating image-specific kernel modeling for the blind SR task.
However, this method does not perform well when test images are downsampled by $k_0$ and $k_1$, as depicted by the significant decrease in PSNR with respect to the oracle model.

Instead of synthesizing realistic LR images, IKC~\cite{sr_ikc} and BlindSR~\cite{sr_blindsr} utilize large-scale synthetic data in which the degradation kernels follow specific shapes.
They first predict the kernel used to generate the given LR image and derive input-dependent SR results within a single network architecture.
As the IKC model only considers isotropic Gaussian kernels, it achieves comparable performance to the oracle models on $k_0$--$k_2$.
However, it does not perform well on the $k_3$ and $k_4$ cases, where the degradation kernels are anisotropic.
While BlindSR~\cite{sr_blindsr} also takes a similar strategy, it is less stable and unable to handle some inputs from $k_1$ and $k_4$, and it also diverges.
As shown in Fig.~\ref{fig:results}~(first row), RRDB that has learned on our training data can reconstruct realistic details from the challenging $\times 4$ SR task.

\input{sections/figure/qualitative}

\subsection{SR on the RealSR-V3 dataset}
\label{ssec:realsr}
Our approach can also be applied to real-world LR images from unknown camera processing pipelines.
For the quantitative evaluation, we utilize RealSR-V3~\cite{sr_realworld} containing 200 well-aligned LR and HR image pairs for two different real-world cameras: Canon and Nikon.
Similar to the description in Section~\ref{ssec:configs}, we divide the dataset by half for each camera model.
The same amounts of images are assigned to $\mathcal{I}_\text{HR}$ and $\mathcal{I}_\text{LR}$ without overlapping.
We learn the corresponding downsampling and SR models by following our pipeline, as described in Section~\ref{ssec:configs}.
As the dataset provides accurately aligned LR and HR examples to train a supervised SR model, the oracle models learn from those image pairs.

Compared to the experiments on synthetic images in Section~\ref{ssec:synth}, real-world cases are more challenging.
\revised{First, a set of LR images may share the similar but not exactly the same degradation process.}
Also, since the dataset mainly consists of indoor scenes and static objects without large motions, training the images may lack the diversity that can hinder generalization.
Table~\ref{tab:ablation_realsr} shows the results of the evaluated SR algorithms on RealSR-V3, where each of the Canon and Nikon split contains 50 test images.
KernelGAN + ZSSR, IKC, and BlindSR do not perform well even compared to EDSR and RRDB learned on bicubic downsampled images.
The primary reason is that numerous constraints in these methods, e.g., Gaussian kernels~\cite{sr_ikc, sr_blindsr} or kernel shape priors~\cite{sr_kernelgan}, do not usually hold for real-world scenes.

In contrast, our downsampling method with different SR backbones (LFL + EDSR, ADL + EDSR, and ADL + RRDB) achieve better results on both cameras at different scales ($\times 2$ and $\times 4$), compared with the other approaches.
Even if the LR images in RealSR-V3 are not formulated from the same kernel, our LFL and ADL can learn an average of all possible downsampling operators and generalize well on the real-world dataset.
We also demonstrate that the larger RRDB model performs better in the realistic case, showing that a better fitting on the generated LR images can help generalization on unseen real-world examples.
Fig.~\ref{fig:results} shows that our approach can reconstruct more visually pleasing results than the existing methods on RealSR-V3.

\input{sections/table/ablation}
\input{sections/figure/kernel_evolution}

\subsection{Ablation study}
\label{ssec:ablation}
To see the contribution of each design component in our LFL and ADL, we conduct extensive ablation studies in this section.
The selected hyperparameters are used throughout the entire experiments in Section~\ref{ssec:synth}, \ref{ssec:realsr}, and \ref{ssec:real}, without any additional adjustments.
We note that an ablation regarding the shape of $\text{LPF}_{m}$ is described in our Appendix~\fakeref{A}.
The stability of our LFL and ADL is described in Appendix~\fakeref{C}.

\Paragraph{Effect of the balancing hyperparameter.}
As we describe in Section~\ref{ssec:data_term}, the predetermined downsampling operator $\mathcal{R}_\text{HR}$ may bias the overall training objective of the downsampler $\mathcal{L}_\mathcal{D}$, especially when the balancing hyperparameter $\alpha$ is large.
To validate that our LFL and ADL terms do not impose negative bias in the learning stage, we analyze how the balance between data and adversarial losses in \eqref{eq:ds_obj} affects the associated SR models.
Table~\ref{tab:ablation_alpha} shows that the SR model with  
LFL and ADL do not perform well when $\alpha = 1$ or $10$, as preserving image content is challenging during the downsampling process.
%
When using relative larger values of $\alpha$, i.e., $\alpha = 100$ or $200$, the baseline SR models with LFL and ADL terms perform reasonably well without making bias.
As such, we choose $\alpha = 200$ for the LFL and $\alpha = 100$ for the ADL to achieve the best performance.

\Paragraph{Effect of the number of training samples.}
In Section~\ref{ssec:synth}, we train the proposed downsampling model on 400 LR images generated from the \emph{same} kernel $k_i$.
Compared with the KernelGAN~\cite{sr_kernelgan} method, which predicts a proper degradation kernel from a single input image, our approach requires more examples to estimate an unknown degradation accurately.
For a fair comparison, we vary the number of LR images to train the downsampling model and analyze the performance of the following SR models.
We use the first 1, 10, 50, and 100 examples from $\mathcal{I}_\text{LR}$, e.g., `\emph{0401.png}' for the $\lvert \mathcal{I}_\text{LR} \rvert = 1$ case, to learn our downsampling model.
The other hyperparameters are fixed unless mentioned otherwise.

Table~\ref{tab:ablation_n} shows how the size of $\mathcal{I}_\text{LR}$ for the downsampler affects the following SR performance.
Our methods (LFL + EDSR and ADL + EDSR) gradually achieve better performance as the number of training samples increases, validating the effectiveness of using large-scale datasets.
Nevertheless, even with a single LR sample, ADL + EDSR outperforms the single-image method.
Table~\ref{tab:ablation_n} also shows that ADL consistently outperforms LFL, especially when the number of training LR images is limited.
Specifically, ADL + EDSR with a single LR image performs equally well as LFL + EDSR with 50 LR images.
In Appendix~\fakeref{D}, we also analyze some opposite cases for fair comparison where KernelGAN is trained with multiple LR and HR images.

\input{sections/figure/perceptual}

\Paragraph{Joint training of downsampling and SR networks.}
Our two-stage (downsampling + SR) pipeline has several advantages. 
First, if the two models are jointly learned, one may affect the other to be suboptimal solution.
For example, the downsampler may generate LR images that can be easily upsampled rather than accurately simulating the desired target.
In addition, connecting the two models increases the algorithmic complexity, making hard to train the whole model.
Finally, the two-stage approach accommodates more effective models and objective functions, as we have a fixed downsampling network and corresponding LR images.

On the other hand, it is possible to optimize the downsampling and SR networks jointly.
Table~\ref{tab:ablation_joint} provides experimental results of joint training with different setups.
To reduce the training time, we optimize downsampling and SR networks together (joint), contrary to the original formulation (two-stage).
We note that the gradient from the SR model does not backpropagate to the downsampler in this configuration.
Interestingly, the joint training approach achieves marginal performance gain to the SR network.
Since we do not fix input of the SR network and keep updating the downsampler, it has similar effects to data augmentation and slightly improves the following SR model.

We also train the downsampling and SR networks together in an end-to-end manner, where backpropagated gradient from the SR model flows to the downsampler (+BP).
However, this approach negatively affects the following SR network in two specific aspects.
First, the downsampler tends to generate images that are easy to be upscaled rather than accurately simulating samples in the target LR distribution.
Second, each color pixel in the generated image is a continuous variable, while pixels in our test samples only have 256 discontinuous values.
We further introduce fake quantization (+BP+FQ) to deal with the second issue, where output of the downsampler are quantized while the gradient flows just as the pixel values are continuous.
Although it brings $+$0.11dB performance gain, the end-to-end learning does not bring any advantage in our framework.
As such, we use the two-stage approach in all the experiments.

\input{sections/table/ablation_adl}

\subsection{Analysis on ADL}
\label{ssec:ablation_adl}
As the estimated kernel $\bar{k}$ in Algorithm~\ref{alg:adl} is derived from the training dataset, the ADL does not make a significant conflict with the adversarial loss $\mathcal{L}_\text{adv}$.
To demonstrate the effectiveness of our adaptive adjustment strategy, we visualize how the estimated kernels $\bar{k}$ on the synthetic DIV2K and RealSR-V3 datasets are updated in Fig.~\ref{fig:akl_kernels}.
We note that the kernels from the RealSR-V3~\cite{sr_realworld} dataset do not appear to be standard Gaussian forms that are preferred in the existing approaches~\cite{sr_ikc, sr_blindsr}.
Since we do not constrain the downsampling network to resemble specific shapes of kernels, our approach can yield better generalizability.

In the synthetic cases, the estimated kernel $\bar{k}$ should be similar to the ground-truth $k_i$ for the following SR training.
Thus, we show the kernel similarity~\cite{rs_deblur_mh} between the retrieved and ground-truth degradation kernels in Table~\ref{tab:ablation_adl_a} to validate that our prediction becomes more accurate as the training proceeds.
It is also demonstrated that the more precise estimation supports the following SR network to perform better, and the performance is maximized at 80 epochs.
Furthermore, Table~\ref{tab:ablation_adl_b} shows that our iterative update prevents the downsampler from being biased toward the fixed kernel $\bar{k}$ and helps with the convergence.
As we describe in Section~\ref{ssec:adl}, ADL is designed to stabilize the potentially noisy downsampling model.
Therefore, the number of samples $N$ for the kernel estimation is also crucial.
Table~\ref{tab:ablation_adl_c} demonstrates that the ADL does not perform well only with a single training image.
However, as the number of examples increases, our method can assist the following SR model to generate better results.

\subsection{SR on the real-world images}
\label{ssec:real}
Our methods assume that the set of available LR images follow the same downsampling process.
In practice, acquiring multiple images from a similar degradation pipeline, such as photos from a fixed camera configuration~\cite{data_dped} or multiple frames in a video, is relatively easier than collecting real-world LR and HR pairs.
We validate the proposed method by using the DPED~\cite{data_dped} dataset, which consists of low-quality photos captured by an iPhone 3GS camera.
To train the downsampler, we assign random 120 images from the DPED~\cite{data_dped} dataset as $\mathcal{I}_\text{LR}$, while DIV2K is used for the HR samples $\mathcal{I}_\text{HR}$.
Since the dataset consists of low-quality examples captured by the iPhone 3GS camera, we remove the unknown noise and artifacts in $\mathcal{I}_\text{LR}$ by applying the off-the-shelf RL-restore~\cite{rl_restore} algorithm as a preprocessing.
Fig.~\ref{fig:results2} compares the SR results of the preprocessed DPED~\cite{data_dped} dataset.
As no ground-truth HR images exist in this dataset, we note that no oracle model is available for the dataset.
Compared with the existing methods, our ADL facilitates RRDB~\cite{sr_esrgan} to reconstruct sharper edges and more detailed textures without introducing visual artifacts.
Moreover, we train the ADL-based downsampler without RL-restore~\cite{rl_restore}.

We also train the perceptual SR network $\mathcal{P}$ using the images from ADL-based downsampler to demonstrate the merit of our approach for real-world SR.
Compared with the PSNR-based model (\textbf{ADL}~$+$~RRDB) trained with \eqref{eq:sr}, the perceptual RRDB model (\textbf{ADL}~$+$~$\mathcal{P}$) from \eqref{eq:sr_percep} reconstructs more realistic and visually pleasing results.
The advantage of the two-stage approach is that we do not require additional training of the downsampler to introduce different optimization objectives for the SR network.
Thus, the only difference between Fig.~\ref{fig:results2_plain} and Fig.~\ref{fig:results2_per} is training loss while the backbone networks are the same.
More additional qualitative SR results are presented in Appendix~\fakeref{G}.

\revised{
In Fig.~\ref{fig:limitation}, we analyze how much our ADL is affected by noise and artifacts.
For comparison, we evaluate the pretrained RealSR~\cite{sr_rwsr} and BSRGAN~\cite{sr_bsrgan} models on the same DPED images.
Since those methods explicitly consider noise and artifacts in LR inputs, their results look robust to such degradation.
In contrast, our ADL yields sharper but a bit noisy outputs. 
Note that when the preprocessing is not applied (\textbf{ADL-n}), it gives blurry SR results since noisy LR samples prevent the discriminator from learning distinguishable features from downsampled images.
}
\input{sections/figure/limitation}

%% file: sections/table/realsr.tex
\begin{table*}[t]
    \centering
    \renewcommand{\arraystretch}{1.05}
    \caption{
        \textbf{Blind super-resolution results on realistic LR images.} \protect\\
        We provide PSNR~(dB) between ground-truth HR and SR results from different methods on the RealSR-V3~\cite{sr_realworld} dataset.
        Since the KernelGAN~\cite{sr_kernelgan} and ZSSR~\cite{sr_zssr} combination is learned on \emph{each} test image, they require $\times 50$ parameters in practice to handle 50 inputs.
    }
    \label{tab:ablation_realsr}
    \tabspace
    \begin{tabularx}{\linewidth}{p{4cm} p{2.5cm} >{\centering\arraybackslash}X >{\centering\arraybackslash}X >{\centering\arraybackslash}X >{\centering\arraybackslash}X >{\centering\arraybackslash}X}
        \toprule
        \multirow{2}{*}{Method} & \multirow{2}{*}{\# Parameters} & \multirow{2}{*}{Training data} & \multicolumn{2}{c}{PSNR$^{\uparrow}$ for $\times 2$ SR} & \multicolumn{2}{c}{PSNR$^{\uparrow}$ for $\times 4$ SR} \\
        & & & Canon & Nikon & Canon & Nikon \\
        \midrule
        EDSR~\cite{sr_edsr} (Bicubic) & 1.5M & \multirow{2}{*}{\makecell{Synthetic \\ (Bicubic $k$)}} & 30.58 & 30.00 & 26.05 & 25.89 \\
        RRDB~\cite{sr_esrgan} (Bicubic) & 16.7M & & $-$ & $-$ & 26.05 & 25.91 \\
        \midrule
        EDSR (Oracle) & 1.5M & \multirow{2}{*}{\makecell{RealSR-V3 \\ (Paired)}} & 32.45 & 31.59 & 27.59 & 27.14 \\
        RRDB (Oracle) & 16.7M & & $-$ & $-$ & 27.90 & 27.39 \\
        \midrule
        KernelGAN~\cite{sr_kernelgan} + ZSSR~\cite{sr_zssr} & 50~$\times$~(0.2M $+$ 0.2M) & A given $\img{LR}$ & 28.79 & 27.54 & 23.68 & 22.46 \\
        \midrule
        IKC~\cite{sr_ikc} & 9.0M & \multirow{2}{*}{\makecell{Synthetic \\ (Multiple $k$)}} & $-$ & $-$ & 25.71 & 25.27 \\
        BlindSR~\cite{sr_blindsr} & 1.1M & & 25.80 & 24.17 & $-$ & $-$ \\
        \midrule
        LFL + EDSR (\textbf{Proposed}) & 0.9M $+$ 1.5M & \multirow{3}{*}{\makecell{RealSR-V3 \\ \textbf{(Unpaired)}}} & \secondbest{31.67} & \secondbest{30.75} & 26.47 & 25.90 \\
        ADL + EDSR (\textbf{Proposed}) & 0.9M $+$ 1.5M & & \best{31.81} & \best{30.99} & \secondbest{26.79} & \secondbest{26.46} \\
        ADL + RRDB (\textbf{Proposed}) & 0.9M $+$ 16.7M & & $-$ & $-$ & \best{26.90} & \best{26.64} \\
        \bottomrule \\
    \end{tabularx}
    \tabxspace
    \vspace{-2mm}
\end{table*}

%% file: sections/figure/qualitative.tex
\begin{figure*}[t]
    \centering
    \renewcommand{\wp}{0.128}
    \subfigure{
        \includegraphics[width=\wp\linewidth]{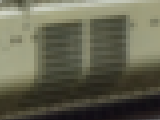}
    }
    \hfill
    \subfigure{
        \includegraphics[width=\wp\linewidth]{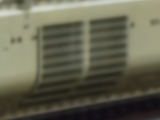}
    }
    \hfill
    \subfigure{
        \includegraphics[width=\wp\linewidth]{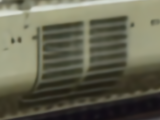}
    }
    \hfill
    \subfigure{
        \includegraphics[width=\wp\linewidth]{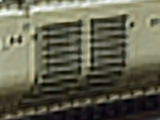}
    }
    \hfill
    \subfigure{
        \includegraphics[width=\wp\linewidth]{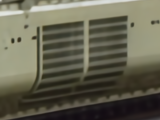}
    }
    \hfill
    \subfigure{
        \includegraphics[width=\wp\linewidth]{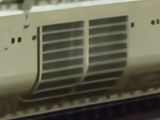}
    }
    \hfill
    \subfigure{
        \includegraphics[width=\wp\linewidth]{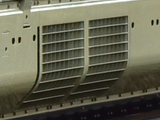}
    }
    \\
    \vspace{-2mm}
    \subfigure{
        \includegraphics[width=\wp\linewidth]{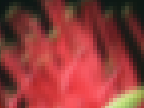}
    }
    \hfill
    \subfigure{
        \includegraphics[width=\wp\linewidth]{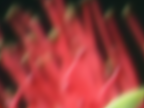}
    }
    \hfill
    \subfigure{
        \includegraphics[width=\wp\linewidth]{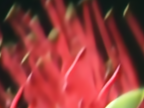}
    }
    \hfill
    \subfigure{
        \includegraphics[width=\wp\linewidth]{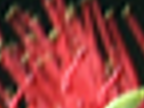}
    }
    \hfill
    \subfigure{
        \includegraphics[width=\wp\linewidth]{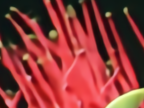}
    }
    \hfill
    \subfigure{
        \includegraphics[width=\wp\linewidth]{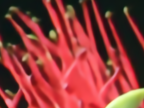}
    }
    \hfill
    \subfigure{
        \includegraphics[width=\wp\linewidth]{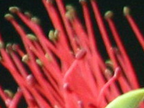}
    }
    \\
    \vspace{-2mm}
    \subfigure{
        \includegraphics[width=\wp\linewidth]{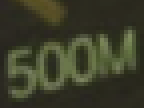}
    }
    \hfill
    \subfigure{
        \includegraphics[width=\wp\linewidth]{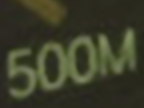}
    }
    \hfill
    \subfigure{
        \includegraphics[width=\wp\linewidth]{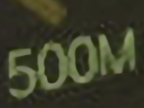}
    }
    \hfill
    \subfigure{
        \includegraphics[width=\wp\linewidth]{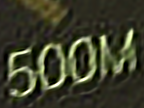}
    }
    \hfill
    \subfigure{
        \includegraphics[width=\wp\linewidth]{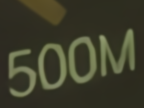}
    }
    \hfill
    \subfigure{
        \includegraphics[width=\wp\linewidth]{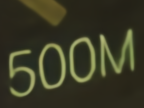}
    }
    \hfill
    \subfigure{
        \includegraphics[width=\wp\linewidth]{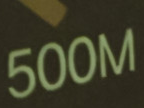}
    }
    \\
    \vspace{-2mm}
    \addtocounter{subfigure}{-21}
    \subfigure[$\img{LR}$~(Input)]{
        \includegraphics[width=\wp\linewidth]{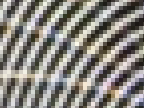}
    }
    \hfill
    \subfigure[RRDB~\cite{sr_esrgan}]{
        \includegraphics[width=\wp\linewidth]{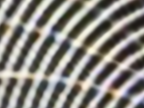}
    }
    \hfill
    \subfigure[IKC~\cite{sr_ikc}]{
        \includegraphics[width=\wp\linewidth]{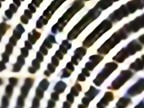}
    }
    \hfill
    \subfigure[\cite{sr_kernelgan}~$+$~\cite{sr_zssr}]{
        \includegraphics[width=\wp\linewidth]{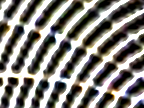}
    }
    \hfill
    \subfigure[\textbf{ADL}~$+$~\cite{sr_esrgan}]{
        \includegraphics[width=\wp\linewidth]{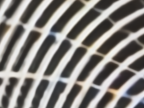}
    }
    \hfill
    \subfigure[Oracle]{
        \includegraphics[width=\wp\linewidth]{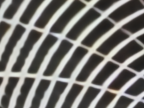}
    }
    \hfill
    \subfigure[$\img{HR}$~(GT)]{
        \includegraphics[width=\wp\linewidth]{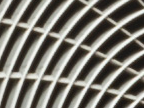}
    }
    \\
    \figspace
    \caption{
        \textbf{Qualitative $\times 4$ SR results on the various datasets.}
        Patches in each row are from the synthetic DIV2K~\cite{data_div2k} dataset `\emph{0820.png}~($k_1$),' `\emph{0853.png}~($k_4$),' RealSR-V3~\cite{sr_realworld} dataset `\emph{Canon/006.png},'  and `\emph{Nikon/041.png},' respectively.
        %
        The RRDB~\cite{sr_esrgan} model is used as a backbone SR architecture for our ADL as well as the Oracle.
    }
    \label{fig:results}
    \figxspace
\end{figure*}

%% file: sections/table/ablation.tex
\begin{table}[t!]
    \centering
    \renewcommand{\arraystretch}{1.05}
    \caption{
        \textbf{Ablation studies on the proposed method.} \protect\\
        We report how different training configurations for the downsampling network affect the SR results on the synthetic DIV2K dataset.
    }
    \label{tab:ablation}
    \tabspace
    \subfigure[Effect of the balancing parameter $\alpha$ in \eqref{eq:ds_obj}. \label{tab:ablation_alpha}]{
        \begin{tabularx}{\linewidth}{p{3.2cm} >{\centering\arraybackslash}X >{\centering\arraybackslash}X >{\centering\arraybackslash}X >{\centering\arraybackslash}X}
            \toprule
            \multirow{2}{*}{Method $\backslash$ $\alpha$} & \multicolumn{4}{c}{PSNR$^{\uparrow}$ for $\times 2$ SR ($k_4$)} \\
            & 1 & 10 & 100 & 200 \\
            \midrule
            LFL + EDSR (\textbf{Proposed}) & 29.36 & 30.70 & \secondbest{31.49} & \best{31.57} \\
            ADL + EDSR (\textbf{Proposed}) & 29.08 & 31.42 & \best{32.05} & \secondbest{32.02} \\
            \bottomrule
        \end{tabularx}
    }
    \\
    \subfigure[Effect of the number of training LR images $\lvert \mathcal{I}_\text{LR} \rvert$. \label{tab:ablation_n}]{
        \begin{tabularx}{\linewidth}{p{3.2cm} >{\centering\arraybackslash}X >{\centering\arraybackslash}X >{\centering\arraybackslash}X >{\centering\arraybackslash}X}
            \toprule
            \multirow{2}{*}{Method $\backslash$ $\lvert \mathcal{I}_\text{LR} \rvert$} & \multicolumn{4}{c}{PSNR$^{\uparrow}$ for $\times 2$ SR~($k_4$)} \\
            & 1 & 10 & 50 & 100 \\
            \midrule
            KernelGAN + ZSSR & 29.12 & $-$ & $-$ & $-$ \\
            \midrule
            LFL + EDSR (\textbf{Proposed}) & 25.43 & 28.70 & 29.86 & 31.21 \\
            ADL + EDSR (\textbf{Proposed}) & 29.84 & 30.20 & 31.36 & 32.10 \\
            \bottomrule
        \end{tabularx}
    }
    \\
    \subfigure[Effect of the joint training. \label{tab:ablation_joint}]{
            \begin{tabularx}{\linewidth}{p{3.2cm} >{\centering\arraybackslash}X c c c}
        \toprule
        \multirow{2}{*}{Method $\backslash$ Configuration} & \multicolumn{4}{c}{PSNR$^\uparrow$~(dB) for $\times 2$ SR ($k_4$)} \\
        & Joint & +BP & +BP+FQ & Two-stage \\
        \midrule
        ADL + EDSR (\textbf{Proposed}) & \best{32.09} & 31.39 & 31.50 & \secondbest{32.05} \\
        \bottomrule
    \end{tabularx}
    }
    \tabxspace
\end{table}

%% file: sections/figure/kernel_evolution.tex
\begin{figure*}[t!]
    \centering
    \renewcommand{\wp}{0.105}
    \makecell{Synthetic \\ DIV2K ($k_3$)}
    \hfill
    \subfigure{\includegraphics[align=c, width=\wp \linewidth]{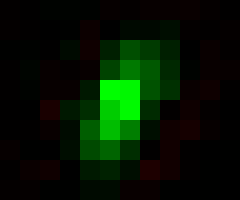}}
    \hfill
    \subfigure{\includegraphics[align=c, width=\wp \linewidth]{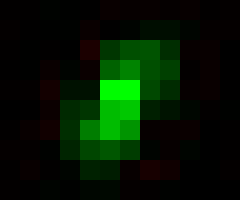}}
    \hfill
    \subfigure{\includegraphics[align=c, width=\wp \linewidth]{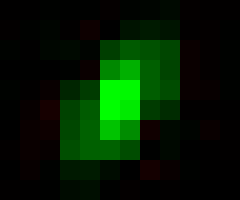}}
    \hfill
    \subfigure{\includegraphics[align=c, width=\wp \linewidth]{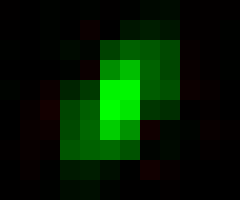}}
    \hfill
    \subfigure{\includegraphics[align=c, width=\wp \linewidth]{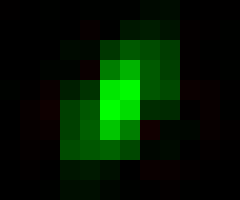}}
    \hfill
    \subfigure{\includegraphics[align=c, width=\wp \linewidth]{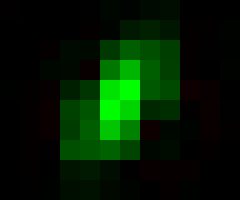}}
    \hfill
    \subfigure{\includegraphics[align=c, width=\wp \linewidth]{supple/akl/d4_80.png}}
    \hfill
    \subfigure{\includegraphics[align=c, width=\wp \linewidth]{supple/lal/gt_d4.png}}
    \\
    \vspace{-2mm}
    \makecell{RealSR-V3 \\ (Nikon)}
    \hfill
    \addtocounter{subfigure}{-8}
    \subfigure[1 epoch]{\includegraphics[align=c, width=\wp \linewidth]{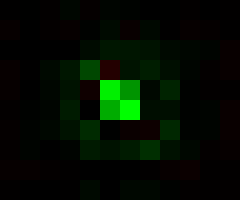}}
    \hfill
    \subfigure[2 epochs]{\includegraphics[align=c, width=\wp \linewidth]{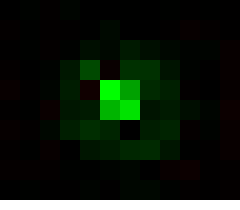}}
    \hfill
    \subfigure[10 epochs]{\includegraphics[align=c, width=\wp \linewidth]{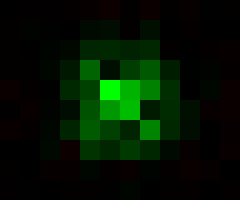}}
    \hfill
    \subfigure[20 epochs]{\includegraphics[align=c, width=\wp \linewidth]{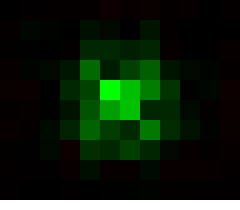}}
    \hfill
    \subfigure[40 epochs]{\includegraphics[align=c, width=\wp \linewidth]{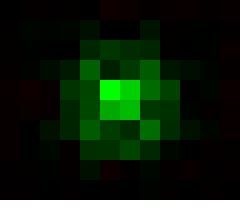}}
    \hfill
    \subfigure[60 epochs]{\includegraphics[align=c, width=\wp \linewidth]{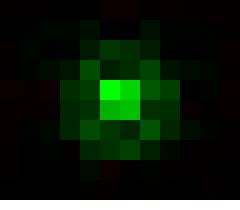}}
    \hfill
    \subfigure[80 epochs]{\includegraphics[align=c, width=\wp \linewidth]{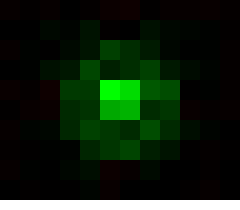}}
    \hfill
    \subfigure[GT kernel]{\includegraphics[align=c, width=\wp \linewidth]{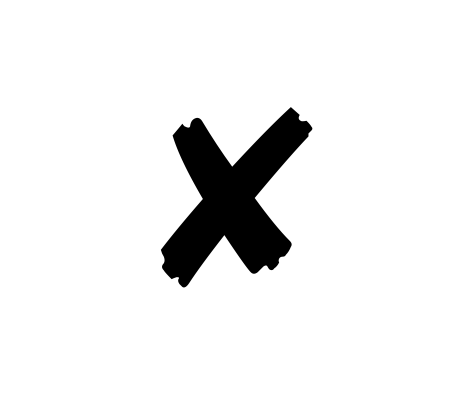}}
    \\
    \figspace
    \caption{
        \textbf{Evolution of the retrieved $\times 2$ degradation kernel $\bar{k}$ in the proposed ADL.}
        We visualize the estimated kernels from \eqref{eq:kernel} on two different datasets.
        For simplicity, we refer 1,741 iterations as one epoch.
        Since we apply the ADL after 10 warm-up epochs, downsampling networks in (a) and (b) are trained under LFL, not ADL.
        Furthermore, (a) and (b) visualize the linear approximations of the learned downsamplers after a certain number of training epochs rather than the approximated kernels $\bar{k}$.
        In the RealSR-V3~\cite{sr_realworld} dataset, no ground-truth kernel is available for the Nikon camera configuration.
        We crop image boundaries for better illustration.
    }
    \label{fig:akl_kernels}
    \figxspace
\end{figure*}

%% file: sections/figure/perceptual.tex
\begin{figure*}[t!]
    \centering
    \renewcommand{\wp}{0.152}
    \subfigure{
        \includegraphics[width=\wp\linewidth]{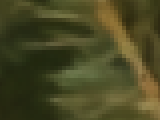}
    }
    \hfill
    \subfigure{
        \includegraphics[width=\wp\linewidth]{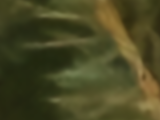}
    }
    \hfill
    \subfigure{
        \includegraphics[width=\wp\linewidth]{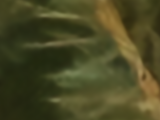}
    }
    \hfill
    \subfigure{
        \includegraphics[width=\wp\linewidth]{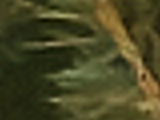}
    }
    \hfill
    \subfigure{
        \includegraphics[width=\wp\linewidth]{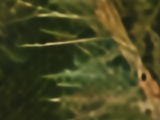}
    }
    \hfill
    \subfigure{
        \includegraphics[width=\wp\linewidth]{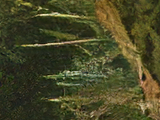}
    }
    \\
    \vspace{-2mm}
    \subfigure{
        \includegraphics[width=\wp\linewidth]{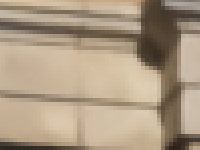}
    }
    \hfill
    \subfigure{
        \includegraphics[width=\wp\linewidth]{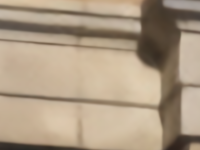}
    }
    \hfill
    \subfigure{
        \includegraphics[width=\wp\linewidth]{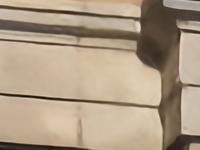}
    }
    \hfill
    \subfigure{
        \includegraphics[width=\wp\linewidth]{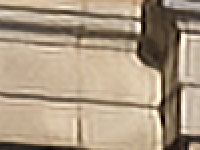}
    }
    \hfill
    \subfigure{
        \includegraphics[width=\wp\linewidth]{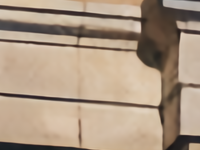}
    }
    \hfill
    \subfigure{
        \includegraphics[width=\wp\linewidth]{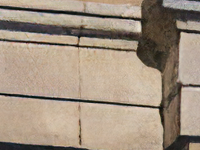}
    }
    \\
    \vspace{-2mm}
    \addtocounter{subfigure}{-12}
    \subfigure[$\img{LR}$~(Input)]{
        \includegraphics[width=\wp\linewidth]{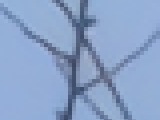}
    }
    \hfill
    \subfigure[RRDB~\cite{sr_esrgan}]{
        \includegraphics[width=\wp\linewidth]{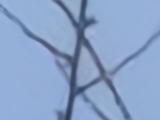}
    }
    \hfill
    \subfigure[IKC~\cite{sr_ikc}]{
        \includegraphics[width=\wp\linewidth]{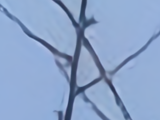}
    }
    \hfill
    \subfigure[\cite{sr_kernelgan}~$+$~\cite{sr_zssr}]{
        \includegraphics[width=\wp\linewidth]{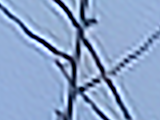}
    }
    \hfill
    \subfigure[\textbf{ADL}~$+$~RRDB \label{fig:results2_plain}]{
        \includegraphics[width=\wp\linewidth]{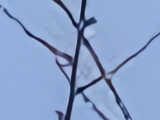}
    }
    \hfill
    \subfigure[\textbf{ADL}~$+$~$\mathcal{P}$ \label{fig:results2_per}]{
        \includegraphics[width=\wp\linewidth]{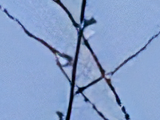}
    }
    \\
    \figspace
    \caption{
        \textbf{Perceptual $\times 4$ SR results on the DPED dataset.}
        We note that no ground-truth HR images exist for the dataset.
        Therefore, quantitative comparison on the DPED images nor the oracle model is not available.
        $\mathcal{P}$ denotes the RRDB~\cite{sr_esrgan} model learned with \eqref{eq:sr_percep} to reconstruct more realistic textures and sharper details.
        From the top, patches are cropped from the DPED-val `\emph{20.png},' `\emph{49.png},' and `\emph{63.png},' respectively.
    }
    \label{fig:results2}
    \figxspace
\end{figure*}

%% file: sections/table/ablation_adl.tex
\begin{table}[t!]
    \centering
    \renewcommand{\arraystretch}{1.05}
    \caption{
        \textbf{Ablation study about the proposed ADL method.} \protect \\
        To evaluate the SR performance, we use ADL + EDSR configuration and the synthetic DIV2K dataset.
        (a) We note that the kernel similarity~\cite{rs_deblur_mh} is measured after the last update, and $\bar{k}$ is a linear approximation of the learned downsampler for each configuration.
        (b) $\infty$ means that the kernel $\bar{k}$ is not updated after the first estimation.
    }
    \label{tab:ablation_adl}
    \tabspace
    \subfigure[Effect of the total training epochs in Algorithm~\ref{alg:adl}. \label{tab:ablation_adl_a}]{
        \begin{tabularx}{\linewidth}{p{3.2cm} >{\centering\arraybackslash}X >{\centering\arraybackslash}X >{\centering\arraybackslash}X >{\centering\arraybackslash}X}
            \toprule
            Total training epochs $T$ & 20 & 40 & 60 & 80 \\
            \midrule
            $\text{Similarity}^{\uparrow} \paren{ \bar{k}, k_4 }$ & 0.9682 & 0.9707 & \secondbest{0.9714} & \best{0.9718} \\
            PSNR$^{\uparrow}$ for $\times 2$ SR ($k_4$) &  31.48 & \secondbest{31.96} & 31.93 & \best{32.05} \\
            \bottomrule
        \end{tabularx}
    }
    \\
    \subfigure[Effect of the iterative adjustment in Algorithm~\ref{alg:adl}. \label{tab:ablation_adl_b}]{
        \begin{tabularx}{\linewidth}{p{3.2cm} >{\centering\arraybackslash}X >{\centering\arraybackslash}X >{\centering\arraybackslash}X >{\centering\arraybackslash}X}
            \toprule
            \multirow{2}{*}{Update interval $t$ (epochs)} & \multicolumn{4}{c}{ADL + EDSR (\textbf{Proposed})} \\
            & 1 & 10 & 20 & $\infty$ \\
            \midrule
            PSNR$^{\uparrow}$ for $\times 2$ SR ($k_4$) & 31.65 & \best{32.05} & \secondbest{32.01} & 31.85 \\
            \bottomrule
        \end{tabularx}
    }
    \\
    \subfigure[Effect of the number of samples for kernel estimation in \eqref{eq:kernel}. \label{tab:ablation_adl_c}]{
        \begin{tabularx}{\linewidth}{p{3.2cm} >{\centering\arraybackslash}X >{\centering\arraybackslash}X >{\centering\arraybackslash}X >{\centering\arraybackslash}X}
            \toprule
            \multirow{2}{*}{The number of samples $N$} & \multicolumn{4}{c}{ADL + EDSR (\textbf{Proposed})} \\
            & 1 & 5 & 10 & 50 \\
            \midrule
            PSNR$^{\uparrow}$ for $\times 2$ SR ($k_4$) & 31.09 & 31.84 & \secondbest{32.01} & \best{32.05} \\
            \bottomrule
        \end{tabularx}
    }
    \tabxspace
\end{table}

%% file: sections/figure/limitation.tex
\begin{figure}[t!]
    \centering
    \renewcommand{\wp}{0.242}
    \subfigure{\includegraphics[width=\wp \linewidth]{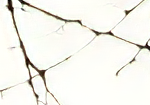}}
    \hfill
    \subfigure{\includegraphics[width=\wp \linewidth]{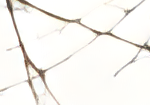}}
    \hfill
    \subfigure{\includegraphics[width=\wp \linewidth]{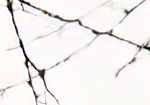}}
    \hfill
    \subfigure{\includegraphics[width=\wp \linewidth]{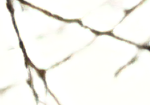}}
    \\
    \vspace{-3mm}
    \subfigure{\includegraphics[width=\wp \linewidth]{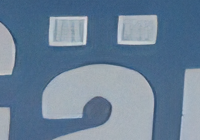}}
    \hfill
    \subfigure{\includegraphics[width=\wp \linewidth]{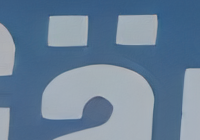}}
    \hfill
    \subfigure{\includegraphics[width=\wp \linewidth]{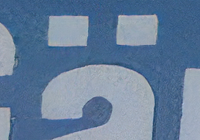}}
    \hfill
    \subfigure{\includegraphics[width=\wp \linewidth]{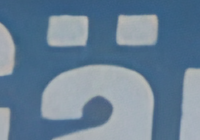}}
    \\
    \vspace{-3mm}
    \addtocounter{subfigure}{-8}
    \subfigure[RealSR]{\includegraphics[width=\wp \linewidth]{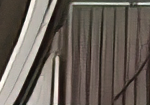}}
    \hfill
    \subfigure[BSRGAN]{\includegraphics[width=\wp \linewidth]{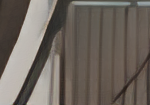}}
    \hfill
    \subfigure[\textbf{ADL}~$+$~$\mathcal{P}$]{\includegraphics[width=\wp \linewidth]{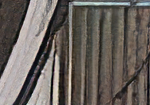}}
    \hfill
    \subfigure[\textbf{ADL-n}~$+$~$\mathcal{P}$]{\includegraphics[width=\wp \linewidth]{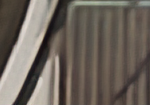}}
    \\  
    \figspace
    \caption{
        \revised{
        \textbf{Effect of preprocessing on the DPED dataset.}
        In ADL, appropriate preprocessing is required for more effective learning.
        Patches are cropped from DPED-val `\emph{45.png},' `\emph{63.png},' and `\emph{84.png},' respectively.
        }
    }
    \label{fig:limitation}
    \figxspace
\end{figure}

%% file: sections/5_conclusion.tex
\section{Conclusions}
We propose a novel unsupervised method to estimate an unknown distribution of LR images using unpaired LR and HR examples.
The proposed LFL and ADL terms facilitate the downsampler to accurately synthesize the LR images with the desired distribution.
Compared to conventional approaches, we do not pose restrictive priors to the learned function in the adversarial training framework.
Consequently, the existing SR models can be trained with our LR images and achieve significant performance gains on synthetic and realistic datasets.
We also demonstrate that our approach can be applied to a set of arbitrary images~\cite{sr_realworld, data_dped} in the wild.
The results verify that the proposed method can be used to handle real-world SR problems.
In the future work, we will extend our approach to estimating a feasible downsampling model with real-world noise jointly.

%% file: bio/bio.tex
%

\vspace{-12mm}
\begin{IEEEbiography}[{\includegraphics[width=1in,height=1.25in,clip,keepaspectratio]{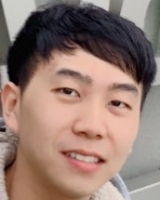}}]{Sanghyun Son}
is a Ph.D. student in the Department of Electrical and Computer Engineering at the Seoul National University (SNU), Seoul, Korea.
He graduated \emph{summa cum laude} from Seoul National University with a B.S. degree in Electrical and Computer Engineering in 2017.
He is interested in real-world computer vision problems including image restoration and enhancement, especially image super-resolution and its practical applications.
\end{IEEEbiography}

\vspace{-12mm}
\begin{IEEEbiography}[{\includegraphics[width=1in,height=1.25in,clip,keepaspectratio]{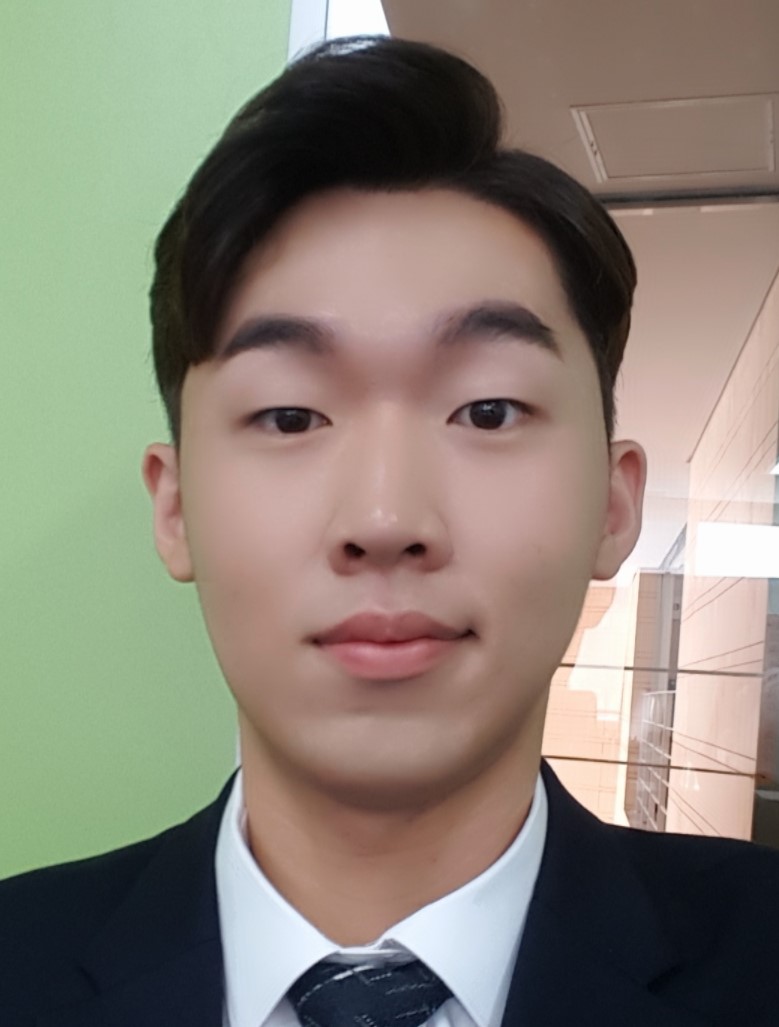}}]{Jaeha Kim}
is a M.S. student in the Department of Electrical and Computer Engineering at the Seoul National University (SNU), Seoul, Korea.
He graduated \emph{summa cum laude} from Korea Advanced Institute of Science and Technology(KAIST), Daejeon, Korea, with a B.S. degree in Electrical Engineering in 2019.
His research field is image enhancment tasks in computer vision, especially real-world image super-resolution. 
\end{IEEEbiography}

\vspace{-12mm}
\begin{IEEEbiography}[{\includegraphics[width=1in,height=1.25in,clip,keepaspectratio]{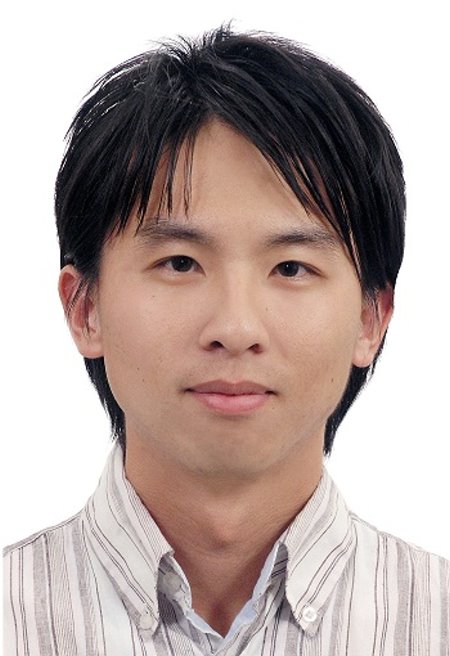}}]{Wei-Sheng Lai}
is a software engineer at Google, CA, USA. He received the B.S. and M.S. degree in Electrical Engineering from the National Taiwan University, Taipei, Taiwan, and his PhD degree in Electrical Engineering and Computer Science at the University of California Merced in 2019.
\end{IEEEbiography}

\vspace{-12mm}
\begin{IEEEbiography}[{\includegraphics[width=1in,height=1.25in,clip,keepaspectratio]{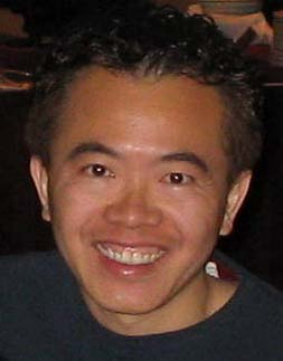}}]{Ming-Hsuan Yang} is affiliated with Google, UC Merced, and Yonsei University.  
Yang serves as a program co-chair of IEEE International Conference on Computer Vision (ICCV) in 2019, program co-chair of Asian Conference on Computer Vision (ACCV) in 2014, and general co-chair of ACCV 2016. Yang served as an associate editor of the IEEE Transactions on Pattern Analysis and Machine Intelligence, and is an associate editor of the International Journal of Computer Vision, Image and Vision Computing and Journal of Artificial Intelligence Research. He received the NSF CAREER award and Google Faculty Award. He is a Fellow of the IEEE.
\end{IEEEbiography}

\vspace{-12mm}
\begin{IEEEbiography}[{\includegraphics[width=1in,height=1.25in,clip,keepaspectratio]{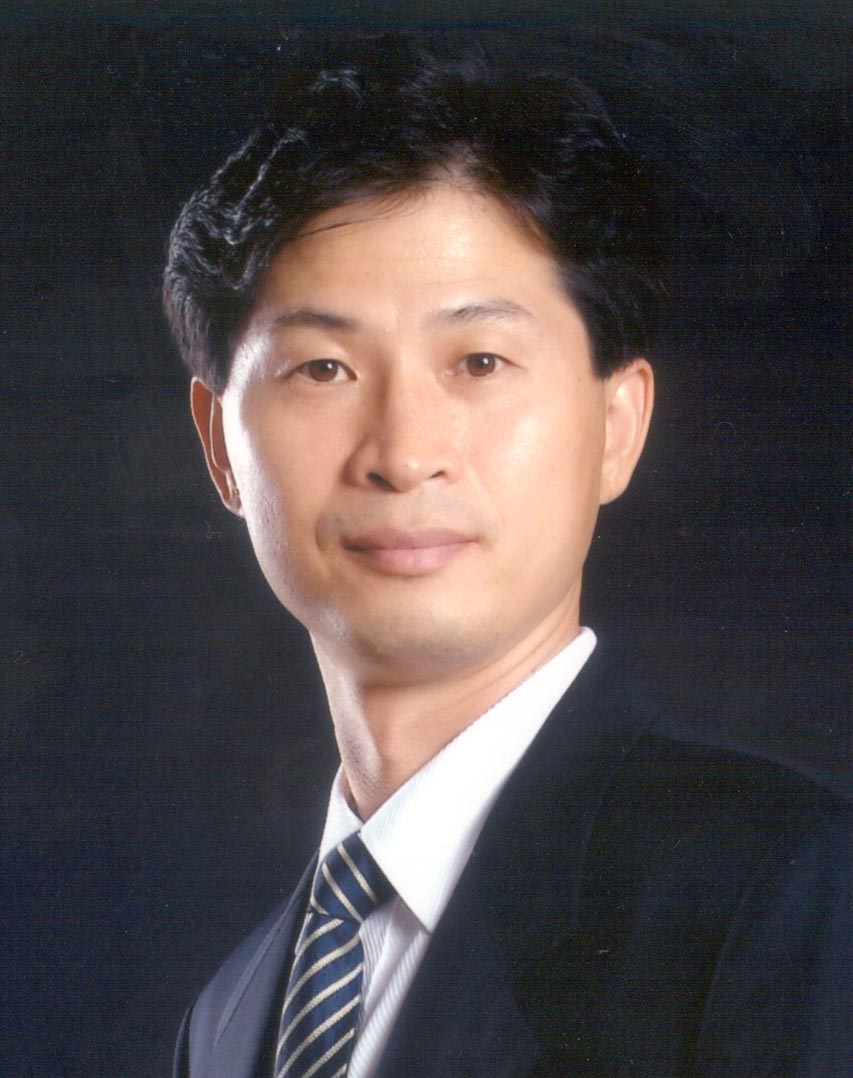}}]{Kyoung Mu Lee}
received the BS and MS degrees in control and instrumentation engineering from Seoul National University (SNU), Seoul, Korea, in 1984 and 1986, respectively, and the PhD degree in electrical engineering from the University of Southern California, in 1993.
He is currently with the Department of ECE, Seoul National University as a professor.
He has received several awards, in particular, the medal of merit and the Scientist of Engineers of the month award from the Korean Government in 2020 and 2018, respectively, the Most Influential Paper over the Decade Award by the IAPR Machine Vision Application in 2009, the ACCV Honorable Mention Award in 2007, the Okawa Foundation Research Grant Award in 2006, the Distinguished Professor Award from the college of Engineering of SNU in 2009, and both the Outstanding Research Award and the Shinyang Engineering Academy Award from the College of Engineering of SNU in 2010.
He has served as an Associate Editor in Chief (AEIC) and an Associate Editor of the IEEE Transactions on Pattern Analysis and Machine Intelligence (TPAMI), an Associate Editor of the Machine Vision Application (MVA) Journal and the IPSJ Transactions on Computer Vision and Applications (CVA), the IEEE Signal Processing Letters (SPL), and an Area Editor of the Computer Vision and Image Understanding (CVIU).
He also has served as a general chair of ICCV2019, ACMMM2018, and ACCV2018, a program chair of ACCV2012, a track chair of ICPR2020 and ICPR2012, and an area char of CVPR, ICCV and ECCV many times.
He was a distinguished lecturer of the Asia-Pacific Signal and Information Processing Association (APSIPA) for 2012-2013.
He is an Advisory Board Member of the Computer Vision Foundation (CVF).
More information can be found on his homepage http://cv.snu.ac.kr/kmlee.
\end{IEEEbiography}





%% file: sections/s_appendix.tex
\clearpage
\setcounter{section}{0}
\setcounter{figure}{0}
\setcounter{table}{0}
\setcounter{equation}{0}

\renewcommand{\thetable}{S\arabic{table}}
\renewcommand{\thesection}{S\arabic{section}}
\renewcommand{\thefigure}{S\arabic{figure}}
\renewcommand{\theequation}{S\arabic{equation}}

\section{Details about the Low-pass Filters}
\label{sec:appendix_lfl}
\Paragraph{Formulation of the low-pass filters.}
We describe specific implementations of low-pass filters in the proposed LFL formulation.
Our LFL utilizes conventional low-pass filters to extract low-frequency components from given images.
Therefore, we replace the term $\text{LPF}_\ast$ in \fakeeqref{4} to a kernel representation $k_\text{HR}$ and $k_\text{Down}$ to simplify the description as follows:
\begin{equation}
    \mathcal{L}_\text{LFL} = \normone{ \paren{ \img{HR} \ast k_\text{HR} }_{\downarrow ms} - \paren{ \img{Down} \ast k_\text{Down} }_{\downarrow m} },
    \label{eq:lfl_filter}    
\end{equation}
where $m$ is a subsampling factor of the LR image, and \eqref{eq:lfl_filter} is equivalent to \fakeeqref{4} in our main manuscript.
For example, our default $\text{LFL}_{m}$ formulation corresponds to a 2D kernel $k_\text{Down}$ of $m \times m$ where $k_\text{Down} \paren{x, y} \equiv \nicefrac{1}{m^2}$.
We note that $m = 16$ is used throughout our studies.
In $\times 2$ downsamping case, a kernel $k_\text{HR}$ for HR images can be expressed as a box filter of $32 \times 32$ with $k_\text{HR} \paren{ x, y } \equiv \nicefrac{1}{32^2}$.
Here, $\paren{x, y}$ describes a coordinate system of the downsampling kernel, including a sub-pixel shift.
Specifically, a center of the $16 \times 16$ kernel, which is not a pixel, corresponds to $k \paren{0, 0}$ and neighboring 4 pixels are represented as $k \paren{\pm 0.5, \pm 0.5}$, respectively.
This formulation is useful for Gaussian kernels on an even-sized grid as follows:
\begin{equation}
    k \paren{x, y} = \frac{1}{Z} \exp \paren{- \frac{x^2}{2 \sigma_{x}^2} - \frac{y^2}{2 \sigma_{y}^2}},
    \label{eq:exact_kernels}
\end{equation}
where $Z$ is a normalization factor so that $\sum_{x, y} k \paren{x, y} \equiv 1$.
For the proposed LFL, only isotropic cases are tested where $\sigma_x$ and $\sigma_y$ are equal.
In the Gaussian cases, we follow a convention and set the kernel grid size to $p \times p$ where $p$ is the nearest power of two from $6\sigma_x$.
While we adopt the filtering-based method for our LFL for simplicity, more complex formulations such as Wavelet can be introduced without losing generality.

\Paragraph{Selection of the low-pass filters.}
An appropriate selection of the low-pass filter in our LFL plays an essential role.
Therefore, we conduct an extensive ablation study to determine the low-pass filter when training the downsampler $\mathcal{D}$.
Table~\ref{tab:ablation_supp} shows how different types and shapes of low-pass filters for the downsampler affect the SR results.
We present the performance evaluation on the synthetic DIV2K dataset with a challenging anisotropic Gaussian kernel $k_4$.
As shown in Table~\fakeref{S2(a)}, a small box filter, e.g., $m = 2$, may bias the training objective and degrade the following SR performance.
On the other hand, a large box filter with $m = 64$ operates as an extremely loose constraint and cannot contribute to preserving image contents across different scales.
In Table~\fakeref{S2(b)}, we have also introduced 2D Gaussian filters for the LFL.
However, simple box filters have demonstrated relatively better performance.
Thus, we use $16 \times 16$ box filters for the low-pass filter $\text{LPF}_m$ and $32\times 32$ for $\text{LPF}_{ms}$ by default throughout our experiments.

\input{sections/table/lpf_specs}
\begin{table}[t!]
    \centering
    \renewcommand{\arraystretch}{1.05}
    \caption{
        \revised{
        \textbf{Ablation study on the shapes and sizes of $\text{LPF}_{m}$.} \protect\\
        We train the LFL-based downsampler and the following SR model on DIV2K $\times 2$ ($k_4$) to observe how different low-pass filters affect the performance of our approach.
        }
    }
    \label{tab:ablation_supp}
    \tabspace
    \subfigure[Box filters for $\text{LPF}_{m}$ \label{tab:ablation_box}]{
        \begin{tabularx}{\linewidth}{p{3.2cm} >{\centering\arraybackslash}X >{\centering\arraybackslash}X >{\centering\arraybackslash}X >{\centering\arraybackslash}X >{\centering\arraybackslash}X >{\centering\arraybackslash}X}
            \toprule
            \multirow{2}{*}{Method $\backslash$ Box size $m$} & \multicolumn{6}{c}{PSNR$^{\uparrow}$(dB) for $\times 2$ SR ($k_4$)}\\
            & 2 & 4 & 8 & 16 & 32 & 64 \\
            \midrule
            LFL + EDSR (\textbf{Proposed}) & 29.51 & 30.17 & 31.06 & \secondbest{31.57} & {\best{31.58}} & 28.11 \\
            \bottomrule
        \end{tabularx}
    }
    \\
    \subfigure[Gaussian filters for $\text{LPF}_{m}$ \label{tab:ablation_gaussian}]{
        \begin{tabularx}{\linewidth}{p{3.2cm} >{\centering\arraybackslash}X >{\centering\arraybackslash}X >{\centering\arraybackslash}X >{\centering\arraybackslash}X >{\centering\arraybackslash}X >{\centering\arraybackslash}X}
            \toprule
            \multirow{2}{*}{Method $\backslash$ Gaussian sigma $\sigma$} & \multicolumn{6}{c}{PSNR$^{\uparrow}$ for $\times 2$ SR ($k_4$)}\\
            & 0.8 & 1.2 & 1.6 & 2.0 & 2.5 & 3.0 \\
            \midrule
            LFL + EDSR (\textbf{Proposed}) & 29.64 & 30.24 & 30.42 & 30.62 & 30.96 & 30.75 \\
            \bottomrule
        \end{tabularx}
    }
    \vspace{-3mm}
\end{table}

\section{Details about the Synthetic Kernels}
\label{sec:appendix_kernels}
We present a formulation of the $\times 2$ synthetic downsampling kernels used for various experiments in our main manuscript.
As we describe in Section~\fakeref{4.1}, $k_0$ denotes a widely-used MATLAB bicubic kernel.
The other kernels, i.e., $k_1 \sim k_4$, are $20 \times 20$ and sampled from a standard 2D Gaussian distribution following \eqref{eq:exact_kernels}.
Table~\ref{tab:exact_kernels} describes the actual parameters used to instantiate our synthetic kernels.
To validate the generalization ability of the proposed method, we do not resort to radial kernels that are relatively easy to model and introduce anisotropic kernels $k_3$ and $k_4$.
The most challenging case $k_4$ further includes rotation of random degrees $\theta$ and have neither vertical nor horizontal symmetries.
For a larger $\times 4$ downsampling factor, we follow an approach from KernelGAN~\cite{sr_kernelgan} and convolve the same kernel twice to generate a larger one.

\begin{table}[t!]
    \centering
    \renewcommand{\arraystretch}{1.05}
    \caption{
        \textbf{Detailed parameters to implement the synthetic Gaussian kernels.}
        Fig.~\fakeref{5} in our main manuscript also visualizes each downsampling kernel in detail.
    }
    \label{tab:exact_kernels}
    \tabspace
    \begin{tabularx}{\linewidth}{>{\centering\arraybackslash}X >{\centering\arraybackslash}X >{\centering\arraybackslash}X >{\centering\arraybackslash}X >{\centering\arraybackslash}X}
        \toprule
        Kernel $k_i$ & $\sigma_{x}$ & $\sigma_{y}$ & $\theta$ & Type \\
        \midrule
        $k_1$ & 1.0 & 1.0 & 0$^\circ$ & Isotropic \\
        $k_2$ & 1.6 & 1.6 & 0$^\circ$ & Isotropic \\
        $k_3$ & 1.0 & 2.0 & 0$^\circ$ & Anisotropic \\
        $k_4$ & 1.0 & 2.0 & 29$^\circ$ & Anisotropic \\
        \bottomrule
    \end{tabularx}
    \tabspace
\end{table}

\section{Stability of the proposed methods}
\begin{figure}
    \centering
    \includegraphics[width=\linewidth]{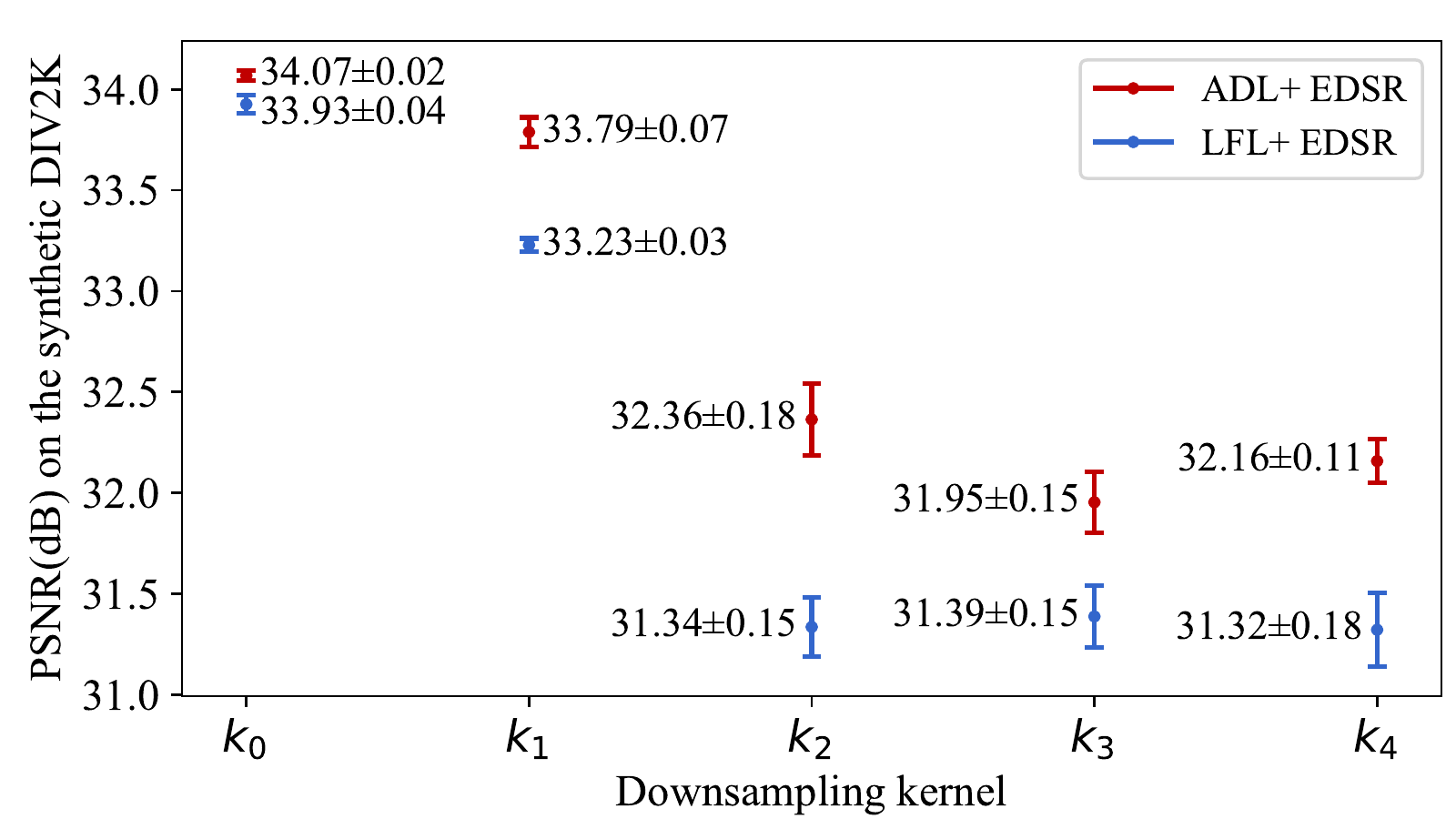}
    \vspace{-6mm}
    \caption{
        \revised{
        \textbf{Stability analysis of the proposed methods.}
        We visualize the average performance of the baseline EDSR $\times 2$ model and standard deviation from five runs on each downsampling kernel.
        Notably, ADL consistently outperforms LFL in all synthetic downsampling kernel configurations.
        }
    }
    \label{fig:errorbar}
\end{figure}

\revised{
Since our LFL and ADL rely on unsupervised adversarial training, stability and reproducibility of the proposed method can be an essential issue.
Therefore, we conduct five independent experiments for each of the five downsampling kernels $k_0 \sim k_4$ at a scale factor of $\times 2$ t analyze the stability of our training scheme.
Fig.~\ref{fig:errorbar} shows the average performance of the following SR model after we train the downsampler using LFL and ADL, across five different kernel configurations on the synthetic DIV2K dataset.
Even in the unsupervised learning framework, the SR model with our LFL and ADL schemes performs consistently with a small variation.
Since the SR model performs stably across different experimental configurations, we report the result from one single run in the other sections.
}

\begin{figure}[t!]
    \centering
    \definecolor{navy}{RGB}{32, 56, 100}
    \definecolor{grass}{RGB}{169, 209, 142}
    \subfigure[Downsampling CNN $\mathcal{D}$]{\includegraphics[width=\linewidth]{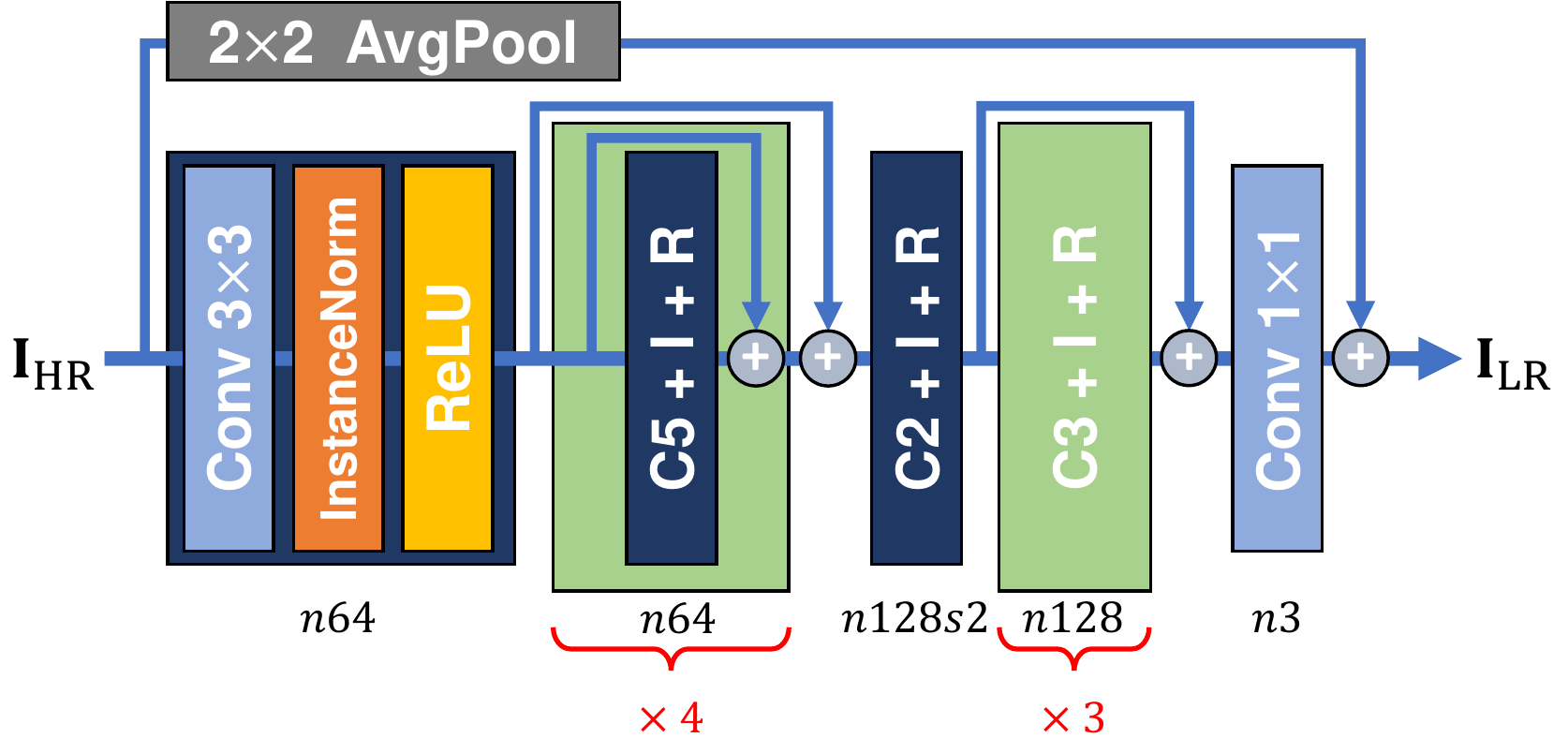}}
    \\
    \subfigure[Discriminator CNN $\mathcal{F}$]{\includegraphics[width=\linewidth]{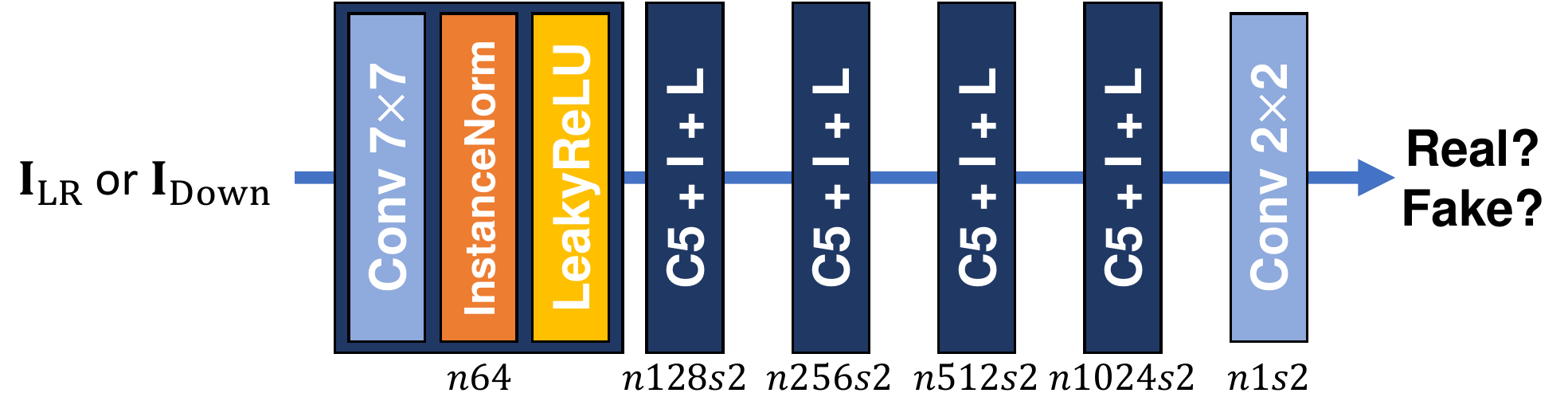}}
    \\
    \vspace{-2mm}
    \caption{\textbf{Our CNN architectures.}
        C$K$ in the \textcolor{navy}{\textbf{navy}} box denotes a $K \times K$ convolutional layer, e.g., C5 for $5 \times 5$.
        For a sequence of the convolutional, instance normalization~\cite{others_in}, and ReLU (or LeakyReLU~\cite{gans_dc}) activation layers, we use the term C$K$ $+$ I $+$ R (or L) for simplicity.
        The \textcolor{grass}{\textbf{green}} block in (a) incorporates a shortcut connection wrapping around the C$K$ $+$ I $+$ R sequence.
        We note that $n$ refers to the number of output channels, and $s$ describes the stride of the convolutional layer with a default value of 1, respectively.
    }
    \label{fig:architecture}
\end{figure}

\section{Detailed comparison with KernelGAN}
\revised{
Table~\fakeref{4} in our main manuscript shows that the proposed ADL + EDSR outperforms the KernelGAN + ZSSR combination by a significant margin even when only one LR image is available for the training.
In this section, we use multiple images to train KernelGAN to demonstrate the advantage of our method when large-scale unpaired images are available.
Table~\ref{tab:kgan} shows extensive experimental results regarding different training datasets of the KernelGAN.
We note that ZSSR is used to reconstruct $\img{SR}$ following KernelGAN by default unless mentioned otherwise.
}

\begin{table*}[t!]
    \centering
    \caption{
        \revised{
        \textbf{Ablation study on using more data for the KernelGAN method.} \protect\\
        HR images correspond to inputs of the $\img{LR}$ generator, i.e., the deep linear generator for KernelGAN experiments and our downsampling network for the ADL configuration, while LR images are \emph{real} samples for the discriminator network.
        We note that Case 0 and ADL correspond to Table~\fakeref{4} in our main manuscript.
        EDSR denotes a baseline version that has 1.4M parameters.
        All evaluations are done using DIV2K `\emph{0801.png}'$\sim$`\emph{0900.png}.'
        }
    }
    \tabspace
    \label{tab:kgan}
    \begin{tabularx}{\linewidth}{c c c c >{\centering\arraybackslash}X >{\centering\arraybackslash}X >{\centering\arraybackslash}X >{\centering\arraybackslash}X >{\centering\arraybackslash}X}
        \toprule
        \multirow{2}{*}{Case} & \multicolumn{2}{c}{Dataset for the $\img{LR}$ Generator} & \multirow{2}{*}{SR model} & \multicolumn{5}{c}{PSNR$^{\uparrow}$(dB) for $\times 2$ SR} \\
        & HR image(s) & LR image(s) & & $k_0$ & $k_1$ & $k_2$ & $k_3$ & $k_4$ \\
        \midrule
        1 & \multicolumn{2}{c}{\multirow{2}{*}{`\emph{0801}'$\sim$`\emph{0900}'}} & ZSSR & 21.54 & 26.41 & 30.55 & 31.18 & 27.68 \\
        2 & \multicolumn{2}{c}{} & EDSR & 16.87 & 18.55 & 29.95 & 31.08 & 23.28 \\
        \midrule
        3 & \multirow{2}{*}{`\emph{0001}'$\sim$`\emph{0400}'} & \multirow{2}{*}{`\emph{0401}'$\sim$`\emph{0800}'} & ZSSR & 20.91 & 26.83 & 29.97 & 28.46 & 27.99 \\
        4 & & & EDSR & 16.11 & 20.35 & 29.26 & 24.85 & 19.68 \\
        \midrule
        0 & $\img{LR}$ & $\img{LR}$ & ZSSR & 22.32 & 26.42 & 30.44 & 29.10 & 29.12 \\
        \midrule
        ADL & `\emph{0001}'$\sim$`\emph{0400}' & `\emph{0401}'$\sim$`\emph{0800}' & EDSR & \textbf{34.07} & \textbf{33.68} & \textbf{32.51} & \textbf{32.08} & \textbf{32.05} \\
        \bottomrule
    \end{tabularx}
    \tabxspace
\end{table*}

\revised{
First, in Cases 1 and 2, we modify KernelGAN to use 100 validation LR images as a training dataset and predict a shared downsampling kernel rather than calculate it for each image.
This configuration demonstrates how KernelGAN operates on large-scale data.
Using the estimated kernel, ZSSR is applied to each image independently.
Compared to the original KernelGAN + ZSSR configuration, i.e., Case 0, using more images for kernel estimation has demonstrated inconsistent performance variations on synthetic kernel experiments $k_0 \sim k_4$ in Case 1.
For the kernels $k_2$ and $k_3$, using more data has brought noticeable performance improvements.
However, with the kernels $k_0$, $k_1$, and $k_4$, using a single image yields better results.
}

\revised{
The capacity of ZSSR is relatively smaller than recent state-of-the-art methods, which may limit the performance of the KernelGAN + ZSSR combination.
Specifically, the model has only 0.2M parameters and uses a single image for training.
Therefore, we introduce a larger EDSR-baseline model as an SR backbone network with 400 training samples in Case 2.
Similar to the proposed LFL and ADL experiments in our main manuscript, we synthesize 400 LR images from DIV2K `\emph{0001.png}'$\sim$`\emph{0400.png}' using the estimated kernel from the KernelGAN model.
The following EDSR is then trained on the synthetic LR-HR pairs.
However, the final SR performance decreases, while EDSR-baseline has a larger capacity than ZSSR.
Unlike our ADL which brings additional performance gains with a larger SR backbone (see ADL + EDSR and ADL + RRDB in Table~\fakeref{4}), such behavior demonstrates that better fitting to the kernel from KernelGAN does not guarantee higher SR performance.
}

\revised{
In Cases 3 and 4, we adopt the same dataset configuration as the proposed LFL and ADL, i.e., 400 HR images with unpaired 400 LR samples, when training KernelGAN.
Therefore, the only difference between our and KernelGAN algorithms is model architectures (nonlinear CNN vs. deep linear generator) and loss functions (adaptive downsampling loss vs. kernel constraints).
We first estimate the shared kernel $k$ by feeding $\img{HR}$ to the deep linear generator, and the discriminator is optimized to distinguish the output of the downsampling model and real LR images $\img{LR}$ synthesized by a ground-truth kernel $k_i$.
In Case 3, ZSSR is applied to `\emph{0801.png}'$\sim$`\emph{0900.png}' independently using a single shared kernel.
In Case 4, we train EDSR similar to Case 2.
We note that the only difference between Cases 1, 2, and Cases 3, 4 is a training dataset for the generator, i.e., downsampling model in KernelGAN.
}

\revised{
To summarize, our ADL + EDSR or ADL + RRDB formulation show consistently better performance compared to KernelGAN regardless of the number of training data and model capacity.
}

\section{Network Architecture}
\label{sec:appendix_models}
Fig.~\ref{fig:architecture} illustrates CNN architectures we use throughout our main manuscript.
The downsampling network adopts the residual connections~\cite{net_residual} and instance normalization~\cite{others_in} strategy for easier optimization.
A global residual connection~\cite{sr_vdsr} with $2 \times 2$ average pooling operation is also introduced to provide a stable starting point.
We note that the $2 \times 2$ average pooling in Figure~\ref{fig:architecture}(a) \emph{does not} operate as a restrictive prior which instabilizes the adversarial training objective, since it does not force the output $\img{Down}$ to be a specific function of the input $\img{HR}$.
Unlike the KernelGAN~\cite{sr_kernelgan} model, our downsampling CNN incorporates nonlinear ReLU activations and thus can learn a more generalized function.
Our discriminator network is fully convolutional~\cite{others_pix2pix} and returns a $2 \times 2$ probability map from a $64 \times 64$ input patch.
Both of the downsampling and discriminator CNNs are initialized with weights of random Gaussian $\mathcal{N} \paren{0, 0.02^2}$.

\section{Details about the Hyperparameters}
\label{sec:appendix_hparams}
We present detailed hyperparameters and experimental configurations that are not described in our main manuscript.
In the downsampling task, i.e., Algorithm~\fakeref{1} and \fakeeqref{2} in our main manuscript, we set the learning rate of $\eta$ of downsampling and discriminator CNNs $\mathcal{D}$ and $\mathcal{F}$ to $5 \times 10^{-5}$.
For each iteration, we use 32 samples of patch size $128 \times 128$ as an input batch and generate LR images of $64 \times 64$.
To train SR models, we use a batch size of 16 with $48 \times 48$ input images.
The learning rate is set to $10^{-4}$, similar to conventional approaches for deep image super-resolution~\cite{sr_edsr, sr_rcan, sr_dbpn}.
The only differences are that we reduce the learning rate by half for every 50 epochs and our baseline EDSR~\cite{sr_edsr} model is trained for 200 epochs, not 300, as validation performance does not change after then.
In all experiments, pixel values are normalized from $\left[ 0, 255 \right]$ to $\left[ -1, 1 \right]$.
We adopt the ADAM~\cite{others_adam} optimizer with $\paren{\beta_1, \beta_2} = \paren{0.9, 0.999}$ and $\epsilon = 10^{-8}$ for all learnable parameters.
It takes about 15 hours to learn the proposed downsampler with a single RTX 2080 Ti GPU on the synthetic DIV2K dataset.
The learning time reduces to about 4 hours on the RealSR-V3 dataset as it contains fewer HR and LR samples.

\section{Additional Qualitative Comparisons}
\label{sec:appendix_results}
We present more qualitative SR results in Figure~\ref{fig:more_synthetic}, \ref{fig:more_realsr}, and \ref{fig:more_dped}.
While the IKC~\cite{sr_ikc} model also reconstructs clean and sharp results for some specific synthetic cases, e.g., $k_1$ and $k_2$, our combination of the ADL + RRDB~\cite{sr_esrgan} generalizes well with a \emph{fixed} hyperparameter configuration, regardless of synthetic or realistic inputs.
As described in our main manuscript, more qualitative results can be found from our project page: \href{https://cv.snu.ac.kr/research/ADL}{https://cv.snu.ac.kr/research/ADL}.

\input{sections/figure/supp_synth}
\input{sections/figure/supp_real}
\input{sections/figure/supp_dped}

%% file: sections/table/lpf_specs.tex
\begin{table}[t]
    \centering
    \renewcommand{\arraystretch}{1.05}
    \caption{
        \revised{
        \textbf{Specifications of low-pass filters we use.} \protect\\
        For box and Gaussian filters, weights are normalized so that their values are summed to 1.
        We note that a subsampling by $ms$ and $m$ follow after $\text{LPF}_{ms}$ and $\text{LPF}_{s}$, respectively, to reduce image resolutions.
        More details about the filters are described in Appendix~\fakeref{A}.
        }
    }
    \label{tab:lpf}
    \tabspace
    %
    \begin{tabularx}{\linewidth}{>{\centering\arraybackslash}X >{\centering\arraybackslash}X >{\centering\arraybackslash}X >{\centering\arraybackslash}X}
        \toprule
        Type & Filter & Size & Shape \\
        \midrule
        \multirow{2}{*}{2D Box} & $\text{LPF}_{ms}$ & $ms \times ms$ & $ms \times ms$ box \\
        & $\text{LPF}_{s}$ & $m \times m$ & $m \times m$ box \\
        \midrule
        \multirow{2}{*}{2D Gaussian} & $\text{LPF}_{ms}$ & $ms \times ms$ & $\sigma_x = \sigma_y = s\sigma$ \\
        & $\text{LPF}_{s}$ & $m \times m$ & $\sigma_x = \sigma_y = \sigma$ \\
        \bottomrule
    \end{tabularx}
    \vspace{-3mm}
\end{table}

%% file: sections/figure/supp_synth.tex
\begin{figure*}[t]
    \centering
    \renewcommand{\wp}{0.128}
    \subfigure{
        \includegraphics[width=\wp\linewidth]{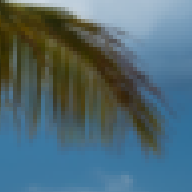}
    }
    \hfill
    \subfigure{
        \includegraphics[width=\wp\linewidth]{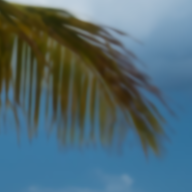}
    }
    \hfill
    \subfigure{
        \includegraphics[width=\wp\linewidth]{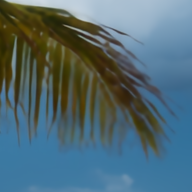}
    }
    \hfill
    \subfigure{
        \includegraphics[width=\wp\linewidth]{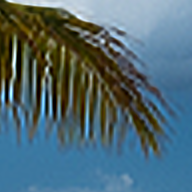}
    }
    \hfill
    \subfigure{
        \includegraphics[width=\wp\linewidth]{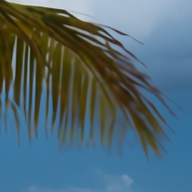}
    }
    \hfill
    \subfigure{
        \includegraphics[width=\wp\linewidth]{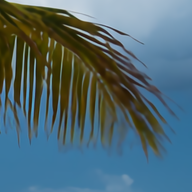}
    }
    \hfill
    \subfigure{
        \includegraphics[width=\wp\linewidth]{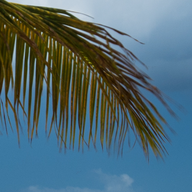}
    }
    \\
    \vspace{-2mm}
    \subfigure{
        \includegraphics[width=\wp\linewidth]{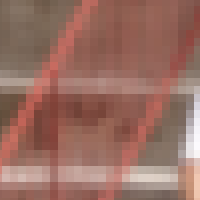}
    }
    \hfill
    \subfigure{
        \includegraphics[width=\wp\linewidth]{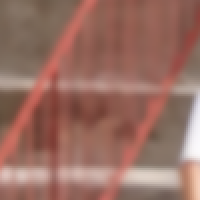}
    }
    \hfill
    \subfigure{
        \includegraphics[width=\wp\linewidth]{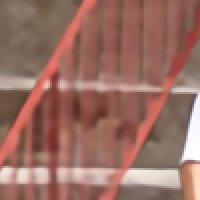}
    }
    \hfill
    \subfigure{
        \includegraphics[width=\wp\linewidth]{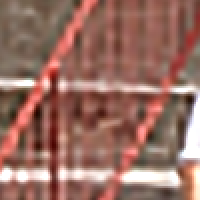}
    }
    \hfill
    \subfigure{
        \includegraphics[width=\wp\linewidth]{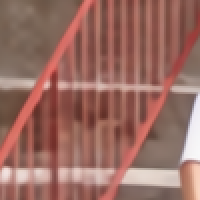}
    }
    \hfill
    \subfigure{
        \includegraphics[width=\wp\linewidth]{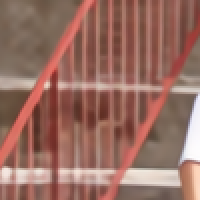}
    }
    \hfill
    \subfigure{
        \includegraphics[width=\wp\linewidth]{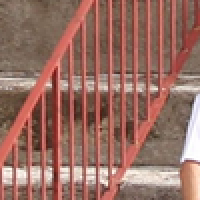}
    }
    \\
        \vspace{-2mm}
    \subfigure{
        \includegraphics[width=\wp\linewidth]{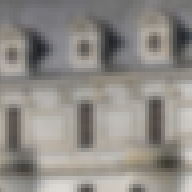}
    }
    \hfill
    \subfigure{
        \includegraphics[width=\wp\linewidth]{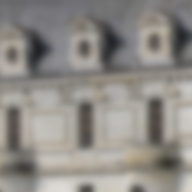}
    }
    \hfill
    \subfigure{
        \includegraphics[width=\wp\linewidth]{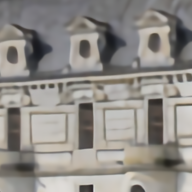}
    }
    \hfill
    \subfigure{
        \includegraphics[width=\wp\linewidth]{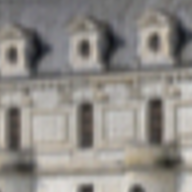}
    }
    \hfill
    \subfigure{
        \includegraphics[width=\wp\linewidth]{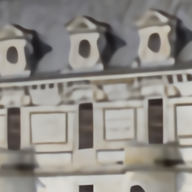}
    }
    \hfill
    \subfigure{
        \includegraphics[width=\wp\linewidth]{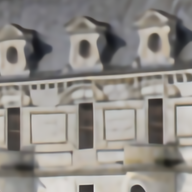}
    }
    \hfill
    \subfigure{
        \includegraphics[width=\wp\linewidth]{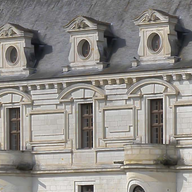}
    }
    \\
    \vspace{-2mm}
    \subfigure{
        \includegraphics[width=\wp\linewidth]{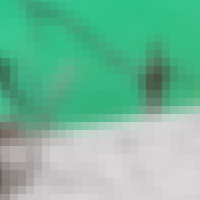}
    }
    \hfill
    \subfigure{
        \includegraphics[width=\wp\linewidth]{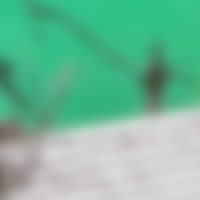}
    }
    \hfill
    \subfigure{
        \includegraphics[width=\wp\linewidth]{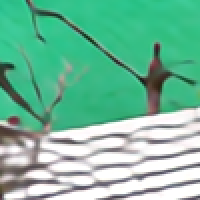}
    }
    \hfill
    \subfigure{
        \includegraphics[width=\wp\linewidth]{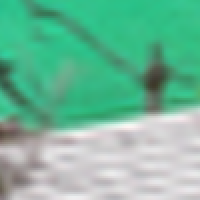}
    }
    \hfill
    \subfigure{
        \includegraphics[width=\wp\linewidth]{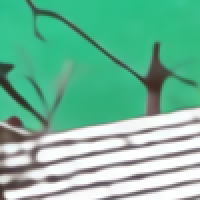}
    }
    \hfill
    \subfigure{
        \includegraphics[width=\wp\linewidth]{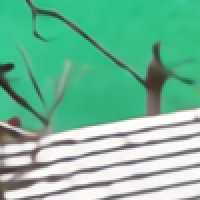}
    }
    \hfill
    \subfigure{
        \includegraphics[width=\wp\linewidth]{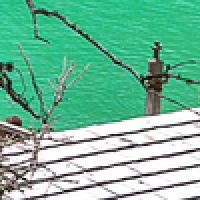}
    }
    \\
    \vspace{-2mm}
    \subfigure{
        \includegraphics[width=\wp\linewidth]{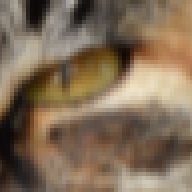}
    }
    \hfill
    \subfigure{
        \includegraphics[width=\wp\linewidth]{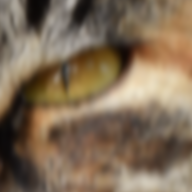}
    }
    \hfill
    \subfigure{
        \includegraphics[width=\wp\linewidth]{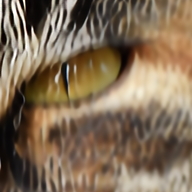}
    }
    \hfill
    \subfigure{
        \includegraphics[width=\wp\linewidth]{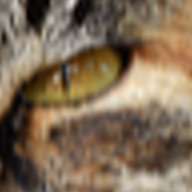}
    }
    \hfill
    \subfigure{
        \includegraphics[width=\wp\linewidth]{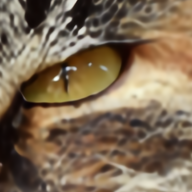}
    }
    \hfill
    \subfigure{
        \includegraphics[width=\wp\linewidth]{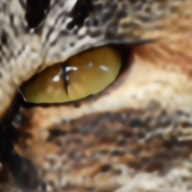}
    }
    \hfill
    \subfigure{
        \includegraphics[width=\wp\linewidth]{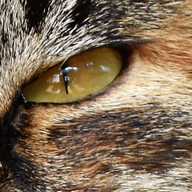}
    }
    \\
    \vspace{-2mm}
    \subfigure{
        \includegraphics[width=\wp\linewidth]{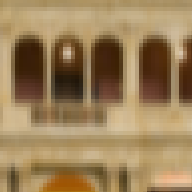}
    }
    \hfill
    \subfigure{
        \includegraphics[width=\wp\linewidth]{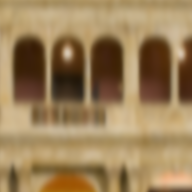}
    }
    \hfill
    \subfigure{
        \includegraphics[width=\wp\linewidth]{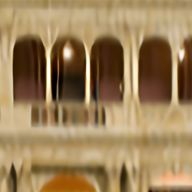}
    }
    \hfill
    \subfigure{
        \includegraphics[width=\wp\linewidth]{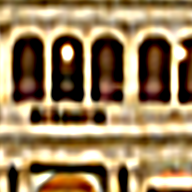}
    }
    \hfill
    \subfigure{
        \includegraphics[width=\wp\linewidth]{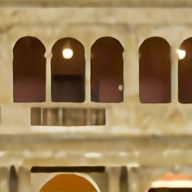}
    }
    \hfill
    \subfigure{
        \includegraphics[width=\wp\linewidth]{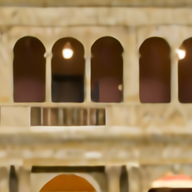}
    }
    \hfill
    \subfigure{
        \includegraphics[width=\wp\linewidth]{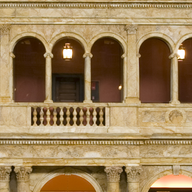}
    }
    \\
    \vspace{-2mm}
    \subfigure{
        \includegraphics[width=\wp\linewidth]{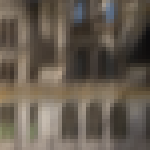}
    }
    \hfill
    \subfigure{
        \includegraphics[width=\wp\linewidth]{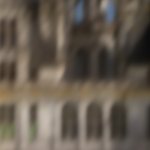}
    }
    \hfill
    \subfigure{
        \includegraphics[width=\wp\linewidth]{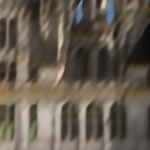}
    }
    \hfill
    \subfigure{
        \includegraphics[width=\wp\linewidth]{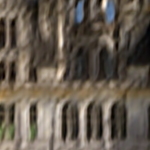}
    }
    \hfill
    \subfigure{
        \includegraphics[width=\wp\linewidth]{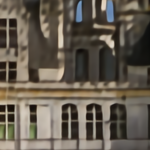}
    }
    \hfill
    \subfigure{
        \includegraphics[width=\wp\linewidth]{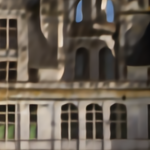}
    }
    \hfill
    \subfigure{
        \includegraphics[width=\wp\linewidth]{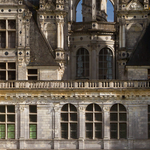}
    }
    \\
    \vspace{-2mm}
    \addtocounter{subfigure}{-49}
    \subfigure[$\img{LR}$~(Input)]{
        \includegraphics[width=\wp\linewidth]{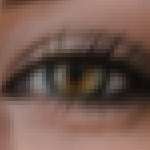}
    }
    \hfill
    \subfigure[RRDB~\cite{sr_esrgan}]{
        \includegraphics[width=\wp\linewidth]{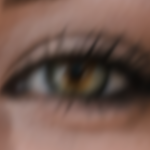}
    }
    \hfill
    \subfigure[IKC~\cite{sr_ikc}]{
        \includegraphics[width=\wp\linewidth]{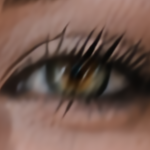}
    }
    \hfill
    \subfigure[\cite{sr_kernelgan}~$+$~\cite{sr_zssr}]{
        \includegraphics[width=\wp\linewidth]{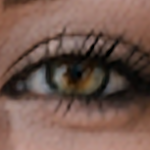}
    }
    \hfill
    \subfigure[\textbf{ADL}~$+$~\cite{sr_esrgan}]{
        \includegraphics[width=\wp\linewidth]{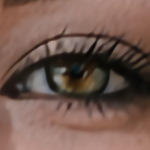}
    }
    \hfill
    \subfigure[Oracle]{
        \includegraphics[width=\wp\linewidth]{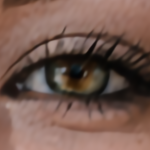}
    }
    \hfill
    \subfigure[$\img{HR}$~(GT)]{
        \includegraphics[width=\wp\linewidth]{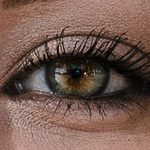}
    }
    \figspace
    \caption{
        \textbf{Additional qualitative $\times 4$ SR results on the synthetic DIV2K~\cite{data_div2k} dataset.}
        From the top, patches are cropped from the DIV2K `\emph{0806.png} ($k_1$),' `\emph{0825.png} ($k_1$),' `\emph{0865.png} ($k_2$),' `\emph{0807.png} ($k_2$),' `\emph{0869.png} ($k_3$),' `\emph{0884.png} ($k_3$),' `\emph{0830.png} ($k_4$),' and `\emph{0855.png} ($k_4$),' respectively, where LR images are synthesized using corresponding downsampling kernels in the parenthesis~$\paren{ \cdot }$.
    }
    \label{fig:more_synthetic}
    \figxspace
\end{figure*}

%% file: sections/figure/supp_real.tex
\begin{figure*}[t]
    \centering
    \renewcommand{\wp}{0.128}
    \subfigure{
        \includegraphics[width=\wp\linewidth]{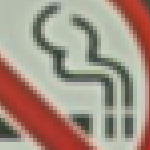}
    }
    \hfill
    \subfigure{
        \includegraphics[width=\wp\linewidth]{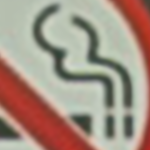}
    }
    \hfill
    \subfigure{
        \includegraphics[width=\wp\linewidth]{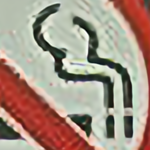}
    }
    \hfill
    \subfigure{
        \includegraphics[width=\wp\linewidth]{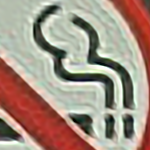}
    }
    \hfill
    \subfigure{
        \includegraphics[width=\wp\linewidth]{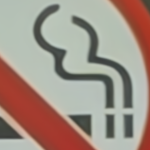}
    }
    \hfill
    \subfigure{
        \includegraphics[width=\wp\linewidth]{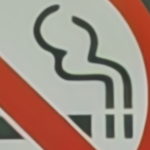}
    }
    \hfill
    \subfigure{
        \includegraphics[width=\wp\linewidth]{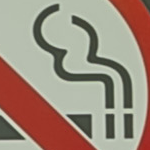}
    }
    \\
    \vspace{-2mm}
    \subfigure{
        \includegraphics[width=\wp\linewidth]{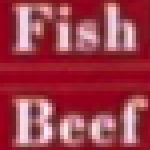}
    }
    \hfill
    \subfigure{
        \includegraphics[width=\wp\linewidth]{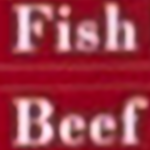}
    }
    \hfill
    \subfigure{
        \includegraphics[width=\wp\linewidth]{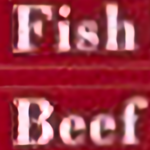}
    }
    \hfill
    \subfigure{
        \includegraphics[width=\wp\linewidth]{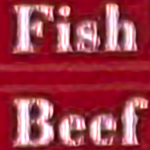}
    }
    \hfill
    \subfigure{
        \includegraphics[width=\wp\linewidth]{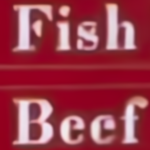}
    }
    \hfill
    \subfigure{
        \includegraphics[width=\wp\linewidth]{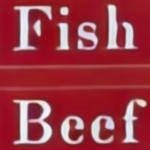}
    }
    \hfill
    \subfigure{
        \includegraphics[width=\wp\linewidth]{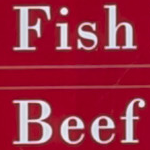}
    }
    \\
    \vspace{-2mm}
    \subfigure{
        \includegraphics[width=\wp\linewidth]{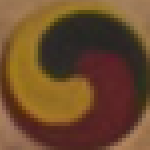}
    }
    \hfill
    \subfigure{
        \includegraphics[width=\wp\linewidth]{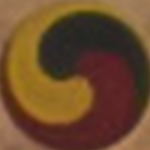}
    }
    \hfill
    \subfigure{
        \includegraphics[width=\wp\linewidth]{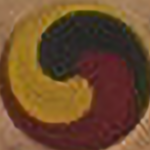}
    }
    \hfill
    \subfigure{
        \includegraphics[width=\wp\linewidth]{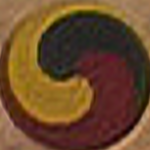}
    }
    \hfill
    \subfigure{
        \includegraphics[width=\wp\linewidth]{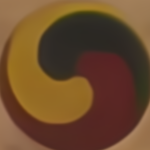}
    }
    \hfill
    \subfigure{
        \includegraphics[width=\wp\linewidth]{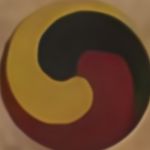}
    }
    \hfill
    \subfigure{
        \includegraphics[width=\wp\linewidth]{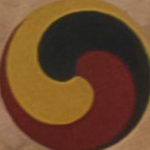}
    }
    \\
    \vspace{-2mm}
    \subfigure{
        \includegraphics[width=\wp\linewidth]{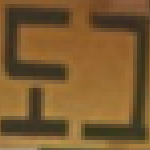}
    }
    \hfill
    \subfigure{
        \includegraphics[width=\wp\linewidth]{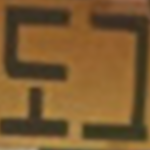}
    }
    \hfill
    \subfigure{
        \includegraphics[width=\wp\linewidth]{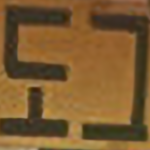}
    }
    \hfill
    \subfigure{
        \includegraphics[width=\wp\linewidth]{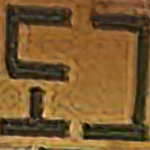}
    }
    \hfill
    \subfigure{
        \includegraphics[width=\wp\linewidth]{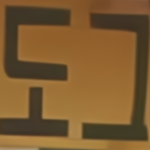}
    }
    \hfill
    \subfigure{
        \includegraphics[width=\wp\linewidth]{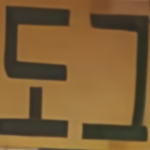}
    }
    \hfill
    \subfigure{
        \includegraphics[width=\wp\linewidth]{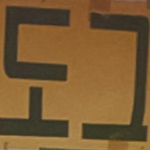}
    }
    \\
    \vspace{-2mm}
    \subfigure{
        \includegraphics[width=\wp\linewidth]{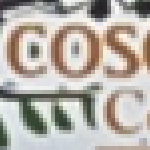}
    }
    \hfill
    \subfigure{
        \includegraphics[width=\wp\linewidth]{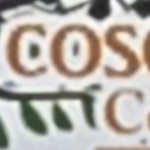}
    }
    \hfill
    \subfigure{
        \includegraphics[width=\wp\linewidth]{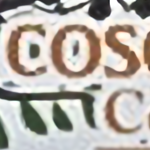}
    }
    \hfill
    \subfigure{
        \includegraphics[width=\wp\linewidth]{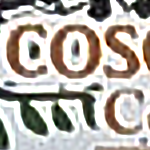}
    }
    \hfill
    \subfigure{
        \includegraphics[width=\wp\linewidth]{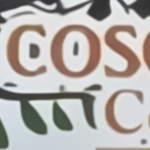}
    }
    \hfill
    \subfigure{
        \includegraphics[width=\wp\linewidth]{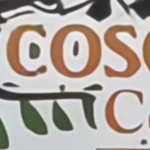}
    }
    \hfill
    \subfigure{
        \includegraphics[width=\wp\linewidth]{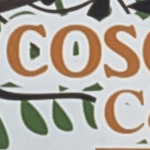}
    }
    \\
    \vspace{-2mm}
    \subfigure{
        \includegraphics[width=\wp\linewidth]{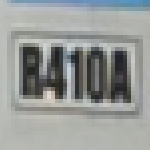}
    }
    \hfill
    \subfigure{
        \includegraphics[width=\wp\linewidth]{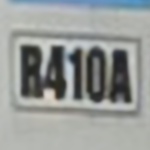}
    }
    \hfill
    \subfigure{
        \includegraphics[width=\wp\linewidth]{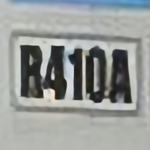}
    }
    \hfill
    \subfigure{
        \includegraphics[width=\wp\linewidth]{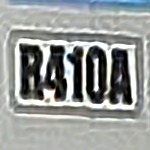}
    }
    \hfill
    \subfigure{
        \includegraphics[width=\wp\linewidth]{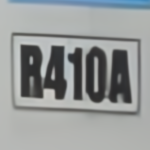}
    }
    \hfill
    \subfigure{
        \includegraphics[width=\wp\linewidth]{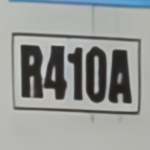}
    }
    \hfill
    \subfigure{
        \includegraphics[width=\wp\linewidth]{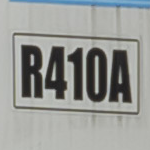}
    }
    \\
    \vspace{-2mm}
    \subfigure{
        \includegraphics[width=\wp\linewidth]{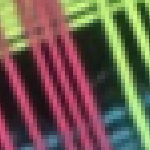}
    }
    \hfill
    \subfigure{
        \includegraphics[width=\wp\linewidth]{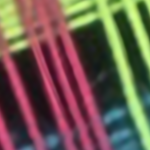}
    }
    \hfill
    \subfigure{
        \includegraphics[width=\wp\linewidth]{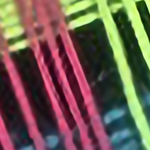}
    }
    \hfill
    \subfigure{
        \includegraphics[width=\wp\linewidth]{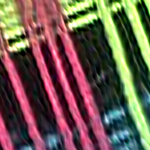}
    }
    \hfill
    \subfigure{
        \includegraphics[width=\wp\linewidth]{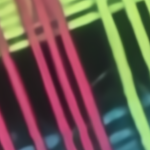}
    }
    \hfill
    \subfigure{
        \includegraphics[width=\wp\linewidth]{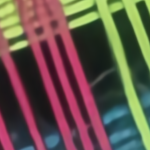}
    }
    \hfill
    \subfigure{
        \includegraphics[width=\wp\linewidth]{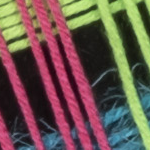}
    }
    \\
    \vspace{-2mm}
    \addtocounter{subfigure}{-49}
    \subfigure[$\img{LR}$~(Input)]{
        \includegraphics[width=\wp\linewidth]{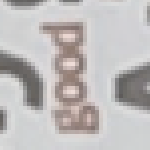}
    }
    \hfill
    \subfigure[RRDB~\cite{sr_esrgan}]{
        \includegraphics[width=\wp\linewidth]{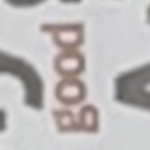}
    }
    \hfill
    \subfigure[IKC~\cite{sr_ikc}]{
        \includegraphics[width=\wp\linewidth]{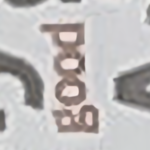}
    }
    \hfill
    \subfigure[\cite{sr_kernelgan}~$+$~\cite{sr_zssr}]{
        \includegraphics[width=\wp\linewidth]{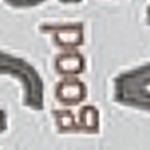}
    }
    \hfill
    \subfigure[\textbf{ADL}~$+$~\cite{sr_esrgan}]{
        \includegraphics[width=\wp\linewidth]{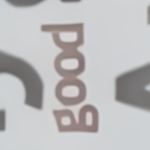}
    }
    \hfill
    \subfigure[Oracle]{
        \includegraphics[width=\wp\linewidth]{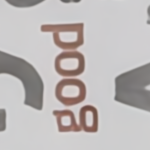}
    }
    \hfill
    \subfigure[$\img{HR}$~(GT)]{
        \includegraphics[width=\wp\linewidth]{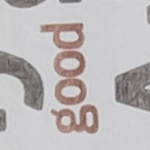}
    }
    \figspace
    \caption{
        \textbf{Additional qualitative $\times 4$ SR results on the RealSR-V3~\cite{sr_realworld} dataset.}
        From the top, patches are cropped from `\emph{Canon/001.png},' `\emph{Canon/003.png},' `\emph{Canon/022.png},' `\emph{Canon/033.png},' `\emph{Nikon/004.png},' `\emph{Nikon/041.png},' `\emph{Nikon/049.png},' and `\emph{Nikon/050.png},' respectively.\
        Our approach (ADL + RRDB~\cite{sr_esrgan}) produces the least upsampling noise and artifacts in output images.
    }
    \label{fig:more_realsr}
    \figxspace
\end{figure*}

%% file: sections/figure/supp_dped.tex
\begin{figure*}
    \centering
    \renewcommand{\wp}{0.148}
    \subfigure{
        \includegraphics[width=\wp\linewidth]{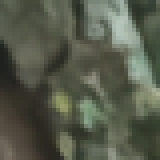}
    }
    \hfill
    \subfigure{
        \includegraphics[width=\wp\linewidth]{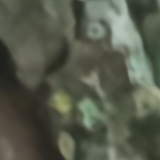}
    }
    \hfill
    \subfigure{
        \includegraphics[width=\wp\linewidth]{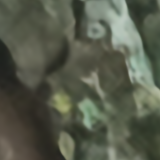}
    }
    \hfill
    \subfigure{
        \includegraphics[width=\wp\linewidth]{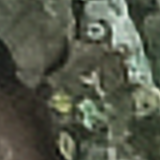}
    }
    \hfill
    \subfigure{
        \includegraphics[width=\wp\linewidth]{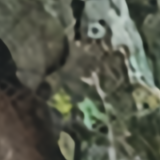}
    }
    \hfill
    \subfigure{
        \includegraphics[width=\wp\linewidth]{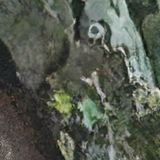}
    }
    \\
    \vspace{-2mm}
    \subfigure{
        \includegraphics[width=\wp\linewidth]{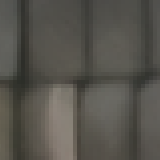}
    }
    \hfill
    \subfigure{
        \includegraphics[width=\wp\linewidth]{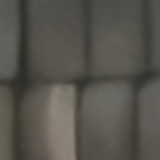}
    }
    \hfill
    \subfigure{
        \includegraphics[width=\wp\linewidth]{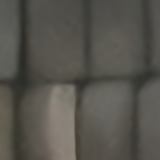}
    }
    \hfill
    \subfigure{
        \includegraphics[width=\wp\linewidth]{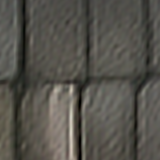}
    }
    \hfill
    \subfigure{
        \includegraphics[width=\wp\linewidth]{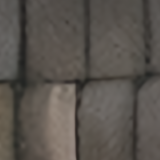}
    }
    \hfill
    \subfigure{
        \includegraphics[width=\wp\linewidth]{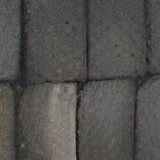}
    }
    \\
    \vspace{-2mm}
    \subfigure{
        \includegraphics[width=\wp\linewidth]{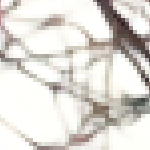}
    }
    \hfill
    \subfigure{
        \includegraphics[width=\wp\linewidth]{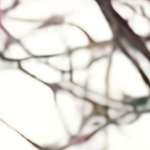}
    }
    \hfill
    \subfigure{
        \includegraphics[width=\wp\linewidth]{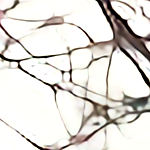}
    }
    \hfill
    \subfigure{
        \includegraphics[width=\wp\linewidth]{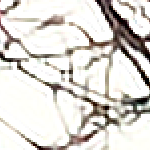}
    }
    \hfill
    \subfigure{
        \includegraphics[width=\wp\linewidth]{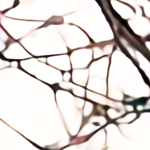}
    }
    \hfill
    \subfigure{
        \includegraphics[width=\wp\linewidth]{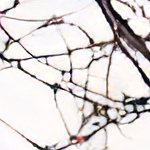}
    }
    \\
    \vspace{-2mm}
    \subfigure{
        \includegraphics[width=\wp\linewidth]{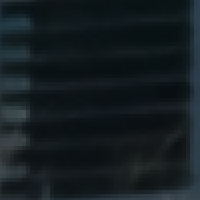}
    }
    \hfill
    \subfigure{
        \includegraphics[width=\wp\linewidth]{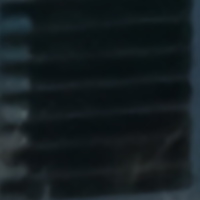}
    }
    \hfill
    \subfigure{
        \includegraphics[width=\wp\linewidth]{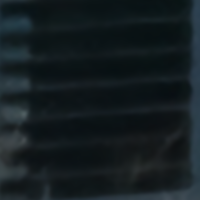}
    }
    \hfill
    \subfigure{
        \includegraphics[width=\wp\linewidth]{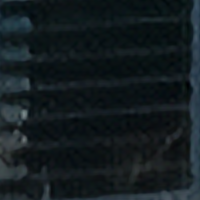}
    }
    \hfill
    \subfigure{
        \includegraphics[width=\wp\linewidth]{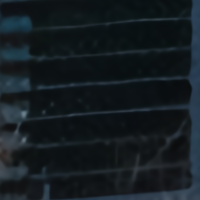}
    }
    \hfill
    \subfigure{
        \includegraphics[width=\wp\linewidth]{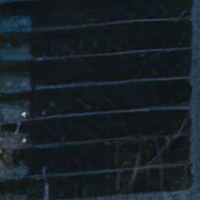}
    }
    \\
    \vspace{-2mm}
    \subfigure{
        \includegraphics[width=\wp\linewidth]{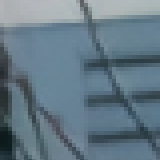}
    }
    \hfill
    \subfigure{
        \includegraphics[width=\wp\linewidth]{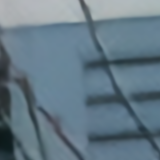}
    }
    \hfill
    \subfigure{
        \includegraphics[width=\wp\linewidth]{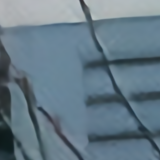}
    }
    \hfill
    \subfigure{
        \includegraphics[width=\wp\linewidth]{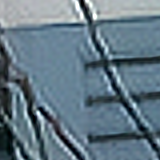}
    }
    \hfill
    \subfigure{
        \includegraphics[width=\wp\linewidth]{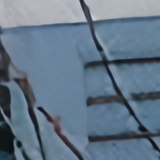}
    }
    \hfill
    \subfigure{
        \includegraphics[width=\wp\linewidth]{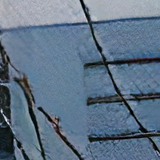}
    }
    \\
    \vspace{-2mm}
    \subfigure{
        \includegraphics[width=\wp\linewidth]{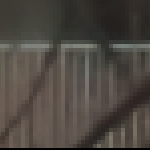}
    }
    \hfill
    \subfigure{
        \includegraphics[width=\wp\linewidth]{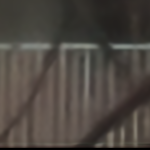}
    }
    \hfill
    \subfigure{
        \includegraphics[width=\wp\linewidth]{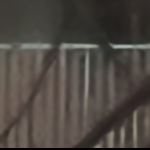}
    }
    \hfill
    \subfigure{
        \includegraphics[width=\wp\linewidth]{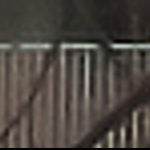}
    }
    \hfill
    \subfigure{
        \includegraphics[width=\wp\linewidth]{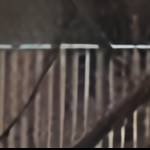}
    }
    \hfill
    \subfigure{
        \includegraphics[width=\wp\linewidth]{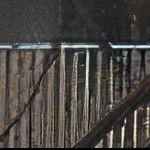}
    }
    \\
    \vspace{-2mm}
    \addtocounter{subfigure}{-36}
    \subfigure[$\img{LR}$~(Input)]{
        \includegraphics[width=\wp\linewidth]{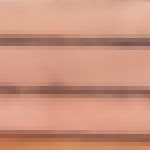}
    }
    \hfill
    \subfigure[RRDB~\cite{sr_esrgan}]{
        \includegraphics[width=\wp\linewidth]{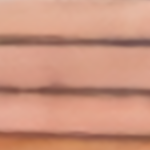}
    }
    \hfill
    \subfigure[IKC~\cite{sr_ikc}]{
        \includegraphics[width=\wp\linewidth]{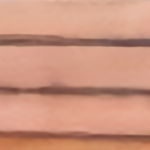}
    }
    \hfill
    \subfigure[\cite{sr_kernelgan}~$+$~\cite{sr_zssr}]{
        \includegraphics[width=\wp\linewidth]{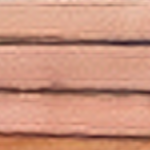}
    }
    \hfill
    \subfigure[\textbf{ADL}~$+$~RRDB~\cite{sr_esrgan}]{
        \includegraphics[width=\wp\linewidth]{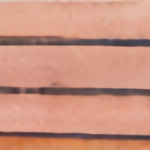}
    }
    \hfill
    \subfigure[\textbf{ADL}~$+ \mathcal{P}$]{
        \includegraphics[width=\wp\linewidth]{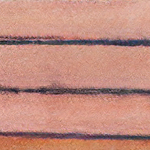}
    }
    \\
    \figspace
    \caption{
        \textbf{Additional Qualitative $\times 4$ SR results on the DPED~\cite{data_dped} dataset.}
        From the top, patches are cropped from the `\emph{DPED-val}' `\emph{14.png},' `\emph{36.png},' `\emph{45.png},' `\emph{58.png} (1),' `\emph{58.png} (2),' `\emph{84.png},' and `\emph{96.png}' respectively.
        We note that $\mathcal{P}$ refers to the perceptual RRDB model trained with \fakeeqref{9} in our main manuscript.
        Compared to the other methods, our approaches (ADL + RRDB and ADL + $\mathcal{P}$) reconstruct sharper edges and detailed structures from given real-world LR images.
    }
    \label{fig:more_dped}
    \figxspace
\end{figure*}

%% file: main_arXiv.bbl
\begin{thebibliography}{10}

\bibitem{sr_srgan}
C.~Ledig, L.~Theis, F.~Huszar, J.~Caballero, A.~Cunningham, A.~Acosta,
  A.~Aitken, A.~Tejani, J.~Totz, Z.~Wang, and W.~Shi, ``Photo-realistic single
  image super-resolution using a generative adversarial network,'' in {\em
  CVPR}, 2017.

\bibitem{sr_explorable}
Y.~Bahat and T.~Michaeli, ``Explorable super resolution,'' in {\em CVPR}, 2020.

\bibitem{wronski2019handheld}
B.~Wronski, I.~Garcia-Dorado, M.~Ernst, D.~Kelly, M.~Krainin, C.-K. Liang,
  M.~Levoy, and P.~Milanfar, ``Handheld multi-frame super-resolution,'' {\em
  ACM TOG}, vol.~38, no.~4, pp.~1--18, 2019.

\bibitem{sr_srcnn}
C.~Dong, C.~C. Loy, K.~He, and X.~Tang, ``Image super-resolution using deep
  convolutional networks,'' {\em TPAMI}, 2016.

\bibitem{sr_lapsrn}
W.-S. Lai, J.-B. Huang, N.~Ahuja, and M.-H. Yang, ``Deep laplacian pyramid
  networks for fast and accurate super-resolution,'' in {\em CVPR}, 2017.

\bibitem{sr_edsr}
B.~Lim, S.~Son, H.~Kim, S.~Nah, and K.~M. Lee, ``Enhanced deep residual
  networks for single image super-resolution,'' in {\em CVPR Workshops}, 2017.

\bibitem{data_div2k}
E.~Agustsson and R.~Timofte, ``{NTIRE} 2017 challenge on single image
  super-resolution: Dataset and study,'' in {\em CVPR Workshops}, 2017.

\bibitem{sr_esrgan}
X.~Wang, K.~Yu, S.~Wu, J.~Gu, Y.~Liu, C.~Dong, Y.~Qiao, and C.~Change~Loy,
  ``{ESRGAN:} enhanced super-resolution generative adversarial networks,'' in
  {\em ECCV Workshops}, 2018.

\bibitem{sr_rcan}
Y.~Zhang, K.~Li, K.~Li, L.~Wang, B.~Zhong, and Y.~Fu, ``Image super-resolution
  using very deep residual channel attention networks,'' in {\em ECCV}, 2018.

\bibitem{sr_dbpn}
M.~Haris, G.~Shakhnarovich, and N.~Ukita, ``Deep back-projection networks for
  super-resolution,'' in {\em CVPR}, 2018.

\bibitem{sr_rdn}
Y.~Zhang, Y.~Tian, Y.~Kong, B.~Zhong, and Y.~Fu, ``Residual dense network for
  image super-resolution,'' in {\em CVPR}, 2018.

\bibitem{sr_zllz}
X.~Zhang, Q.~Chen, R.~Ng, and V.~Koltun, ``Zoom to learn, learn to zoom,'' in
  {\em CVPR}, 2019.

\bibitem{sr_camera}
C.~Chen, Z.~Xiong, X.~Tian, Z.-J. Zha, and F.~Wu, ``Camera lens
  super-resolution,'' in {\em CVPR}, 2019.

\bibitem{sr_realworld}
J.~Cai, H.~Zeng, H.~Yong, Z.~Cao, and L.~Zhang, ``Toward real-world single
  image super-resolution: A new benchmark and a new model,'' in {\em ICCV},
  2019.

\bibitem{sr_cdc}
P.~Wei, Z.~Xie, H.~Lu, Z.~Zhan, Q.~Ye, W.~Zuo, and L.~Lin, ``Component
  divide-and-conquer for real-world image super-resolution,'' in {\em ECCV},
  2020.

\bibitem{sr_srmd}
K.~Zhang, W.~Zuo, and L.~Zhang, ``Learning a single convolutional
  super-resolution network for multiple degradations,'' in {\em CVPR}, 2018.

\bibitem{sr_ikc}
J.~Gu, H.~Lu, W.~Zuo, and C.~Dong, ``Blind super-resolution with iterative
  kernel correction,'' in {\em CVPR}, 2019.

\bibitem{sr_blindsr}
V.~Cornill{\`e}re, A.~Djelouah, W.~Yifan, O.~Sorkine-Hornung, and C.~Schroers,
  ``Blind image super-resolution with spatially variant degradations,'' {\em
  ACM Transactions on Graphics (TOG)}, vol.~38, no.~6, pp.~1--13, 2019.

\bibitem{sr_kernelgan}
S.~Bell-Kligler, A.~Shocher, and M.~Irani, ``Blind super-resolution kernel
  estimation using an internal-gan,'' in {\em NeurIPS}, 2019.

\bibitem{sr_zssr}
A.~Shocher, N.~Cohen, and M.~Irani, ``“{Z}ero-{S}hot” super-resolution
  using deep internal learning,'' in {\em CVPR}, 2018.

\bibitem{sr_deg}
T.~Zhao, W.~Ren, C.~Zhang, D.~Ren, and Q.~Hu, ``Unsupervised degradation
  learning for single image super-resolution,'' {\em arXiv}, 2018.

\bibitem{sr_unpaired_pesudo}
S.~Maeda, ``Unpaired image super-resolution using pseudo-supervision,'' in {\em
  CVPR}, 2020.

\bibitem{sr_unsupervised}
A.~Lugmayr, M.~Danelljan, and R.~Timofte, ``Unsupervised learning for
  real-world super-resolution,'' {\em arXiv}, 2019.

\bibitem{sr_gandeg}
A.~Bulat, J.~Yang, and G.~Tzimiropoulos, ``To learn image super-resolution, use
  a {GAN} to learn how to do image degradation first,'' in {\em ECCV}, 2018.

\bibitem{gans}
I.~Goodfellow, J.~Pouget-Abadie, M.~Mirza, B.~Xu, D.~Warde-Farley, S.~Ozair,
  A.~Courville, and Y.~Bengio, ``Generative adversarial nets,'' in {\em NIPS},
  2014.

\bibitem{sr_vdsr}
J.~Kim, J.~K. Lee, and K.~M. Lee, ``Accurate image super-resolution using very
  deep convolutional networks,'' in {\em CVPR}, 2016.

\bibitem{sr_carn}
N.~Ahn, B.~Kang, and K.-A. Sohn, ``Fast, accurate, and lightweight
  super-resolution with cascading residual network,'' in {\em ECCV}, 2018.

\bibitem{sr_msrn}
J.~Li, F.~Fang, K.~Mei, and G.~Zhang, ``Multi-scale residual network for image
  super-resolution,'' in {\em ECCV}, 2018.

\bibitem{sr_espcn}
W.~Shi, J.~Caballero, F.~Huszar, J.~Totz, A.~P. Aitken, R.~Bishop, D.~Rueckert,
  and Z.~Wang, ``Real-time single image and video super-resolution using an
  efficient sub-pixel convolutional neural network,'' in {\em CVPR}, 2016.

\bibitem{lai2018fast}
W.-S. Lai, J.-B. Huang, N.~Ahuja, and M.-H. Yang, ``Fast and accurate image
  super-resolution with deep laplacian pyramid networks,'' {\em TPAMI},
  vol.~41, no.~11, pp.~2599--2613, 2018.

\bibitem{wang2018fully}
Y.~Wang, F.~Perazzi, B.~McWilliams, A.~Sorkine-Hornung, O.~Sorkine-Hornung, and
  C.~Schroers, ``A fully progressive approach to single-image
  super-resolution,'' in {\em CVPRW}, 2018.

\bibitem{sr_dense}
T.~Tong, G.~Li, X.~Liu, and Q.~Gao, ``Image super-resolution using dense skip
  connections,'' in {\em ICCV}, 2017.

\bibitem{ir_rdn}
Y.~Zhang, Y.~Tian, Y.~Kong, B.~Zhong, and Y.~Fu, ``Residual dense network for
  image restoration,'' {\em TPAMI}, 2020.

\bibitem{sr_drcn}
J.~Kim, J.~K. Lee, and K.~M. Lee, ``Deeply-recursive convolutional network for
  image super-resolution,'' in {\em CVPR}, 2016.

\bibitem{sr_ebrn}
Y.~Qiu, R.~Wang, D.~Tao, and J.~Cheng, ``Embedded block residual network: A
  recursive restoration model for single-image super-resolution,'' in {\em
  ICCV}, 2019.

\bibitem{sr_feedback}
Z.~Li, J.~Yang, Z.~Liu, X.~Yang, G.~Jeon, and W.~Wu, ``Feedback network for
  image super-resolution,'' in {\em CVPR}, 2019.

\bibitem{sr_san}
T.~Dai, J.~Cai, Y.~Zhang, S.-T. Xia, and L.~Zhang, ``Second-order attention
  network for single image super-resolution,'' in {\em CVPR}, 2019.

\bibitem{sr_han}
B.~Niu, W.~Wen, W.~Ren, X.~Zhang, L.~Yang, S.~Wang, K.~Zhang, X.~Cao, and
  H.~Shen, ``Single image super-resolution via a holistic attention network,''
  in {\em ECCV}, 2020.

\bibitem{sr_idn}
Z.~Hui, X.~Wang, and X.~Gao, ``Fast and accurate single image super-resolution
  via information distillation network,'' in {\em CVPR}, 2018.

\bibitem{sr_pi}
W.~Lee, J.~Lee, D.~Kim, and B.~Ham, ``Learning with privileged information for
  efficient image super-resolution,'' in {\em ECCV}, 2020.

\bibitem{sr_lattice}
X.~Luo, Y.~Xie, Y.~Zhang, Y.~Qu, C.~Li, and Y.~Fu, ``Lattice{N}et: Towards
  lightweight image super-resolution with lattice block,'' in {\em ECCV}, 2020.

\bibitem{others_perceptual}
J.~Johnson, A.~Alahi, and L.~Fei-Fei, ``Perceptual losses for real-time style
  transfer and super-resolution,'' in {\em ECCV}, 2016.

\bibitem{sr_enhancenet}
M.~S.~M. Sajjadi, B.~Scholkopf, and M.~Hirsch, ``Enhance{N}et: Single image
  super-resolution through automated texture synthesis,'' in {\em ICCV}, 2017.

\bibitem{per_contextual}
R.~Mechrez, I.~Talmi, and L.~Zelnik-Manor, ``The contextual loss for image
  transformation with non-aligned data,'' in {\em ECCV}, 2018.

\bibitem{feat_deep}
R.~Zhang, P.~Isola, A.~A. Efros, E.~Shechtman, and O.~Wang, ``The unreasonable
  effectiveness of deep features as a perceptual metric,'' in {\em CVPR}, 2018.

\bibitem{sr_rank}
W.~Zhang, Y.~Liu, C.~Dong, and Y.~Qiao, ``Rank{SRGAN}: Generative adversarial
  networks with ranker for image super-resolution,'' in {\em ICCV}, 2019.

\bibitem{sr_srobb}
M.~S. Rad, B.~Bozorgtabar, U.-V. Marti, M.~Basler, H.~K. Ekenel, and J.-P.
  Thiran, ``{SROBB}: Targeted perceptual loss for single image
  super-resolution,'' in {\em ICCV}, 2019.

\bibitem{sr_correction}
S.~A. Hussein, T.~Tirer, and R.~Giryes, ``Correction filter for single image
  super-resolution: Robustifying off-the-shelf deep super-resolvers,'' in {\em
  CVPR}, 2020.

\bibitem{sr_bsrgan}
K.~Zhang, J.~Liang, L.~Van~Gool, and R.~Timofte, ``Designing a practical
  degradation model for deep blind image super-resolution,'' {\em arXiv}, 2021.

\bibitem{tsr_blindsr}
T.~Michaeli and M.~Irani, ``Nonparametric blind super-resolution,'' in {\em
  CVPR}, 2013.

\bibitem{rs_deblurring_l0}
J.~Pan, Z.~Hu, Z.~Su, and M.-H. Yang, ``Deblurring text images via
  l0-regularized intensity and gradient prior,'' in {\em CVPR}, 2014.

\bibitem{rs_usrnet}
K.~Zhang, L.~Val~Gool, and R.~Timofte, ``Deep unfolding network for image
  super-resolution,'' in {\em CVPR}, 2020.

\bibitem{sr_udvd}
Y.-S. Xu, S.-Y.~R. Tseng, Y.~Tseng, H.-K. Kuo, and Y.-M. Tsai, ``Unified
  dynamic convolutional network for super-resolution with variational
  degradations,'' in {\em CVPR}, 2020.

\bibitem{sr_kmsr}
R.~Zhou and S.~Susstrunk, ``Kernel modeling super-resolution on real
  low-resolution images,'' in {\em ICCV}, 2019.

\bibitem{others_cycle}
J.-Y. Zhu, T.~Park, P.~Isola, and A.~A. Efros, ``Unpaired image-to-image
  translation using cycle-consistent adversarial networks,'' in {\em ICCV},
  2017.

\bibitem{sr_rwsr}
X.~Ji, Y.~Cao, Y.~Tai, C.~Wang, J.~Li, and F.~Huang, ``Real-world
  super-resolution via kernel estimation and noise injection,'' in {\em CVPRW},
  2020.

\bibitem{sr_real_raw}
X.~Xu, Y.~Ma, and W.~Sun, ``Towards real scene super-resolution with raw
  images,'' in {\em CVPR}, 2019.

\bibitem{gans_dc}
A.~Radford, L.~Metz, and S.~Chintala, ``Unsupervised representation learning
  with deep convolutional generative adversarial networks,'' {\em arXiv}, 2015.

\bibitem{net_vgg}
K.~Simonyan and A.~Zisserman, ``Very deep convolutional networks for
  large-scale image recognition,'' in {\em ICLR}, 2015.

\bibitem{others_pix2pix}
P.~Isola, J.-Y. Zhu, T.~Zhou, and A.~A. Efros, ``Image-to-{I}mage translation
  with conditional adversarial networks,'' in {\em CVPR}, 2017.

\bibitem{others_in}
D.~Ulyanov, A.~Vedaldi, and V.~Lempitsky, ``Instance normalization: The missing
  ingredient for fast stylization,'' {\em arXiv}, 2016.

\bibitem{rs_deblur_mh}
Z.~Hu and M.-H. Yang, ``Good regions to deblur,'' in {\em ECCV}, 2012.

\bibitem{data_dped}
A.~Ignatov, N.~Kobyshev, R.~Timofte, K.~Vanhoey, and L.~Van~Gool,
  ``{DSLR}-quality photos on mobile devices with deep convolutional networks,''
  in {\em ICCV}, 2017.

\bibitem{rl_restore}
K.~Yu, C.~Dong, L.~Lin, and C.~Change~Loy, ``Crafting a toolchain for image
  restoration by deep reinforcement learning,'' in {\em CVPR}, 2018.

\bibitem{net_residual}
K.~He, X.~Zhang, S.~Ren, and J.~Sun, ``Deep residual learning for image
  recognition,'' in {\em CVPR}, 2016.

\bibitem{others_adam}
D.~P. Kingma and J.~Ba, ``Adam: A method for stochastic optimization,'' {\em
  arXiv}, 2014.

\end{thebibliography}
